\documentclass[english,pra, eqsecnum, reprint, superscriptaddress, showkeys, showpacs, nofootinbib, raggedbottom]{revtex4-1}
\usepackage[T1]{fontenc}
\usepackage[latin9]{inputenc}
\usepackage{xcolor}
\usepackage{pdfcolmk}
\usepackage{booktabs}
\usepackage{units}
\usepackage{mathrsfs}
\usepackage{amsmath}
\usepackage{amsthm}
\usepackage{amssymb}
\usepackage{graphicx}
\usepackage{wasysym}
\PassOptionsToPackage{normalem}{ulem}
\usepackage{ulem}

\makeatletter

\providecommand{\tabularnewline}{\\}
\providecolor{lyxadded}{rgb}{1,0,0}
\providecolor{lyxdeleted}{rgb}{1,0,0}

\usepackage{graphicx,url,layouts,xcolor}
\definecolor{blueviolet}{rgb}{0.2, 0.2, 0.6}
\usepackage{etoolbox} 
\usepackage[pdftex,          
    bookmarks=false,          
    colorlinks=true,
    allcolors=blueviolet,
    pdfstartview={FitH},  
    ]{hyperref}
\usepackage{multirow}

\makeatother

\usepackage{babel}
\begin{document}
\global\long\def\half{\frac{1}{2}}
\global\long\def\a{\alpha}
\global\long\def\b{\beta}
\global\long\def\g{\gamma}
\global\long\def\c{\chi}
\global\long\def\d{\delta}
\global\long\def\o{\omega}
\global\long\def\m{\mu}
\global\long\def\s{\sigma}
\global\long\def\n{\nu}
\global\long\def\z{\zeta}
\global\long\def\l{\lambda}
\global\long\def\k{\kappa}
\global\long\def\x{\chi}
\global\long\def\r{\rho}
\global\long\def\t{\theta}
\global\long\def\G{\Gamma}
\global\long\def\D{\Delta}
\global\long\def\O{\Omega}
\global\long\def\pr{\prime}
\global\long\def\k{\kappa}

\global\long\def\dg{\dagger}
\global\long\def\dgt{\ddagger}
\global\long\def\tr{\text{Tr}}
\global\long\def\Tr{\textsc{Tr}}
\global\long\def\id{\mathcal{I}}
\global\long\def\e{\epsilon}
\global\long\def\nb{\bar{n}}
\global\long\def\ph{\hat{n}}
\global\long\def\aa{\hat{a}}
\global\long\def\bb{\hat{b}}
\global\long\def\al{\mathsf{A}}
\global\long\def\bo{\mathsf{B}}

\global\long\def\rgkp{\text{AGKP}}
\global\long\def\A{\mathcal{A}}
\global\long\def\L{\mathcal{N}}
\global\long\def\K{\mathcal{K}}
\global\long\def\KP{\dot{\mathcal{K}}}
\global\long\def\kp{\dot{K}}
\global\long\def\H{\mathcal{H}}
\global\long\def\S{\mathcal{S}}
\global\long\def\R{\mathcal{R}}
\global\long\def\E{\mathcal{E}}
\global\long\def\bra{\langle}
\global\long\def\ket{\rangle}
\global\long\def\br{\bra\!\bra}
\global\long\def\ke{\ket\!\ket}

\global\long\def\pc{P_{\!\textnormal{\ensuremath{\mathtt{cat}}}}}
\global\long\def\pl{P_{\!\textnormal{\ensuremath{\mathtt{gkp}}}}}
\global\long\def\pb{P_{\textnormal{\ensuremath{\mathtt{bin}}}}}
\global\long\def\pg{P_{\!\textnormal{\ensuremath{\mathtt{gkps}}}}}
\global\long\def\pn{P_{\!\textnormal{\ensuremath{\mathtt{num}}}}}
\global\long\def\pp{P_{\!\textnormal{\ensuremath{\mathtt{code}}}}}
\global\long\def\gkp{\textnormal{\ensuremath{\mathtt{gkps}}}}
\global\long\def\cat{\textnormal{\ensuremath{\mathtt{cat}}}}
\global\long\def\cc{\textnormal{\ensuremath{\mathtt{code}}}}
\global\long\def\pt{\textnormal{\ensuremath{\mathtt{perm}}}}
\global\long\def\pa{\textnormal{\ensuremath{\mathtt{perm}}}^{\prime}}
\global\long\def\tw{\textnormal{\ensuremath{\mathtt{bin2}}}}
\global\long\def\noon{\textnormal{\ensuremath{\mathtt{noon}}}}
\global\long\def\sh{\textnormal{\ensuremath{\mathtt{leung}}}}
\global\long\def\num{\textnormal{\ensuremath{\mathtt{num}}}}
\global\long\def\bin{\textnormal{\ensuremath{\mathtt{bin}}}}
\global\long\def\lat{\textnormal{\ensuremath{\mathtt{gkp}}}}
\global\long\def\oone{\textnormal{\ensuremath{\mathtt{one}}}}
\global\long\def\ttwo{\textnormal{\ensuremath{\mathtt{two}}}}
\global\long\def\xx{\hat{x}}
\global\long\def\ha{D_{\E}}
\global\long\def\fe{F_{\E}}
\global\long\def\fet{F_{\widetilde{\E}}}
\global\long\def\lt{\widetilde{\L}}
\global\long\def\fq{F_{\E}^{\text{QR}}}
\global\long\def\frgkp{F_{\E}^{\text{AGKP}}}
\global\long\def\qr{\text{QR}}
\global\long\def\Z{\mathbb{Z}}
\global\long\def\kkk#1{\left|#1\right\rangle }
\global\long\def\nmax{N_{\text{max}}}
\global\long\def\Re{\text{Re}}
\global\long\def\Im{\text{Im}}
\global\long\def\lp{\ell^{\prime}}
\global\long\def\su{\mathfrak{su}}
\global\long\def\ee{\textsf{E}}
\global\long\def\err{\mathscr{E}}
\global\long\def\h{\mathcal{F}_{\text{0}}}
\global\long\def\ho{\mathcal{F}_{1}}

\DeclareRobustCommand{\one}{{\raisebox{.5pt}{\textcircled{\raisebox{-.9pt}{1}}}}}
\DeclareRobustCommand{\two}{{\raisebox{.5pt}{\textcircled{\raisebox{-.9pt}{2}}}}}

\title{Performance and structure of single-mode bosonic codes}

\author{Victor~V.~Albert}

\thanks{Equal contribution.}

\affiliation{Yale Quantum Institute, Departments of Applied Physics and Physics,
Yale University, New Haven, Connecticut 06520, USA}

\affiliation{Walter Burke Institute for Theoretical Physics and Institute for
Quantum Information and Matter, California Institute of Technology,
Pasadena, California 91125, USA}

\author{Kyungjoo~Noh}

\thanks{Equal contribution.}

\affiliation{Yale Quantum Institute, Departments of Applied Physics and Physics,
Yale University, New Haven, Connecticut 06520, USA}

\author{Kasper~Duivenvoorden}

\thanks{Equal contribution.}

\affiliation{JARA Institute for Quantum Information, RWTH Aachen University, Aachen
52056, Germany}

\author{Dylan~J.~Young}

\affiliation{Yale Quantum Institute, Departments of Applied Physics and Physics,
Yale University, New Haven, Connecticut 06520, USA}

\author{R.~T.~Brierley}

\affiliation{Yale Quantum Institute, Departments of Applied Physics and Physics,
Yale University, New Haven, Connecticut 06520, USA}

\author{Philip~Reinhold}

\affiliation{Yale Quantum Institute, Departments of Applied Physics and Physics,
Yale University, New Haven, Connecticut 06520, USA}

\author{Christophe~Vuillot}

\affiliation{JARA Institute for Quantum Information, RWTH Aachen University, Aachen
52056, Germany}

\author{Linshu~Li}

\affiliation{Yale Quantum Institute, Departments of Applied Physics and Physics,
Yale University, New Haven, Connecticut 06520, USA}

\author{Chao~Shen}

\affiliation{Yale Quantum Institute, Departments of Applied Physics and Physics,
Yale University, New Haven, Connecticut 06520, USA}

\author{S.~M.~Girvin}

\affiliation{Yale Quantum Institute, Departments of Applied Physics and Physics,
Yale University, New Haven, Connecticut 06520, USA}

\author{Barbara~M.~Terhal}

\affiliation{JARA Institute for Quantum Information, RWTH Aachen University, Aachen
52056, Germany}

\author{Liang~Jiang}

\affiliation{Yale Quantum Institute, Departments of Applied Physics and Physics,
Yale University, New Haven, Connecticut 06520, USA}
\begin{abstract}
The early Gottesman, Kitaev, and Preskill (GKP) proposal for encoding
a qubit in an oscillator has recently been followed by cat- and binomial-code
proposals. Numerically optimized codes have also been proposed, and
we introduce new codes of this type here. These codes have yet to
be compared using the same error model; we provide such a comparison
by determining the entanglement fidelity of all codes with respect
to the bosonic pure-loss channel (i.e., photon loss) after the optimal
recovery operation. We then compare achievable communication rates
of the combined encoding-error-recovery channel by calculating the
channel's hashing bound for each code. Cat and binomial codes perform
similarly, with binomial codes outperforming cat codes at small loss
rates. Despite not being designed to protect against the pure-loss
channel, GKP codes significantly outperform all other codes for most
values of the loss rate. We show that the performance of GKP and some
binomial codes increases monotonically with increasing average photon
number of the codes. In order to corroborate our numerical evidence
of the cat/binomial/GKP order of performance occurring at small loss
rates, we analytically evaluate the quantum error-correction conditions
of those codes. For GKP codes, we find an essential singularity in
the entanglement fidelity in the limit of vanishing loss rate. In
addition to comparing the codes, we draw parallels between binomial
codes and discrete-variable systems. First, we characterize one- and
two-mode binomial as well as multi-qubit permutation-invariant codes
in terms of spin-coherent states. Such a characterization allows us
to introduce check operators and error-correction procedures for binomial
codes. Second, we introduce a generalization of spin-coherent states,
extending our characterization to qudit binomial codes and yielding
a new multi-qudit code.
\end{abstract}


\date{\today}

\maketitle
\begin{figure*}[!t]
\includegraphics[width=1\textwidth]{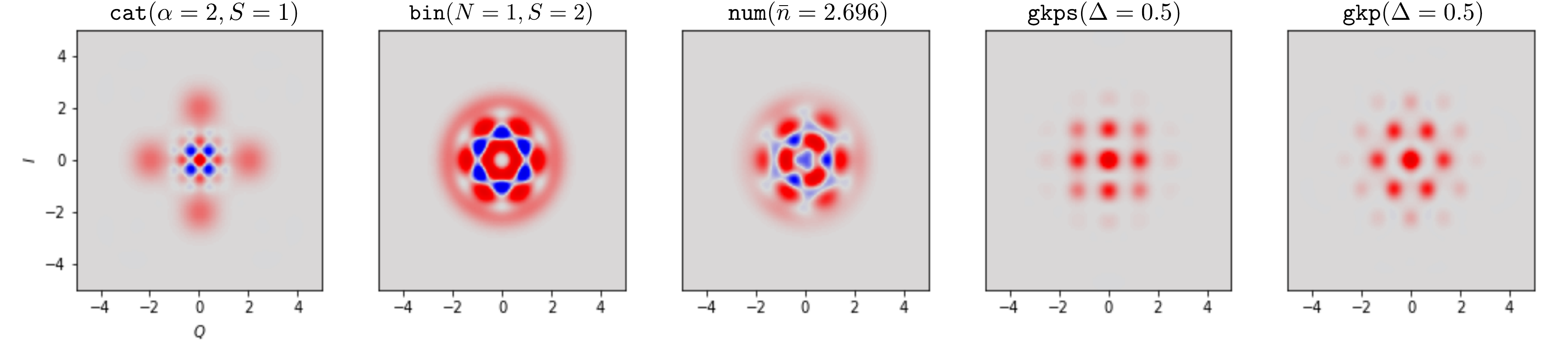}\caption{\label{f:codes}Wigner function plots for maximally mixed logical
states $\protect\half\protect\pp$ for $\protect\cc$ being $\protect\cat$
(\ref{eq:cat}), $\protect\bin$ (\ref{eq:bin}), $\protect\num$
(Appx. \ref{sec:details}), $\protect\gkp$ (\ref{eq:gkpcoherent}),
and $\protect\lat$ (\ref{eq:gkp}), evaluated for given values of
the respective parameters of the codes. On the axes, $Q=\protect\half\protect\bra\protect\aa+\protect\aa^{\protect\dg}\protect\ket$
and $I=\frac{i}{2}\protect\bra\protect\aa^{\protect\dg}-\protect\aa\protect\ket$;
color scales are not the same for all plots.}
\end{figure*}

\section{Introduction and problem setup}

Continuous-variable (CV) systems \cite{Braunstein2005,cvbook,Weedbrook2012,serafinibook}
continue to gain applications in quantum information processing and
communication. The fundamental ``moving part'' of discrete-variable
(DV) systems is one physical qubit, and one has to have a multitude
of such qubits to construct a reliable logical qubit. By contrast,
one may cleverly utilize the infinite-dimensional space of an oscillator
or mode \textemdash{} the fundamental ``moving part'' of CV systems
\textemdash{} in order to realize a comparably reliable logical qubit
out of fewer moving parts. While many current linear-optical CV encodings
use two modes per qubit in a ``dual-rail'' scheme \cite{Chuang1995,Kok2007}
and CV logical qubits may consist of several modes in the long-term,
here we focus on a single mode since its theoretical limitations are
not yet well-understood and since it is useful for communication.

There have been several error-correcting CV encoding schemes proposed
to-date, formulated in terms of superpositions of either position/momentum
eigenstates \cite{Lloyd1998,Braunstein1998,Gottesman2001,Menicucci2014,Hayden2016,Ketterer2016},
coherent states \cite{Cochrane1999,Niset2008,Leghtas2013b,Lacerda2016,paircat},
or Fock states \cite{Chuang1997,Knill2001,Ralph2005,Wasilewski2007,Bergmann2016a,bin,Niu2017}
(see also other hybrid CV-DV schemes \cite{Lee2013,Kapit2016}). Besides
the rich variety of quantum codes, there are two prevailing CV noise
models: classical (i.e., Gaussian or displacement) noise and pure
loss (more generally, thermal noise) \cite{cvbook}. Classical noise
is modeled by a distribution of phase-space \textit{displacements}
while pure loss contracts phase space to the vacuum and is best understood
in terms of \textit{losses}. Due to the differing physical pictures
and mathematical formalisms of these noise models, codes designed
to protect against one may or may not protect against the other. However,
it is often difficult to rigorously prove protection from noise against
which a code wasn't designed to protect. It is also difficult to study
CV codes using the conventional stabilizer formalism because the noise
model operators are not as well-behaved. This manuscript closes these
gaps by applying tools from qubit-based quantum error-correction to
CV codes which were not analyzed in this manner before.

\subsection{Codes and error model\label{subsec:Codes-and-error}}

The code classes we consider are
\begin{equation}
\cc\in\{\cat,\bin,\num,\gkp,\lat\}\,.\label{eq:codes}
\end{equation}
The logical states for the first code class \textemdash{} the $\cat$-codes
(\ref{eq:cat}) \textemdash{} consist of superpositions of coherent
states which are evenly distributed around a circle in phase space
\cite{Cochrane1999,Leghtas2013b,cats}. The second class of codes,
the recently developed $\bin$omial codes (\ref{eq:bin}), are designed
to protect exactly against errors consisting of powers of raising/lowering
operators up to some maximum order \cite{bin}. Here we show that
$\bin$ codes are spin-coherent states embedded in an oscillator.
We also include $\num$erically optimized codes (some from Ref.~\cite{bin}
and the rest developed here) that were obtained by minimizing the
photon number of the code states subject to the constraints of protecting
exactly against the first few errors powers of the lowering operator.
The last class consists of GKP codes \cite{Gottesman2001} which are
the $+1$ eigenspace of two commuting phase-space displacement operators;
since the codespace is invariant under both displacements, the codespace
makes a lattice in phase space. The $\gkp$ (\ref{eq:gkp}) class,
with $\texttt{s}$ standing for square, corresponds to a square lattice.
The $\lat$ (\ref{eq:gkpnum}) class consists of GKP codes built out
of both the square and other non-rectangular lattices as well as codes
whose lattice is shifted by half a lattice spacing from the origin,
thus subsuming the $\gkp$ class. The $\gkp$ codes are presented
separately in order to quantify any advantages of other lattices.

A single-mode qubit CV code is a two-dimensional subspace of the bosonic
Hilbert space picked to be able to protect quantum information against
errors.\footnote{While there is a more general definition for multi-qubit subsystem
codes \cite{Kribs}, we stick with the simpler definition since to
our knowledge there are no single-mode CV subsystem codes. } It is unambiguously represented by the corresponding orthogonal projection
onto the subspace, 
\begin{equation}
\pp=|0_{\cc}\ket\bra0_{\cc}|+|1_{\cc}\ket\bra1_{\cc}|\,,
\end{equation}
where $\cc$ is picked from Eq.~(\ref{eq:codes}) and $|\m_{\cc}\ket$
(for $\m\in\{0,1\}$) is any orthonormal basis for the code subspace.
The maximally mixed state $\half\pp$ thus provides a concise basis-independent
fingerprint for each code; we plot the Wigner function of this state
in Fig.~\ref{f:codes}.

We deal exclusively with codes representing a single qubit and are
guided by the question:
\[
\text{\textit{Which code best protects against the pure-loss channel?}}
\]
We are interested in the pure-loss channel because it is a model for
broadband-line and free-space communication \cite{Weedbrook2012}
and it is the most common \textit{incoherent} error process in optical
and microwave cavities \cite{bin}. The second most common error is
cavity dephasing, which is caused by fluctuations in the cavity frequency.
Optical cavities have to be actively stabilized to fix the frequency,
but the effects of such fluctuations are small relative to effects
of energy loss, particularly in microwave cavities. There are also
other \textit{coherent} error processes, such as a Kerr nonlinearity
\cite{Ofek2016}, which we briefly consider in Sec.~\ref{sec:Addition-of-a}.

The pure-loss bosonic channel (also known as bosonic amplitude damping
or, more simply, as the lossy channel \cite{cvbook}) is Markovian:
$\L=\exp(\chi{\cal D})$ with superoperator ${\cal D}(\cdot)=\aa\cdot\aa^{\dg}-\half\{\ph,\cdot\}$,
where $\aa/\aa^{\dg}$ is the lowering/raising operator for the bosonic
mode and $\ph\equiv\aa^{\dg}\aa$. The dimensionless damping parameter
equals $\chi=\k t$ for microwave or optical cavities (with excitation
loss rate $\k$ and time interval $t$) or $\chi=l/l_{\text{att}}$
for optical fibers (with fiber length $l$ and attenuation length
$l_{\text{att}}$). It is convenient to use the dimensionless \textit{loss
rate}
\begin{equation}
\g\equiv1-e^{-\chi}\label{eq:gamma}
\end{equation}
to quantify the severity of the error channel, denoted as $\L\equiv\L_{\g}$
from now on. This channel can be expressed via unraveling \cite{Ueda1989,Lee1994,Chuang1997,Vitali1998}
or Lie-algebraic \cite{klimov_book} techniques in the Kraus representation,
$\L_{\g}(\cdot)=\sum_{\ell=0}^{\infty}E_{\ell}\cdot E_{\ell}^{\dg}$,
with Kraus operators
\begin{equation}
E_{\ell}\equiv\left(\frac{\g}{1-\g}\right)^{\ell/2}\frac{\aa^{\ell}}{\sqrt{\ell!}}\left(1-\g\right)^{\ph/2}\,.\label{eq:krausloss}
\end{equation}
To leading order in $\g$, expansions of the first two Kraus operators
suffice,
\begin{equation}
E_{0}=I-\half\g\ph\,\,\,\,\,\,\,\,\text{and}\,\,\,\,\,\,\,\,E_{1}=\sqrt{\g}\aa\,.\label{eq:kraussmalldelta}
\end{equation}
This channel can also be derived by introducing an environment mode
$\bb$, coupling our oscillator with the vacuum state of this mode
via a beam-splitter interaction
\begin{equation}
\aa\rightarrow\sqrt{1-\g}\,\aa+\sqrt{\g}\,\bb\,,\label{eq:env}
\end{equation}
and tracing out the $\bb$-mode \cite{Ivan2011,Weedbrook2012}. The
$K$-mode channel $\L_{\g}^{\otimes K}$ reduces to the multi-qubit
amplitude damping channel when restricted to the first two Fock states
of each mode and reduces to the erasure channel when restricted to
the single-excitation subspace. 

Notice that this channel does not contain the identity as a Kraus
operator for $\g\neq0$. This is due to the \textit{backaction} or
\textit{damping} term $\left(1-\g\right)^{\ph/2}$ in Eq.~(\ref{eq:krausloss}),
which reduces the probabilities of being in Fock states $|n>0\ket$
such that the only state remaining in the $\g\rightarrow1$ limit
is the vacuum Fock state $|n=0\ket$. Thus, when no losses are recorded
(i.e., if $E_{\ell>0}$ has not yet acted on the state), there is
still a redistribution of probability caused by $E_{0}$. Colloquially
for $\g>0$, if one hasn't lost any photons, then one likely did not
have many photons to begin with.

\subsection{Channel fidelity and recovery optimization}

The combined quantum channel we consider consists of an encoding step
$\S_{\cc}$, action of the pure-loss channel $\L_{\g}$, a recovery
channel $\R$, and a perfect decoding step $\S_{\cc}^{-1}$. The encoding
step maps the quantum information from the qubit \textit{source space}
$\al$ \cite{Fletcher2007} into the code subspace of the bosonic
Hilbert space; this step is represented by $\S_{\cc}$. More precisely,
$\S_{\cc}(\r)=S\r S^{-1}$ where $S=|0_{\cc}\ket\bra0_{\al}|+|1_{\cc}\ket\bra1_{\al}|$,
$\r$ is a qubit density matrix in $\al$, and $|0_{\al}/1_{\al}\ket$
is a basis for $\al$. Therefore, $SS^{-1}=\pp$ and, if $\{|0_{\cc}\ket,|1_{\cc}\ket\}$
are orthonormal, $S^{-1}S=I_{\al}$, the identity on $\al$. The combined
channel 
\begin{equation}
\E\equiv\S_{\cc}^{-1}\circ\R\circ\L_{\g}\circ\S_{\cc}\label{eq:chan}
\end{equation}
thus maps density matrices in $\al$ back to $\al$. In contrast,
$\R\circ\L_{\g}$ is a map from the bosonic space to the code subspace.
The form of $\E$ depends on the $\cc$, the loss rate $\g$, and
the recovery $\R$. The channel can be written in the Kraus representation,
$\E(\cdot)=\sum_{k}A_{k}\cdot A_{k}^{\dg}$, or in the \textit{matrix
or Liouville representation} \textemdash{} as a $4\times4$ matrix
with elements 
\begin{equation}
\E_{kl}={\textstyle \half}\tr\{\s_{k}\E(\s_{\ell})\}\,,\label{eq:matrep}
\end{equation}
using the three Pauli matrices $\s_{1,2,3}$ and identity $I_{\al}\equiv\s_{0}$
(e.g., \cite{Caves1999}, Sec.~2.2). Composition ``$\circ$'' in
Eq.~(\ref{eq:chan}) is equivalent to matrix multiplication in the
matrix representation, so we omit the symbol.

None of the codes we consider protect against all errors in $\L_{\g}$,
so we have to consider approximate quantum error correction \cite{Leung1997,Crepeau2005}.
We compare the codes using the \textit{channel fidelity} $\fe$ \cite{Reimpell2005}
\textemdash{} a specific case of entanglement fidelity \cite{schumacher}
that is motivated as follows. Let the source qubit $\al$ be in a
maximally entangled state with ancillary qubit $\bo$, i.e., $|\varPsi\ket=(|0_{\al}0_{\bo}\ket+|1_{\al}1_{\bo}\ket)/\sqrt{2}$.
Qubit $\bo$ is left alone (i.e., acted on by the identity superoperator
$\id$) while the source qubit is acted on by the channel $\E$ in
Eq.~(\ref{eq:chan}). The channel fidelity $\fe$ is simply the overlap
between the initial state $|\varPsi\ket$ and the final state 
\begin{equation}
\r_{\E}\equiv\E\otimes\id(|\varPsi\ket\bra\varPsi|)\,,\label{eq:choi}
\end{equation}
(which we define to be the Choi matrix\footnote{\label{fn:choi}Note that $\id\otimes\E(|\varPsi\ket\bra\varPsi|)$
is used for the Choi matrix in \cite{Konig2009,tomamichel}. Our convention
\cite{Fletcher2007a,Fletcher2007} yields $\r_{\E}=\sum_{k}|A_{k}\ke\br A_{k}|$
for vectorized Kraus operators $|A_{k}\ke$ of $\E$, but unfortunately
makes Alice and Bob switch places.} of $\E$):
\begin{equation}
\fe\equiv\bra\varPsi|\r_{\E}|\varPsi\ket\,.\label{eq:fid0}
\end{equation}
Remembering that $\tr_{\bo}\{|\varPsi\ket\bra\varPsi|\}$ is the maximally
mixed state $\half I_{\al}$ of qubit $\al$, a few simple manipulations
yield 
\begin{equation}
\fe=\frac{1}{4}\sum_{k=1}^{4}\left|\tr\{A_{k}\}\right|^{2}=\frac{1}{4}\Tr\{\E\}\,,\label{eq:fid}
\end{equation}
where $\Tr$ is the trace in the matrix representation (\ref{eq:matrep}).

Besides clearly being an intrinsic property of $\E$ that is invariant
under unitary rotations, several other properties of $\fe$ make the
quantity both meaningful and practically useful. We first mention
the property that is crucial for our task, listing the remaining properties
in Appx.~\ref{sec:The-many-faces}. It turns out that the recovery
$\R$ which gives the optimal $\fe$ is computable via a semi-definite
program \cite{Audenaert2002} (see also \cite{Reimpell2005,Kosut2009}).
This allows us to quickly obtain the highest possible $\fe$ using
a laptop (given reasonable $\nb_{\cc}$) and without having to design
a recovery for each code. This procedure was applied to the multi-qubit
context by Fletcher, Shor, and Win \cite{Fletcher2007a} (see also
\cite{Fletcher2007} and references therein), and our benchmarking
is in some sense a counterpart to that work in the oscillator context.
From now on and unless otherwise noted, we let the recovery piece
$\R$ of $\E$ (\ref{eq:chan}) be one which gives the highest $\fe$,
given a member of the $\cc$ family and a loss rate $\g$.

\subsection{Outline of this paper}

In Sec.~\ref{sec:Take-home-messages}, we state our main numerical
code comparison results and summarize the supporting analytical calculations.
In Sec.~\ref{sec:The-hashing-bound}, we numerically analyze communication
rates of our codes by calculating the hashing bound of $\E$. In Sec.~\ref{sec:QEC matrix},
we review the quantum error-correction conditions. In Secs.~\ref{sec:Cat-codes},
\ref{sec:bin-codes}, and \ref{sec:GKP-codes}, we calculate these
conditions for the $\cat$, $\bin$, and $\lat$ codes, respectively.
In Sec.~\ref{subsec:Relation-to-spin-coherent}, we characterize
single-qubit $\bin$ codes in terms of spin-coherent states and relate
them to two-mode binomial codes and multi-qubit permutation-invariant
codes. In Sec.~\ref{sec:Additional-nonlinearityfeatures}, we analyze
code performance after a Kerr interaction is added to the pure loss
channel and briefly study the effect of tracking the photon number
parity. We summarize our results and discuss future directions in
Sec.~\ref{sec:Conclusion}.

\section{Take-home messages\label{sec:Take-home-messages}}

Here we summarize the results related to code performance, but start
off by mentioning two caveats to our primary numerical comparison.
Results relating the structure of $\bin$ codes to spin-coherent states
and other multi-qubit codes are summarized in Sec.~\ref{subsec:Code-structure}.

Caveat \one~is that \textit{the encoding, recovery, and decoding
are all assumed perfect}, meaning that there are no other errors besides
$\L_{\g}$ incurred by the state. Therefore, the results of this section
should be interpreted as theoretical bounds on code capabilities and
not as practical suggestions on the best experimental design. Moreover,
optimal recovery procedures are not created equal in the eyes of current
technologies: $\cat$ code error correction has already been performed
\cite{Ofek2016} while $\lat$ states have yet to be realized. We
briefly investigate one additional imperfection in Sec.~\ref{sec:Additional-nonlinearityfeatures}
by including a nonlinearity. There, we also address the consequences
of being able to perfectly track the photon number parity. 

Caveat \two~has to do with how we quantify the ``size'' of the
codes. Namely, \textit{we organize the codes by mean occupation number}
\begin{equation}
\nb_{\cc}\equiv\tr\left\{ \pp\ph\right\} /2\,.\label{eq:nbar}
\end{equation}
While $\nb_{\cc}$ is proportional to the average energy required
to construct a code state, it does not describe the spread or variance
in Fock space, $\s_{\cc}^{2}\equiv\half\tr\{\pp\ph^{2}\}-\nb_{\cc}^{2}$.
While $\cat$ and $\bin$ codes follow approximately Poisson and binomial
distributions in Fock space, respectively, we will show that $\lat$
codes are \textit{geometrically (i.e., thermally) distributed} and
thus have much larger ``tails'' in Fock space at higher $\nb$.
Therefore, for the same $\nb_{\cc}$, $\lat$ codes utilize much more
of the Fock space than $\cat/\bin$ and are therefore ``larger''
(in the same sense that multi-qubit codes constructed out of ten physical
qubits are larger than those constructed out of five). A simple energy
parameter does not quantify such a notion of size.

\begin{figure}[t]
\includegraphics[width=1\columnwidth]{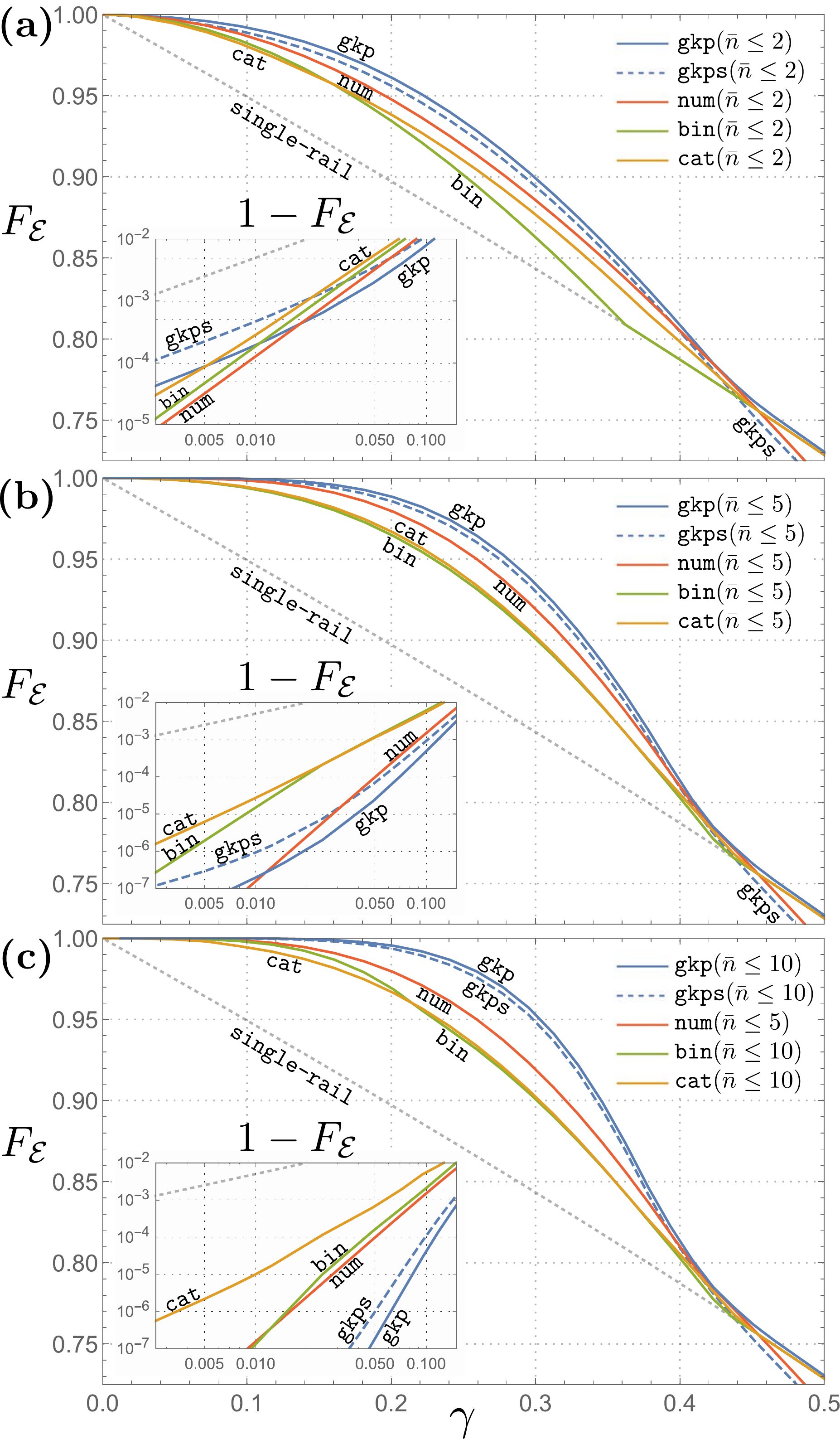}\caption{\label{f:num}Channel fidelity $\protect\fe$ (\ref{eq:fid}) given
an optimal recovery operation and optimized over all instances of
each code given an occupation number constraint $\protect\nb_{\protect\cc}\leq$
2\textbf{ (a)}, 5 \textbf{(b)}, or 10 \textbf{(c)}. The dotted diagonal
line, drawn for reference, is the optimal $\protect\fe$ for \textit{single-rail
encoding} \cite{Kok2007} (whose logical states are the Fock states
$|0\protect\ket,|1\protect\ket$). While $\protect\lat$ codes perform
worse than the other codes for sufficiently small $\protect\g$ (see
insets), they outperform all other codes as $\protect\g$ is increased
despite not being designed to protect against the pure-loss channel.
We were not able to obtain significantly better $\protect\num$ codes
with $\protect\nb_{\protect\num}>5$ due to a large set of parameters
to be optimized over, so red curves in \textbf{(b)} and \textbf{(c)}
are identical. Parameters for all of the codes used are in Table~\ref{tab:numdet}.}
\end{figure}

\subsection{Numerical comparison}

The procedure we use to determine the channel fidelities shown in
Fig.~\ref{f:num} is as follows. Recall that each code family (\ref{eq:codes})
contains multiple instances of codes. For example, a member of the
$\bin$ code family is parameterized by the number of dephasing and
loss errors it can correct ($N$ and $S$, respectively; more details
are in Sec.~\ref{sec:QEC matrix}). For each loss rate $\g$ in Eq.~(\ref{eq:gamma}),
we calculate the optimal $\fe$ for all instances of each code family
subject to the energy constraint that $\nb_{\cc}\leq2,5,10$ {[}shown
in Figs.~\ref{f:num}(a), (b), and (c), respectively{]}. Then, we
pick the highest $\fe$ out of all members of the code family and
plot it in the figure. We repeat for other values of $\g\leq0.5$
\textemdash{} the point at which the one-way channel capacity of $\L_{\g}$
becomes zero (see Sec.~\ref{sec:The-hashing-bound} for details).
As a result of this simultaneous optimization over each code family
and over the recovery for a given member of the family, the code which
gives the highest $\fe$ may change with $\g$ and curves in Fig.~\ref{f:num}
may display discontinuous derivatives. 

Let us first focus on the $\nb_{\cc}\leq2$ case in Fig.~\ref{f:num}(a)
and examine the infidelity ($1-\fe$) shown in the log-plot inset.
For $\g\leq0.025$, specific members of $\num$, $\bin$, and $\cat$
perform the best (in that order), showing similar scaling with $\g$.
All three of these codes were designed to deal with errors $\{I,\aa\}$,
the dominant terms in $E_{\ell=0,1}$ at small $\g$ (\ref{eq:kraussmalldelta}).
The $\num$ and $\bin$ codes show quadratic scaling versus small
$\g$: in a polynomial fit up to order two for $\g\leq0.025$, $c_{0}+c_{1}\g+c_{2}\g^{2}$,
$\num/\bin$ codes have negligible coefficients $c_{0}$, $c_{1}\approx10^{-4}$,
and a $c_{2}$ of $1.3/1.8$, respectively. The $\cat$ codes have
negligible $c_{0}$, $c_{1}\approx10^{-3}$, and $c_{2}\approx2.2$.
Following these codes, $\gkp$ and $\lat$ perform the worst for small
$\g$, underperforming the other codes as $\g\rightarrow0$. This
should be expected since these codes were designed to protect against
small displacements and not loss events. It is also reasonable that
$\lat$ should slightly outperform $\gkp$ due to the idea that non-square
lattices allow for tighter packing than square lattices \cite{Gottesman2001}.
The main unexpected behavior for $\nb_{\cc}\leq2$ occurs for $\g\geq0.025$.
There, we see that $\gkp$ and $\lat$ actually \textit{outperform}
the rest of the codes (this will be discussed later).

We remark here that, for each $\g$, the amplitude $\a$ of the coherent
states making up $\cat$ codes {[}see Fig.~\ref{f:codes}(a){]} for
the optimal $\cat$ code is at a fine-tuned ``sweet spot'' $\a_{\star}(\g)$
which balances the backaction due to the difference in the mean occupation
number of the logical states (significant for small $\a$ but zero
at $\a\rightarrow\infty$) against the probability that a photon will
be lost (zero at $\a\rightarrow0$). We discuss this effect more in
Sec.~\ref{sec:Cat-codes}, noting that it has also been studied elsewhere
\cite{bin,Li2016}.

Continuing to $\nb_{\cc}\leq5$ in Fig.~\ref{f:num}(b), we see substantial
increases in performance for all codes. We list notable infidelities
for selected $\g$ and $\nb_{\cc}\leq5$ in Table~\ref{tab:values}(b).
For example, for the relatively lossy channel having $\g=0.0952$,
there exist codes in \textit{all five families} which have a channel
fidelity higher than 99.4\%. Moreover, all such codes have only five
photons in them on average, so they could be within reach even with
noisy intermediate-scale quantum (NISQ) technologies \cite{Preskill2018}.
At small $\g$ {[}inset of Fig.~\ref{f:num}(b){]}, we again see
polynomial scaling for the $\cat/\bin/\num$ codes, which are able
to deal with loss errors $\{I,\aa,\aa^{2}\}$. We also once again
see $\lat$ outperform the other codes for $\g\geq0.05$ and underperform
as $\g\rightarrow0$. The $\cat$ and $\bin$ code performances are
almost identical, with the exception of small $\g$, where $\bin$
performs slightly better. Expounding on this behavior in Sec.~\ref{subsec:Error-correction-criteria-for},
we show that $\bin$ allows for better error suppression than $\cat$
in certain ranges of $\nb$. One $\num$ code performs the best for
all $\g\leq0.4$ \textemdash{} the code with $\nb_{\num}=4.15$ (see
Appx.~\ref{sec:details}).

\begin{table}
\begin{raggedright}
\textbf{(a) }$1-\fe(\nb\leq2)$
\par\end{raggedright}
\begin{raggedright}
\begin{tabular}{cccccc}
\toprule 
$\g$~~ & $\cat$ & $\bin$ & $\num$ & $\gkp$ & $\lat$\tabularnewline
\midrule
0.01~~ & $4.2\mathrm{e}10^{-4}$ & $2.9\mathrm{e}10^{-4}$ & $2.0\mathrm{e}10^{-4}$ & $6.0\mathrm{e}10^{-4}$ & $2.5\mathrm{e}10^{-4}$\tabularnewline
0.05~~ & $5.3\mathrm{e}10^{-3}$ & $4.3\mathrm{e}10^{-3}$ & $3.1\mathrm{e}10^{-3}$ & $3.4\mathrm{e}10^{-3}$ & $1.9\mathrm{e}10^{-3}$\tabularnewline
0.10~~ & $1.8\mathrm{e}10^{-2}$ & $1.6\mathrm{e}10^{-2}$ & $1.2\mathrm{e}10^{-2}$ & $1.0\mathrm{e}10^{-2}$ & $7.1\mathrm{e}10^{-3}$\tabularnewline
0.20~~ & $6.2\mathrm{e}10^{-2}$ & $6.6\mathrm{e}10^{-2}$ & $5.3\mathrm{e}10^{-2}$ & $4.5\mathrm{e}10^{-2}$ & $3.9\mathrm{e}10^{-2}$\tabularnewline
0.31~~ & $1.3\mathrm{e}10^{-1}$ & $1.5\mathrm{e}10^{-1}$ & $1.2\mathrm{e}10^{-1}$ & $1.2\mathrm{e}10^{-1}$ & $1.1\mathrm{e}10^{-1}$\tabularnewline
\bottomrule
\end{tabular}
\par\end{raggedright}
\begin{raggedright}
\medskip{}
\par\end{raggedright}
\begin{raggedright}
\textbf{(b) }$1-\fe(\nb\leq5)$
\par\end{raggedright}
\begin{raggedright}
\begin{tabular}{cccccc}
\toprule 
$\g$~~ & $\cat$ & $\bin$ & $\num$ & $\gkp$ & $\lat$\tabularnewline
\midrule
0.01~~ & $4.4\mathrm{e}10^{-5}$ & $2.8\mathrm{e}10^{-5}$ & $3.7\mathrm{e}10^{-7}$ & $1.4\mathrm{e}10^{-6}$ & $3.2\mathrm{e}10^{-7}$\tabularnewline
0.05~~ & $1.1\mathrm{e}10^{-3}$ & $1.1\mathrm{e}10^{-3}$ & $8.8\mathrm{e}10^{-5}$ & $6.3\mathrm{e}10^{-5}$ & $2.2\mathrm{e}10^{-5}$\tabularnewline
0.10~~ & $4.9\mathrm{e}10^{-3}$ & $5.4\mathrm{e}10^{-3}$ & $1.2\mathrm{e}10^{-3}$ & $7.6\mathrm{e}10^{-4}$ & $3.9\mathrm{e}10^{-4}$\tabularnewline
0.20~~ & $3.4\mathrm{e}10^{-2}$ & $3.6\mathrm{e}10^{-2}$ & $2.1\mathrm{e}10^{-2}$ & $1.5\mathrm{e}10^{-2}$ & $1.2\mathrm{e}10^{-2}$\tabularnewline
0.31~~ & $1.1\mathrm{e}10^{-1}$ & $1.1\mathrm{e}10^{-1}$ & $9.2\mathrm{e}10^{-2}$ & $8.2\mathrm{e}10^{-2}$ & $7.7\mathrm{e}10^{-2}$\tabularnewline
\bottomrule
\end{tabular}
\par\end{raggedright}
\begin{raggedright}
\medskip{}
\par\end{raggedright}
\begin{raggedright}
\textbf{(c) }$1-\fe(\nb\leq10)$
\par\end{raggedright}
\begin{raggedright}
\begin{tabular}{ccccc}
\toprule 
$\g$~~ & $\cat$ & $\bin$ & $\gkp$ & $\lat$\tabularnewline
\midrule
0.01~~ & $1.7\mathrm{e}10^{-5}$ & $3.7\mathrm{e}10^{-7}$ & $3.0\mathrm{e}10^{-10}$ & $1.0\mathrm{e}10^{-11}$\tabularnewline
0.05~~ & $6.3\mathrm{e}10^{-4}$ & $1.5\mathrm{e}10^{-4}$ & $8.2\mathrm{e}10^{-7}$ & $1.5\mathrm{e}10^{-7}$\tabularnewline
0.10~~ & $4.9\mathrm{e}10^{-3}$ & $1.7\mathrm{e}10^{-3}$ & $7.9\mathrm{e}10^{-5}$ & $2.9\mathrm{e}10^{-5}$\tabularnewline
0.20~~ & $3.4\mathrm{e}10^{-2}$ & $3.1\mathrm{e}10^{-2}$ & $6.5\mathrm{e}10^{-3}$ & $4.6\mathrm{e}10^{-3}$\tabularnewline
0.31~~ & $1.1\mathrm{e}10^{-1}$ & $1.1\mathrm{e}10^{-1}$ & $6.3\mathrm{e}10^{-2}$ & $5.9\mathrm{e}10^{-2}$\tabularnewline
\bottomrule
\end{tabular}
\par\end{raggedright}
\caption{\label{tab:values}Channel infidelity $1-\protect\fe$ from Figs.~\ref{f:num}(a-c)
at selected loss rates $\protect\g$ (with $\protect\g$ rounded to
the nearest hundredth).}
\end{table}

Let us now consider the case $\nb_{\cc}\leq10$ in Fig.~\ref{f:num}(c).
While these codes may be difficult to create and correct experimentally
in the near future, it is nevertheless interesting to see whether
doubling the occupation number constraint allows for any improvements
of the code. The most noticeable difference between $\nb_{\cc}\leq10$
and $\nb_{\cc}\leq5$ is that $\lat$ pulls away from the other codes
for \textit{all values} of $\g$ sampled. While it is believable that
the other codes will still scale more favorably for sufficiently small
$\g$, this occurs only at $\g<0.01$. For larger $\g$, $\lat$ codes
demonstrate $\fe\geq0.99$ even at $\g=0.2$ (see Table~\ref{tab:values}).
Looking at Table~\ref{tab:numdet}, the best codes in those families
for most $\g$ are those which also have $\nb_{\lat}\approx10$. In
other words, $\lat$ performs better with increasing $\nb$. A similar
monotonic increase in performance occurs for subsets $\bin(N,S\approx\xi N)$
of binomial codes (with $\xi$ dependent on $\g$) when the $\nb_{\bin}$
constraints are relaxed (see Sec.~\ref{subsec:Removing-energy-constraints}).
We explain the $\bin$ increase in performance in Secs.~\ref{subsec:Error-correction-criteria-for}-\ref{subsec:Removing-energy-constraints},
revealing that they have a larger set of approximately correctable
errors than previously thought. This behavior is not seen in $\cat$
codes, which do not perform much better than those in Fig.~\ref{f:num}(b)
and work best at some finite $\nb$. This idea that increasing $\nb$
does not lead to better $\cat$ code performance has been observed
before in different contexts \cite{Mirrahimi2016,Zakiprivate,Li2016}.
By contrast, the ideal $\lat$ codes have infinite $\nb$, so it seems
reasonable that increasing $\nb$ should improve performance. These
results support the conjecture that the ordering of the codes with
respect to $\fe$ is $\lat>\bin>\cat$ when there is no energy constraint.

The numerical results show that codes designed to work well at small
$\g$ do not perform well for large $\g$, and vice versa. More specifically,
extensions of the ideas used to correct dominant errors at small $\g$
do not necessarily lead to good codes at larger $\g$. For instance,
the $\cat$ and $\bin$ codes protect exactly from the first few loss
errors by making sure there is adequate \textit{Fock state spacing}
$S$ between the states. As an example, an $S=2$ $\cat/\bin$ code
uses superpositions of Fock states $|0\ket,|6\ket,\cdots$ for $|0_{\cc}\ket$
and $|3\ket,|9\ket,\cdots$ for $|1_{\cc}\ket$. This guarantees that
loss events $E_{\ell=1,2}$, which lower each Fock state by either
1 or 2, do not cause the logical states to overlap with each other.
Both $\cat$ and $\bin$ allow one to increase $S$ arbitrarily, while
$\lat$ codes have $S\in\{0,1\}$, depending on whether their lattice
is shifted from the origin or not. Figure \ref{f:qec} shows that,
for sufficiently large $\g$ and $\nb_{\cc}$, correcting a few errors
exactly with spacing (done by $\cat$ and $\bin$) is not as helpful
as suppressing all errors approximately (done by $\lat$).

\subsection{Analytical results}

To summarize, the family of $\lat$ codes outperforms all of the other
codes for most $\g$, with the exception of small $\g$ (which gets
even smaller as the energy constraint is loosened). We see similar
behavior analyzing the optimal codes from an information-theoretic
perspective in Sec.~\ref{sec:The-hashing-bound}. Since all of the
other codes were specifically designed to protect against loss errors
and $\lat$ codes were designed to protect against displacement errors,
$\lat$ have apparently outperformed all of the other codes ``at
their own game'' (albeit a game whose rules were set by caveats \one~and
\two). To understand this effect, we have undertaken extensive analytical
calculations to determine the quantum error-correction conditions
for the pure-loss channel for $\lat$ codes {[}see Eq.~(\ref{eq:gkpz}){]}
as well as how $1-\fe$ scales with $\g$. As noted in the previous
subsection, while other codes protect against errors (to some order
in $\g$) exactly, $\lat$ protect against all errors \textit{approximately}.
In other words, other codes protect against the first few errors exactly,
but have low fidelity when there is a large probability of an unprotected
error occurring. By contrast, $\lat$ codes do not protect against
most errors exactly, but the contributions from \textit{all errors}
to the infidelity is small. 

In order to bound the scaling of $\fe(\lat)$ and since there is no
analytic expression for the optimal $\R$, we calculate a lower bound
on the channel fidelity $\frgkp$ using a recovery $\R^{\rgkp}$ \eqref{eq:agkp}
which consists of phase-insensitive amplification, followed by conventional
$\lat$ recovery consisting of displacement measurements and corrections.
The recovery is based on the idea that, after fine-tuning some of
the channel parameters,
\[
\text{\textit{pure loss}}+\text{\textit{amplification}}=\text{\textit{Gaussian noise},}
\]
where Gaussian noise corresponds to uniform diffusion in phase space
and its channel has displacements for its Kraus operators (\cite{cvbook},
Sec.~2.3; see also \cite{Caruso2006}). In other words, amplification
(with gain $\frac{1}{1-\g}$) exactly compensates the contractive
effect of pure loss (with loss rate $\g$) while at the same time
adding noise that, in this context, reduces to Gaussian noise (with
variance $\frac{\g}{1-\g}$). This idea has been considered in the
context of communication schemes \cite{Gottesman2001a,Garcia-Patron2012};
we apply it to bosonic error-correction by noting that Gaussian noise
is \textit{exactly} the type of noise that $\lat$ was designed for.
That way, we can use earlier tools \cite{Gottesman2001,Harrington2001}
developed to quantify $\lat$ performance against such noise. We consider
single-mode behavior here, noting that, in the multimode case, $\lat$
codes can be used to communicate efficiently across more general Gaussian
channels \cite{Noh2018}. Note that the related idea of ``\textit{amplication$+$pure
loss$=$Gaussian noise}'' has been recently applied to determine
bounds on quantum capacities of more general Gaussian channels \cite{Rosati,Noh2018,Sharma2017}.

The probability of failure of $\R^{\rgkp}$ gives a lower bound on
$\frgkp$, which in turn bounds $\fe(\lat)$ (see Sec.~\ref{subsec:Removing-energy-constraints-1}).
The bound contains an essential singularity at $\g=0$,
\begin{equation}
\fe(\lat)>1-\exp\left(-\frac{\pi}{4c}\frac{1-\g}{\g}\right)\,,
\end{equation}
where $c$ is a constant determined by the lattice used to construct
the code. This exponential suppression of infidelity explains the
non-trivial scaling of $\lat$ codes at small $\g$ and accounts for
their breakaway performance at higher $\g$. Due to the non-analyticity,
having different lattices becomes important for $\g\ll1$, where $\lat$
outperforms $\gkp$ by an order of magnitude {[}see Table~\ref{tab:values}(c){]}.

\begin{figure}
\includegraphics[width=0.95\columnwidth]{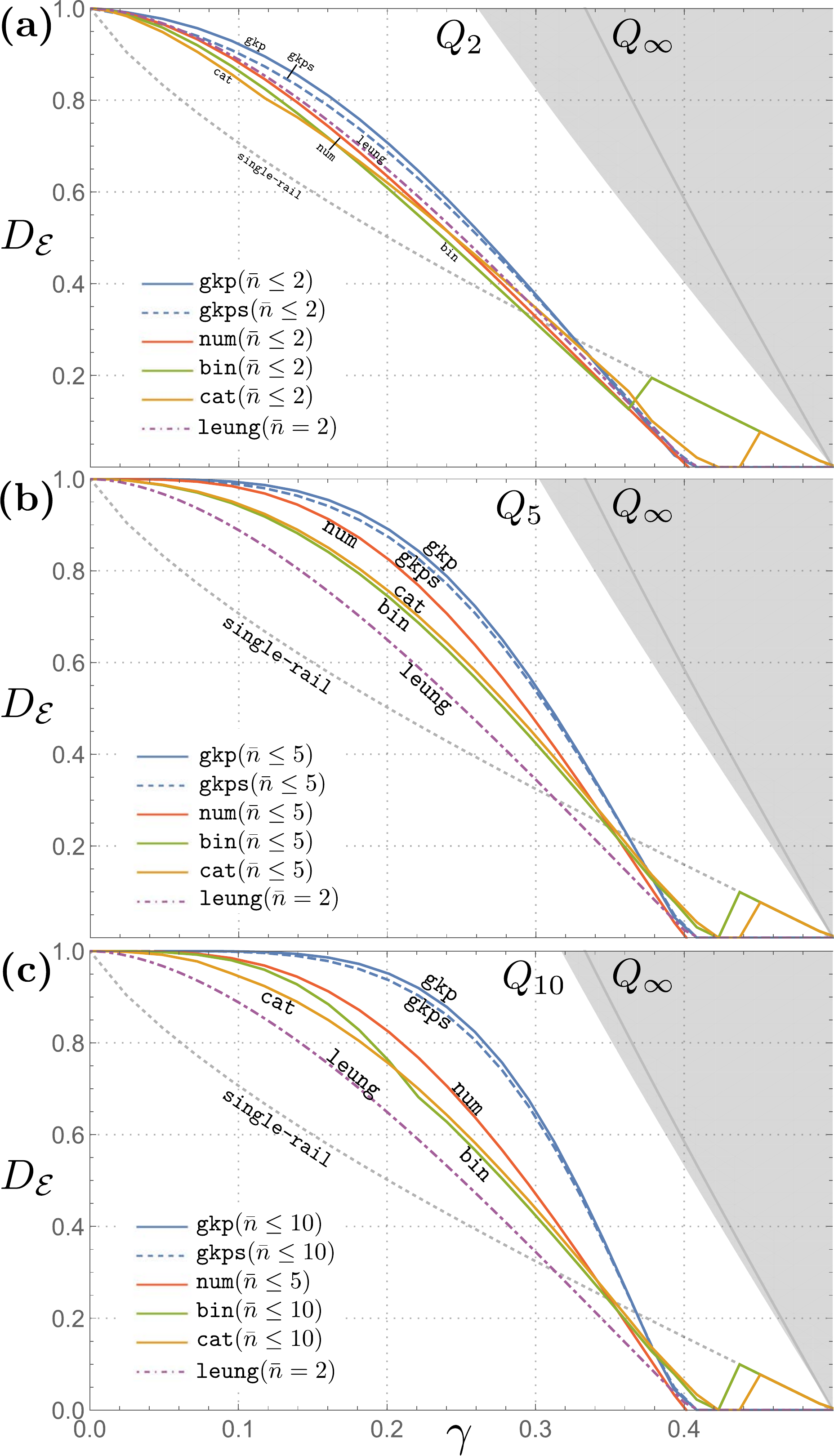}\caption{\label{f:cap}Hashing bound $\protect\ha$ (\ref{eq:hash}) of the
codes which optimize $\protect\fe$ for a given $\protect\g$ and
given constraints $\protect\nb_{\protect\cc}\leq$ 2\textbf{ (a)},
5 \textbf{(b)}, or 10 \textbf{(c)}. The boundary of the gray region
is $Q_{\protect\nb=2,5,10}$ (\ref{eq:cap}), the capacity of $\protect\L_{\protect\g}$
given the energy constraint of $\protect\nb$. The gray line is the
unconstrained capacity $Q_{\protect\nb\rightarrow\infty}$ (\ref{eq:capinf}).
The thin dotted diagonal line is $\protect\ha$ for the single-rail
encoding, which has the highest $\protect\ha$ at large $\protect\g$.
Recall that $\protect\cat/\protect\bin$ codes include the single-rail
encoding {[}i.e., this encoding is also $\protect\cat(\protect\a=S=0)$
and $\protect\bin(N=S=0)${]}, so $\protect\bin/\protect\cat$ eventually
jump to match the dotted line. We also show $\protect\ha$ for the
four-qubit $\protect\sh$ code (\ref{eq:leung}) vs. qubit amplitude
damping (i.e., the pure-loss channel restricted to the first two Fock
states of four modes). There are no $\protect\num$ codes with $\protect\nb_{\protect\num}\geq5$,
so red curves in \textbf{(b)} and \textbf{(c)} are identical.}
\end{figure}

\section{The hashing bound of $\protect\E$\label{sec:The-hashing-bound}}

Since channel fidelity provides a measure for entanglement preservation,
it is also of interest to examine these results from an information-theoretic
perspective. The question we aim to answer in this section is:
\[
\text{\textit{What is the achievable communication rate of \ensuremath{\E}?}}
\]
As with the $\fe$, this question also pre-supposes caveats \one~and
\two~from Sec.~\ref{sec:Take-home-messages}. In other words, we
assume recovery operations $\R$ can be done perfectly and organize
the codes by mean occupation number $\nb_{\cc}$, ignoring their ``size''
in terms of the number of Fock states necessary to express the logical
states.

The quantum communication rate of $\E=\S_{\cc}^{-1}\R\L_{\g}\S_{\cc}$
(\ref{eq:chan}) is ultimately limited by its most destructive link
\textemdash{} the pure-loss channel $\L_{\g}$ (\ref{eq:krausloss}).
Given an energy constraint of maximum $\nb$, the quantum capacity
of $\L_{\g}$ is given by \{\cite{Wilde2012a}, Thm. 8; see also \cite{Wilde2016},
Eq.~(12), and \cite{Wilde2012}\}
\begin{equation}
Q_{\nb}=\max\left\{ 0,g\left((1-\g)\nb\right)-g\left(\g\nb\right)\right\} \,,\label{eq:cap}
\end{equation}
where $g\left(\nb\right)=\left(\nb+1\right)\log_{2}\left(\nb+1\right)-\nb\log_{2}\nb$
is the von Neumann entropy of a thermal state with $\nb$. Note that
in the limit of $\nb\rightarrow\infty$, this capacity approaches
the unconstrained quantum capacity\footnote{Equation (\ref{eq:capinf}) was proven in Ref.~\cite{Wolf2007},
which built on Ref.~\cite{Holevo1999} (see also remark 17 in Ref.~\cite{Wilde2016}).
A standalone derivation is, e.g., in Sec. III of Ref.~\cite{Noh2018}.} 
\begin{equation}
Q_{\infty}=\max\left\{ 0,\log_{2}\left(\frac{1-\g}{\g}\right)\right\} \,.\label{eq:capinf}
\end{equation}

We see how close the codes giving the best $\fe$ (see Fig.~\ref{f:num})
come to $Q_{\nb}$ for $\nb\leq2,5,10$ by calculating a lower bound
on the capacity of $\E$ for each $\g$ and each code listed in Table~\ref{tab:numdet}.
The lower bound we use is known as the \textit{hashing bound} $\ha$
of $\E$ \cite{wildebook} \textemdash{} the (reverse$^{\ref{fn:choi}}$
\cite{Garcia-Patron2009}) coherent quantum information of $\E$'s
Choi matrix $\r_{\E}$ (\ref{eq:choi}), where $|\varPsi\ket=(|0_{\al}0_{\bo}\ket+|1_{\al}1_{\bo}\ket)/\sqrt{2}$
is a maximally entangled state of $\al$ and $\bo$ in an arbitrary
basis:
\begin{equation}
\ha\equiv H(\tr_{\bo}\{\r_{\E}\})-H(\r_{\E})\,,\label{eq:hash}
\end{equation}
where $H(\r)=-\tr\{\r\log_{2}\r\}$. This one-shot (i.e., with one
application of the channel) coherent information of $\r_{\E}$ provides
an achievable rate of quantum communication and entanglement distillation,
assuming many copies of $\E$ are available (\cite{Garcia-Patron2009,Pirandola2009,plob15};
see also Corr. 21.2.1 and Thm. 23.9.1 in \cite{wildebook}). Therefore,
$\ha$ \textit{does not} supply an achievable rate of any one oscillator
$\cc$, but instead gives an achievable rate of concatenation schemes
of the oscillator $\cc$ with other (outer) codes with the restriction
that $\R$ is used as the recovery for the (inner) oscillator $\cc$.

The first term in $\ha$ can be simplified to yield an expression
only in terms of $\E$,
\begin{equation}
\ha=H\left(\left\{ {\textstyle \frac{1}{2}}\pm{\textstyle \frac{1}{2}}\sqrt{{\textstyle \frac{1}{2}}|\!|\E(I_{\al})|\!|^{2}-1}\right\} \right)-H(\r_{\E})\,,
\end{equation}
where $H\left(\left\{ x\right\} \right)=-\sum_{x}x\log_{2}x$ for
a set of variables $\left\{ x\right\} $, $|\!|O|\!|^{2}\equiv\tr\{O^{\dg}O\}$
is the Frobenius norm of an operator $O$, and $I_{\al}$ is the qubit
identity. Derivation of the first term was done by determining the
reduced qubit density matrix $\tr_{\bo}\{\r_{\E}\}$ in terms $\E$'s
matrix representation (\ref{eq:matrep}) and then diagonalizing to
yield the two eigenvalues in the term's argument. Note that this term
is maximized when $\E$ is unital {[}$\E(I_{\al})=I_{\al}${]}. The
second term \textemdash{} the von Neumann entropy of the Choi matrix
$\r_{\E}$ \textemdash{} increases with the minimal number of Kraus
operators needed to express $\E$ and is zero when $\E$ is a unitary
channel.

Incidentally, the analytical formula for $\ha$ provides an easily
calculable lower bound on the quantum capacity $Q_{\E}$ of any qubit
channel $\E$. It turns out that $\ha$ is quite close to $Q_{\E}$
for a family of two-Kraus operator qubit channels \{\cite{Wolf2007a},
Eq.~(5)\} which includes the dephasing and amplitude damping channel:
we have checked numerically that the difference $Q_{\E}-\ha<0.005$.

\subsection{Hashing bound for codes giving optimal $\protect\fe$}

In Fig.~\ref{f:cap}, we plot the hashing bound $\ha$ for all of
the codes which produce the optimal channel fidelity $\fe$ in Fig.~\ref{f:num}
for the three energy constraints $\nb_{\cc}\leq2,5,10$. In other
words, this plot is not an optimization of $\ha$ over all codes,
but merely a plot of $\ha$ for the codes which give optimal $\fe$.
Recall that $\ha$ is a lower bound on the entanglement that is theoretically
distillable from using unlimited instances of $\E$ and one-way classical
communication from $\bo$ to $\al$. By contrast, $\fe$ is an overlap
which gauges how well entanglement was transmitted over just one instance
of $\E$. While $\fe$ bounds one of the terms in $\ha$ {[}see Eq.~(\ref{eq:bound}){]}
and the two yield a similar order of performance of the codes, there
is no guarantee that codes giving the optimal value of $\fe$ should
also give \textit{optimal} values for $\ha$. In other words, even
if $\fe(\oone)>\fe(\ttwo)$ given two codes $\oone$ and $\ttwo$,
there could still exist an entanglement distillation scheme which
extracts more entanglement from the unlimited instances of $\E$ from
code $\ttwo$. This is true for our codes at large $\g$. Let us for
example consider $\nb\leq10$. At around $\g\approx0.37$, all of
the codes begin to have a lower $\ha$ than the single-rail $\{|0\ket,|1\ket\}$
Fock state encoding {[}thin black line in Fig.~\ref{f:cap}(c){]}.
By contrast, the point at which the codes begin to have a lower $\fe$
than the single-rail encoding is $\g\approx0.42$ {[}see Fig.~\ref{f:num}(c){]}.
The $\cat/\bin$ codes include this encoding {[}i.e., single-rail
is also $\cat(\a=S=0)$ and $\bin(N=S=0)${]}, so we can say for certain
that the $\cat/\bin$ codes which optimize $\fe$ are not those which
optimize $\ha$ at $0.37\leq\g\leq0.42$.

For $\g\lesssim0.3$, we once again see similar behavior of the codes
(relative to each other) as with $\fe$. For $\nb_{\cc}\leq10$, $\lat$
codes break from the pack and bridge the gap with $Q_{\nb}$ most
rapidly. For example, at the high loss rate $\g=0.3$, $\ha[\lat(\nb\leq10)]\approx0.63$
bits is about twice that of the naive Fock state code. Moreover, $\ha[\lat(\nb)]$
approaches roughly $\half Q_{\nb}$ for large $\g$. In addition,
$\bin$ codes exhibit better performance with increasing $\nb$ in
the $\g\lesssim0.2$ range. This begs the question of how close $\ha$
for $\lat$ and $\bin$ comes to $Q_{\nb}$ when one encodes more
than a qubit's worth of information and when one utilizes two- or
higher-mode generalizations of the codes. Such a question is outside
the scope of this work, but is being investigated for a subsequent
publication.

Since the pure-loss bosonic channel reduces to qubit amplitude damping
when restricted to the Fock states $|0\ket$ and $|1\ket$, one interesting
question to ask is whether the $\nb_{\cc}$ photons, which so far
are concentrated in one mode, will produce a better rate when distributed
the first two Fock states of multiple modes. While comparing single-mode
codes to the various discrete-variable codes specialized to protect
against qubit amplitude damping (e.g., \cite{Leung1997,gottesman_thesis,Lang2007,Fletcher2008,Duan2010,Shor2011,Han2014,Ouyang2014,Jackson2016,Ouyang2016,Ouyang2016a})
is outside the scope of this work, we do provide a reference $\ha$
for one specialized code \textemdash{} the four-qubit $\sh$ code
\cite{Leung1997} \textemdash{} that is the smallest known discrete-variable
code to protect against one amplitude damping error.\footnote{Interestingly, this code can also protect against one erasure \cite{Grassl1997}
and be used as an error-detecting code for other errors \cite{Vaidman1996}.} Each of the four physical qubits in the $\sh$ code,
\begin{equation}
|\pm_{\sh}\ket=\frac{1}{2}\left(|00\ket\pm|11\ket\right)^{\otimes2}\,,\label{eq:leung}
\end{equation}
correspond to the first two Fock states of four oscillators. We can
then apply $\L_{\g}^{\otimes4}$ (which reduces to amplitude damping
within the $|0\ket,|1\ket$ Fock state subspace), optimize $\fe$
to yield $\E$, and calculate $\ha$ via the same procedure as with
the rest of the codes. A simple calculation yields a total occupation
number of $\nb_{\sh}=2$ photons, which in this case are distributed
over the first two Fock states of four modes. The $\sh$ code performs
similar to the $\cat/\bin/\num(\nb\leq2)$ codes from Fig.~\ref{f:cap}(a),
but is outperformed by the $\lat(\nb\leq2)$ codes for $\g\leq0.35$.
This suggests that, at least for intermediate $\g$ and all else being
equal, \textit{it is better to encode two photons in a single-mode
$\lat$ state than to spread them out over four modes}. The $\sh$
code is outperformed at almost all $\g$ by all codes considered with
$\nb_{\cc}\leq5$, but this is not a fair comparison since those codes
use more photons.

\section{Primer: the QEC matrix\label{sec:QEC matrix}}

In order to analyze errors for the codes, we consider the quantum
error-correction (QEC) conditions \cite{Bennett1996,Knill1997} (see
also \cite{nielsen_chuang}, Thm. 10.1). The errors we consider are
the Kraus operators $E_{\ell}$ (\ref{eq:krausloss}), where $\ell$
denotes the number of photons lost after application of the error.
Calculating the effect of the error $E_{\ell}^{\dg}E_{\lp}$ on the
codespace yields a $2\times2$ matrix $\e_{\ell\lp}$,
\begin{equation}
P_{\cc}E_{\ell}^{\dg}E_{\lp}P_{\cc}=\e_{\ell\lp}^{\cc}\in\text{Mat}_{2\times2}\,.\label{eq:qec}
\end{equation}
We write $\e_{\ell\lp}^{\cc}$ as a superposition of $P_{\cc}$ and
matrices\begin{subequations}
\begin{align}
Z_{\cc} & =|0_{\cc}\ket\bra0_{\cc}|-|1_{\cc}\ket\bra1_{\cc}|\label{eq:logz}\\
X_{\cc} & =|0_{\cc}\ket\bra1_{\cc}|+|1_{\cc}\ket\bra0_{\cc}|\label{eq:logx}\\
Y_{\cc} & =|1_{\cc}\ket\bra0_{\cc}|-|0_{\cc}\ket\bra1_{\cc}|\,.\label{eq:logy}
\end{align}
\end{subequations}We define our matrix basis as such because both
$P_{\cc}$ and $E_{\ell}$ are real for our codes, so the \textit{QEC
matrix} $\e^{\cc}$ is real, symmetric, and $2N$-dimensional (with
$N\rightarrow\infty$ being the dimension of the oscillator). Expanding
each $2\times2$ subblock yields
\begin{equation}
\e_{\ell\lp}^{\cc}=c_{\ell\lp}^{\cc}P_{\cc}+x_{\ell\lp}^{\cc}X_{\cc}+y_{\ell\lp}^{\cc}Y_{\cc}+z_{\ell\lp}^{\cc}Z_{\cc}\,,
\end{equation}
with coefficients denoted by 
\begin{equation}
[c,x,y,z]_{\ell\lp}^{\cc}=\frac{1}{2}\tr\left\{ [P,X,Y,Z]_{\cc}E_{\ell}^{\dg}E_{\lp}\right\} \,.\label{eq:coeff}
\end{equation}
For $E_{\ell}$ to be perfectly correctable, one must satisfy the
QEC condition 
\begin{equation}
\e_{\ell\lp}^{\cc}=c_{\ell\lp}^{\cc}P_{\cc}
\end{equation}
(equivalently, $\bra\m_{\cc}|E_{\ell}^{\dg}E_{\lp}|\n_{\cc}\ket=c_{\ell\lp}^{\cc}\d_{\m\n}$
for $\m,\n\in\{0,1\}$). In words, a correctable error must act as
the identity within the code subspace (equivalently, the effect of
the error must be the same on both code states). Therefore, the coefficient
$c_{\ell\lp}^{\cc}$ represents the correctable part of $\e_{\ell\lp}^{\cc}$
while $\{x_{\ell\lp}^{\cc},y_{\ell\lp}^{\cc},z_{\ell\lp}^{\cc}\}$
represent various uncorrectable parts corresponding to bit, phase,
and joint bit-phase flips, respectively. Since not all errors $E_{\ell}$
can be corrected, we proceed to analyze the magnitude of the uncorrectable
parts \textemdash{} the $2N$-dimensional matrix $\e^{\cc}-c^{\cc}$
\textemdash{} with $\e_{\ell\ell^{\pr}}^{\cc},c_{\ell\ell^{\pr}}^{\cc}$
being $2\times2$ submatrices of $\e^{\cc},c^{\cc}$, respectively.

The QEC matrix block $\e_{\ell\ell^{\prime}}^{\cc}$ can also be interpreted
(\cite{nielsen_chuang}, Fig.~10.5) as a matrix of overlaps between
the two error subspaces spanned by $\{E_{\ell}|\m_{\cc}\ket\}_{\m=0}^{1}$
and $\{E_{\lp}|\m_{\cc}\ket\}_{\m=0}^{1}$, i.e., the range of $E_{\ell}P_{\cc}$
and $E_{\ell^{\prime}}P_{\cc}$. We call these subspaces $E_{\ell}P_{\cc}$
and $E_{\lp}P_{\cc}$ for short. When no loss events are occurring,
the code state undergoes the backaction-induced evolution corresponding
to the subspace $E_{0}P_{\cc}$. As $\ell$ loss events occur, one's
ability to detect them hinges on the orthogonality between $E_{0}P_{\cc}$
and $E_{\ell}P_{\cc}$, the latter being the space to which a state
has gone after losing $\ell$ photons. The $\e_{0\ell}^{\cc}$ and
$\e_{\ell0}^{\cc}$ parts of the QEC matrix thus correspond to the
ability to distinguish between $\ell$ losses and no losses, making
their satisfaction similar to the satisfaction of the \textit{error-detection
conditions }$\d_{\ell}^{\cc}=P_{\cc}E_{\ell}P_{\cc}\propto P_{\cc}$
\cite{preskillnotes}. While the backaction in $E_{0}$ makes $\e_{0\ell}^{\cc}\neq\d_{\ell}^{\cc}$,
the two converge to each other as $\g\rightarrow0$. Since $\bin$
and $\cat$ codes satisfy both $\e_{0\ell}^{\cc},\d_{\ell}^{\cc}\propto P_{\cc}$
exactly up to some $\ell\leq S$, uncorrectable parts in the QEC matrix
blocks $\e_{0\ell}^{\cc}$ quantify how well $\ell$-photon losses
can be detected for those codes.

Uncorrectable parts $\{x_{\ell\ell}^{\cc},y_{\ell\ell}^{\cc},z_{\ell\ell}^{\cc}\}$
for ``diagonal'' errors $E_{\ell}^{\dg}E_{\ell}$ represent \textit{distortion}
of the quantum information within the subspace $E_{\ell}P_{\cc}$
and limit how well one is able to correct the error $E_{\ell}$ after
detection. Since our codespace can become distorted even when there
are no loss events, we have to also consider backaction-induced distortion
captured by $\e_{00}^{\cc}$. The loss event probability distribution
is governed by $c_{\ell\ell}^{\cc}$ and depends on both $\g$ and
$\nb_{\cc}$. For a fixed $\g$ and sufficiently large $\nb_{\cc}$,
we will see that $c_{\ell\ell}^{\cc}$ for $\cat$ ($\gkp$) is a
Poisson (geometric) distribution having mean $\g\nb_{\cc}$. In such
cases, we can interpret $\g\nb_{\cc}$ as the average number of photons
lost, and only when $\g\nb_{\cc}\ll1$ can we say that $E_{0}$ is
the most likely error for a code.

\section{Cat codes\label{sec:Cat-codes}}

Cat code logical states are coherent states projected onto subspaces
of occupation number modulo $2(S+1)$:
\begin{equation}
|\m_{\cat}\ket=\frac{\Pi_{(S+1)\m}|\a\ket}{\sqrt{N_{\a}^{(S+1)\m}}}\,,\label{eq:cat}
\end{equation}
with $\a$ real (for simplicity), $\m\in\{0,1\}$, and normalization
\begin{equation}
N_{\a}^{(S+1)\m}=\bra\a|\Pi_{(S+1)\m}|\a\ket\,.
\end{equation}
The projections $\{\Pi_{0},\Pi_{S+1}\}$ used to define the code states
belong to the family (for $r\in\{0,1,\cdots,2S+1\}$)\footnote{\label{fn:shift}One can also consider shifted $\cat$ and $\bin$
codes by picking subspaces of Fock states $n=s$ and $n=S+1+s$ modulo
$2(S+1)$ for shift parameter $s\in\{0,1,\cdots,S\}$. Sampling some
of the ``sweet spots'' for shifted $\cat$ codes \cite{Li2016}
did not alter the qualitative behavior of $\cat$ relative to the
other codes.}
\begin{equation}
\Pi_{r}=\sum_{n=0}^{\infty}|2n(S+1)+r\ket\bra2n(S+1)+r|\,.
\end{equation}
In the large $\a$ limit, i.e., when
\begin{equation}
2\a\sin\left(\frac{\pi}{S+1}\right)\gg1\,,\label{eq:limit-1}
\end{equation}
$\cat$-code states become equal superpositions of coherent states
$\{|\a e^{i\frac{\pi}{S+1}k}\ket\}_{k=0}^{2(S+1)-1}$ distributed
equidistantly on a circle of radius $\a$ in phase space. In that
limit, the seemingly bothersome normalization factors approach the
same constant, while when $\a\lesssim S$, they become distinct in
order to account for the various overlaps between the coherent states.
Expressing the normalization factor in terms of such overlaps \{\cite{pub011},
Eq.~(3.22)\}, we have
\begin{equation}
N_{\a}^{(S+1)\m}=\frac{1}{2(S+1)}\sum_{s=0}^{2S+1}\left(-1\right)^{\m s}\bra\a|\a e^{i\frac{\pi}{S+1}s}\ket\,.\label{eq:norm}
\end{equation}
Since the $\cat$ codes which produce the optimal $\fe$ have $\a\lesssim S$,
we have to consider the factors $N_{\a}^{(S+1)\m}$ at intermediate
$\a$ in order to explain Fig.~\ref{f:num}.

Due to properties of coherent states and for $\a\neq0$, $\cat$ logical
states satisfy
\begin{equation}
\left(\aa/\a\right)^{2(S+1)}|\m_{\cat}\ket=|\m_{\cat}\ket\,,
\end{equation}
and $\cat(S)$ {[}and $\bin(S)$, as we shall see{]} can detect exactly
$S$ photon loss events using the check operator
\begin{equation}
C_{\cat}=C_{\bin}=\exp\left(i\frac{2\pi\hat{n}}{S+1}\right)\,.\label{eq:disrot}
\end{equation}
Its square root, $\exp(i\frac{\pi\hat{n}}{S+1})$, makes for a logical
$Z$-operator within $\cat$ and $\bin$ code subspaces. We call this
a check operator (and not a stabilizer) since it has other eigenvalues
which are not modulus one and is not part of a set of commuting such
operators which is used to construct the code states.

The $S=0$ $\cat$ code was proposed in Ref.~\cite{Cochrane1999}
for coherent-state quantum computation \cite{Jeong2002,Ralph2003,Glancy2004,Lund2008}.
This code cannot detect loss events since losing a photon causes a
logical bit-flip, but it of course can be concatenated with, e.g.,
a bit-flip code \cite{Ralph2003}. The $\cat(S=1,2)$ codes were studied
first in Refs.~\cite{Leghtas2013b,cats}, followed by investigations
into $\cat(S\geq2)$ and qudit extensions \cite{albert2015,Bergmann2016,Li2016}.
There are several theoretical and experimental schemes for $\cat$
state preparation \cite{Andersen2015}. Schemes designed to protect
against loss errors \cite{Ofek2016} (for $S=1$) and backaction errors
\cite{Leghtas2014} (for $S=0$), respectively, were realized using
microwave cavities coupled to superconducting qubits.

\subsection{A simple example}

As an example of the utility of $\cat$ codes, consider the simplest
non-trivial code family $\cat(S=1)$ whose logical states are
\begin{equation}
|\m_{\cat}^{S=1}\ket=\frac{|\a\ket+\kkk{-\a}+\left(-1\right)^{\m}\left(|i\a\ket+\kkk{-i\a}\right)}{4\sqrt{2[\cosh\a^{2}+\left(-1\right)^{\m}\cos\a^{2}]}}\,.\label{eq:cats1}
\end{equation}
In the Fock basis, the above $|0_{\cat}\ket$ is a superposition of
Fock states $|0\ket,|4\ket,|8\ket,\cdots$ while $|1_{\cat}\ket$
is supported by $|2\ket,|6\ket,|10\ket,\cdots$. Therefore, there
are exactly $S=1$ Fock states separating the Fock states supporting
$|0_{\cat}\ket$ from those supporting $|1_{\cat}\ket$, so we call
$S$ the \textit{spacing} between logical states. Due to this spacing,
$\e_{01}^{\cat}=0$ \textemdash{} one loss event is always detectable.
However, such an event is not always \textit{correctable} since $\e_{11}^{\cat}$
contains uncorrectable parts at generic values of $\a$.

\begin{figure}[t]
\includegraphics[width=1\columnwidth]{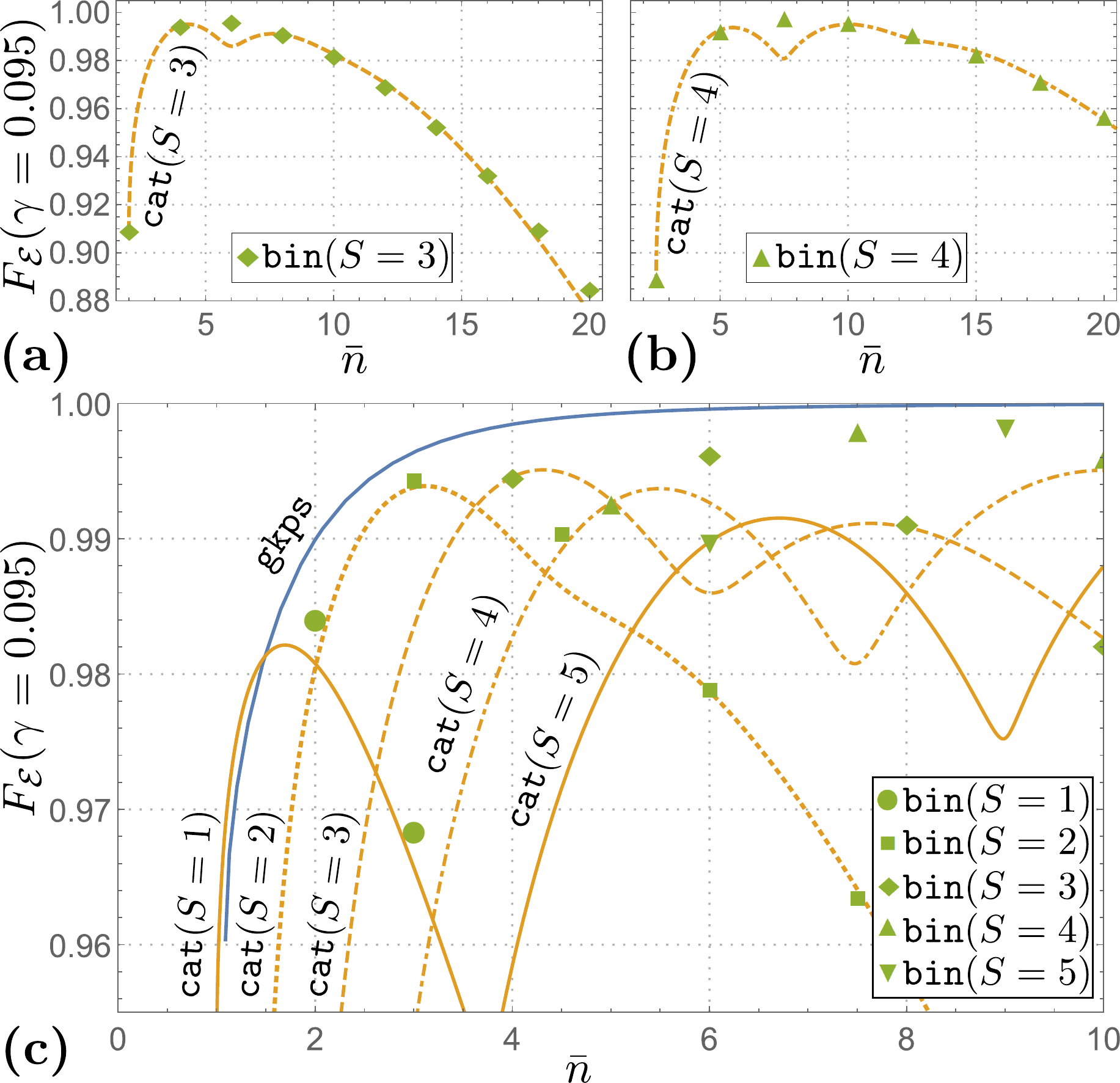}\caption{\label{f:nbar}Channel fidelity $\protect\fe(\protect\g=0.095)$ (\ref{eq:fid})
vs. mean occupation number $\protect\nb$ (\ref{eq:nbar}) for $\protect\cat$
and $\protect\bin$ at spacing \textbf{(a)} $S=3$ and \textbf{(b)}
$S=4$. Note that $\protect\cat(S)$ depends continuously on $\protect\nb$
and $\protect\bin$ exists only at discrete values. For a given $S$,
$\protect\cat$ codes perform best at specific $\protect\nb$, corresponding
to ``sweet-spot'' values of $\protect\a=\protect\a_{\star}$. However,
due to the oscillatory nature of the errors in $\protect\cat(S\apprge3)$,
the codes also develop troughs in $\protect\fe$. On the other hand,
$\protect\bin(S)$ performs similar to $\protect\cat(S)$ at small
and large $\protect\nb$ but does not suffer from troughs at intermediate
$\protect\nb$. \textbf{(c)} Plots of $\protect\cat$ and $\protect\bin$
for $S\in\{1,2,3,4,5\}$ along with $\protect\gkp$, which turns out
to outperform both $\protect\cat$ and $\protect\bin$ and increase
monotonically with $\protect\nb$. Similar trends are observed for
smaller $\protect\g$.}
\end{figure}

In general, $\e_{\ell\ell}^{\cat}\neq c_{\ell\ell}P_{\cat}$ because
of the backaction term $(1-\g)^{\ph/2}$ in $E_{\ell}$. We will see
that uncorrectable parts in $\e_{\ell\ell}^{\cat}$ are well-suppressed
as $\a\rightarrow\infty$, but in that limit the code consists of
large-amplitude coherent states and there is a high chance of losing
more than one photon (i.e., uncorrectable parts in $\e_{02}^{\cat}$
become very large). Therefore, for more general codes with a given
spacing $S$ and loss rate $\g$, there is an optimal or ``sweet
spot'' value $\a=\a_{\star}(\g)$ that balances the backaction with
the loss errors. This is exactly what we see in Figs.~\ref{f:nbar}(a-c),
where $\fe$ is plotted vs $\nb_{\cat}$ for various $\cat(S)$ and
at a fixed $\g$. We can see that $\fe[\cat(S)]$ is maximized at
certain $\nb_{\cat}$ {[}which in turn corresponds to a certain $\a_{\star}(\g)${]}
and decays with sufficiently large $\a$. Note that there can be multiple
$\a_{\star}$ for a given $S$. By contrast, $\gkp$ code fidelities
increase monotonically with $\nb_{\gkp}$.

We now show how the sweet spot can be analytically determined for
$\cat(S=1)$ and in the limit $\g\rightarrow0$, discussing general
$\g$ in the next subsection. Recalling the form of the Kraus operators
(\ref{eq:kraussmalldelta}) for small $\g$, to suppress distortion
due to backaction, one has to make sure that both code states have
the same occupation number:
\begin{equation}
\d\nb_{\cat}={\textstyle \half}\tr\{Z_{\cat}\ph\}=0\,.
\end{equation}
While $\d\ph_{\cat}\rightarrow0$ as $\a\rightarrow\infty$, there
are certain fine-tuned $\a$ at which the occupation numbers of the
two logical states also coincide due to the oscillatory nature of
the normalization factors $N_{\a}$ (\ref{eq:norm}). Solving the
above equation using $|\m_{\cat}^{S=1}\ket$ (\ref{eq:cats1}) yields
the transcendental equation 
\begin{equation}
\tan\a^{2}=-\tanh\a^{2}\,,
\end{equation}
whose solution is $\a_{\star}(\g\rightarrow0)\approx1.538$ (corresponding
to $\nb_{\cc}\approx2.324$). Thus, in the small $\g$ limit and at
this fine-tuned $\a_{\star}$, a single loss event is both detectable
and correctable. This was about the value used in a recent $\cat$-code
error-correction experiment \cite{Ofek2016}.

\subsection{QEC matrix for cat codes}

Let us briefly elaborate on the discussion regarding sweet spots using
the QEC matrix for $\cat$ codes for finite (but still small) $\g$.
Due to the spacing of the codes, QEC matrix subblocks $\e_{\ell\lp}$
(\ref{eq:qec}) contain uncorrectable parts only for certain values
of $\ell,\lp$. Let us first study the distortion due to errors $\e_{\ell\ell}^{\cat}$,
which we can easily calculate \cite{Li2016,thesis} by using the representation
(\ref{eq:cat}) of the $\cat$ states, observing that $\Pi_{r}\aa=\aa\Pi_{r+1}$,
and using Eq.~(\ref{eq:iden3}):
\begin{equation}
\epsilon_{\ell\ell}^{\cat}=\frac{(\g\a^{2})^{\ell}e^{-\g\a^{2}}}{\ell!}\begin{pmatrix}\frac{N_{\a\sqrt{1-\g}}^{-\ell}}{N_{\a}^{0}} & 0\\
0 & \frac{N_{\a\sqrt{1-\g}}^{S+1-\ell}}{N_{\a}^{S+1}}
\end{pmatrix}\,,\label{eq:ecat}
\end{equation}
where $N_{\a\sqrt{1-\g}}$ are damped versions of the normalization
factors in Eq.~(\ref{eq:norm}). Expansion yields the correctable
part (\ref{eq:coeff}),
\begin{align}
c_{\ell\ell}^{\cat} & \equiv\half\frac{(\g\a^{2})^{\ell}e^{-\g\a^{2}}}{\ell!}\left(\frac{N_{\a\sqrt{1-\g}}^{-\ell}}{N_{\a}^{0}}+\frac{N_{\a\sqrt{1-\g}}^{S+1-\ell}}{N_{\a}^{S+1}}\right)\,,\label{eq:ccat}
\end{align}
and only one uncorrectable part (\ref{eq:coeff}),
\begin{align}
z_{\ell\ell}^{\cat} & \equiv\half\frac{(\g\a^{2})^{\ell}e^{-\g\a^{2}}}{\ell!}\left(\frac{N_{\a\sqrt{1-\g}}^{-\ell}}{N_{\a}^{0}}-\frac{N_{\a\sqrt{1-\g}}^{S+1-\ell}}{N_{\a}^{S+1}}\right)\,.\label{eq:zcat}
\end{align}
The correctable part represents the probability of losing $\ell$
photons. This distribution is Poisson in the large $\a$ limit, in
which the ratios of normalization factors inside the parentheses both
go to one exponentially with $-(1-\g)\a^{2}$. The uncorrectable part
$z_{\ell\ell}^{\cat}$ represents the inability to correct against
$\ell$ loss events. It is suppressed as $\a\rightarrow\infty$, but
is also zero at certain ``sweet-spots'' $\a_{\star}$, which we
discuss using another example.

Consider $\cat(S=2)$ for $\g\leq0.0124$ and $\nb_{\cat}\leq5$.
This is a case when $\g\nb\ll1$ and so we only have to consider distortion
due to backaction $\e_{00}^{\cat}$. We see from Table~\ref{tab:numdet}
that the $\cat$ code which achieves the highest $\fe$ out of all
$\cat$ codes with $\nb\leq5$ is $\cat(\a=1.739,S=2)$. This is exactly
the sweet spot at which the distortion $z_{00}^{\cat}$ is approximately
zero: $\a_{\star}(\g=0.005)\approx1.739$. More generally, $\a_{\star}(\g)$
at which $z_{00}^{\cat}\approx0$ decreases as $\g\rightarrow0$ to
the value $\a_{\star}(\g=0)\approx1.737$. {[}The reason that $\cat(\a=1.739)$
and not $\cat(\a=1.737)$ is the optimal code at $\g<0.005$ is because
the resolution in our sampling of $\a$ is not sufficient to resolve
the difference.{]} As mentioned in our discussion of $\cat(S=1)$
in the previous subsection, in the limit $\g\rightarrow0$, $z_{00}^{\cat}$
becomes proportional to
\begin{equation}
\d\nb_{\cat}=\half\tr\{Z_{\cat}\ph\}=\frac{\a^{2}}{2}\left(\frac{N_{\a}^{-1}}{N_{\a}^{0}}-\frac{N_{\a}^{S}}{N_{\a}^{S+1}}\right)\,.\label{eq:dncat}
\end{equation}
This $\d\nb_{\cat}$ tells us how well we are able to correct against
distortion due to no loss events, the dominant error when $\g\nb_{\cat}\ll1$.
It is zero at exactly the ``sweet spot'' $\a_{\star}(\g=0)$, and
we can similarly relate $\a_{\star}$ to $\d\nb_{\cat}$ for $\cat$
at other values of $S$. Having only covered $\cat(S=2)$, we refer
the reader to Ref.~\cite{Li2016} for such calculations.

\begin{figure*}[t]
\includegraphics[width=0.9\textwidth]{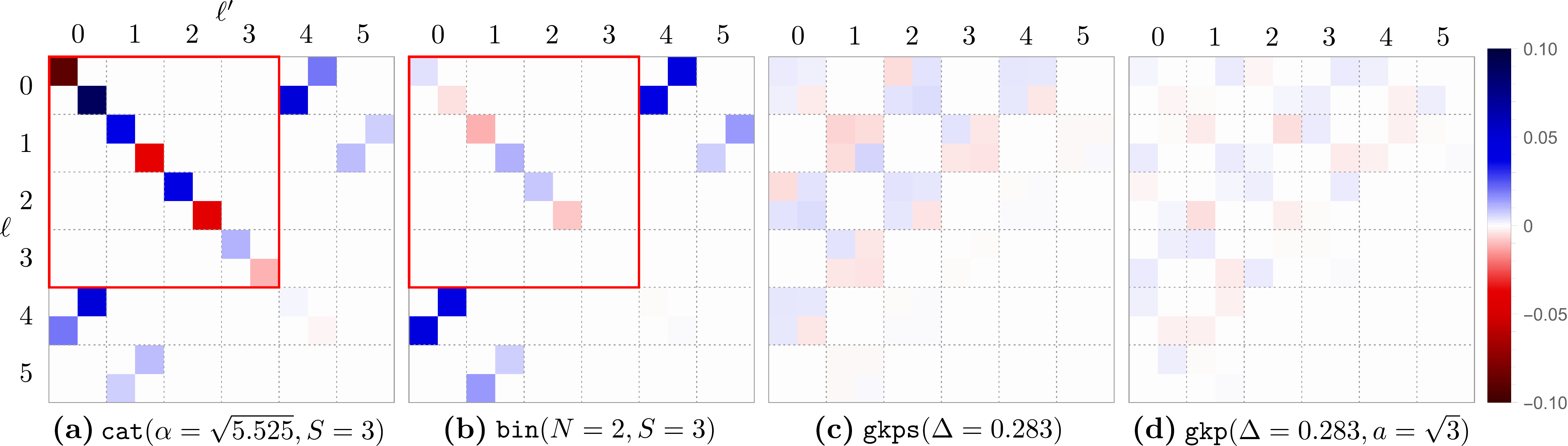}

\caption{\label{f:qec}Uncorrectable parts of the QEC matrices, $\protect\e^{\protect\cc}-c^{\protect\cc}$
(see Sec.~\ref{sec:QEC matrix}), for $\protect\cc$ being \textbf{(a)}
$\protect\cat$ (\ref{eq:cat}), \textbf{(b)} $\protect\bin$ (\ref{eq:bin}),
\textbf{(c)} $\protect\gkp$ (\ref{eq:gkp}), and \textbf{(d)} $\protect\lat$
(\ref{eq:gkpnum}) with parameters such that $\protect\nb_{\protect\cc}\approx6$
for all codes and given $\protect\g=0.095$. Recall that $\protect\e^{\protect\cc}$
is a block matrix consisting of $2\times2$ matrices $\protect\e_{\ell\protect\lp}^{\protect\cc}$
that quantify the overlap between error subspaces $E_{\ell}P_{\protect\cc}$
and $E_{\protect\lp}P_{\protect\cc}$. These $2\times2$ matrices
are delineated by dotted lines. The four entries in $\protect\e_{\ell\protect\lp}^{\protect\cc}$
are presented as colored squares; note that $\protect\e_{\ell\protect\lp}^{\protect\cc}$
has no imaginary part (all $E_{\ell}P_{\protect\cc}$ are real). From
Fig.~\ref{f:nbar}(c), we see that $\protect\fe(\protect\cat)<\protect\fe(\protect\bin)<\protect\fe(\protect\gkp)$
at $\protect\nb_{\protect\cc}\approx6$, and the above QEC plots nicely
corroborate that order of performance. Since $\protect\cat$ and $\protect\bin$
have spacing $S=3$, there are no off-diagonal errors for $\ell\leq3$
(inside the red square). However, both codes suffer from distortion
on the diagonal portions of $\protect\e_{\ell\protect\lp}^{\protect\cc}$,
with $\protect\bin$ suffering noticeably less at this particular
$\protect\g$. On the other hand, $\protect\gkp$ (square lattice)
and $\protect\lat$ (shifted hexagonal lattice) barely suffer from
\textit{any} noticeable errors. Since $\protect\gkp$ codes have spacing
$S=1$ due to a symmetry of their underlying lattice, $\protect\e_{\ell\protect\lp}$
does not contain off-diagonal errors for $\ell\leq1$ (more generally,
for $\ell-\protect\lp$ odd). Shifted $\protect\lat$ codes do not
have that symmetry, but perform comparably.}
\end{figure*}

We have seen that at $\a<S$, backaction-induced errors are not suppressed
due to $\a$ not being sufficiently large {[}see Eq.~(\ref{eq:limit-1}){]}.
So why are $\cat$ codes with high values of $\a$ not optimal? This
is because for $\a\approx S$, the fraction $\g\nb_{\cat}\approx\g\a^{2}$
of photons lost yields a large probability of losing $S+1$ photons,
an uncorrectable error. More technically, recall that due to spacing
$S$, the effect of $\ell\leq S$ loss events is zero, $\e_{0\ell}^{\cat}=0$.
However, the first uncorrectable loss at $\ell=S+1$ produces an error
that scales unfavorably with $\a$, prohibiting $\a$ from getting
too large. A calculation yields
\begin{align}
\e_{0,S+1}^{\cat} & =\frac{(\g\a^{2})^{\frac{S+1}{2}}e^{-\g\a^{2}}}{\sqrt{\left(S+1\right)!N_{\a}^{0}N_{\a}^{S+1}}}\begin{pmatrix}0 & N_{\a\sqrt{1-\g}}^{0}\\
N_{\a\sqrt{1-\g}}^{S+1} & 0
\end{pmatrix}\,.
\end{align}
Once again, the ratios of normalization factors go to one as $\a\rightarrow\infty$.
In that limit, this error becomes proportional to a pure bit flip,
$\e_{0,S+1}^{\cat}\sim x_{0,S+1}^{\cat}X_{\cat}$ (where we use the
mathematician's definition of ``$\sim$''), with
\begin{equation}
x_{0,S+1}^{\cat}\sim\sqrt{c_{00}^{\cat}c_{S+1,S+1}^{\cat}}\label{eq:scaling}
\end{equation}
independently of $\g$. In other words, given even a very small $\g$,
the above equation is still satisfied for a sufficiently large $\a$.
We claim that this is the worst possible scaling, making such errors
completely undetectable (and therefore uncorrectable). Recall that
since the QEC matrix (\ref{eq:qec}) is positive semidefinite ($\epsilon\geq0$),
in the absence of other errors for a given $\ell,\lp$, all bit-flip
errors are bounded by $x_{\ell\lp}\leq\sqrt{c_{\ell\ell}c_{\lp\lp}}$.
Ideally we would like to have no error ($x_{\ell\lp}=0)$, and in
the worst case the inequality is saturated ($x_{\ell\lp}=\sqrt{c_{\ell\ell}c_{\lp\lp}}$).
In Eq.~(\ref{eq:scaling}), we see that the inequality is saturated
as $\a\rightarrow\infty$. Because the code states are eigenstates
of $\aa^{2(S+1)}$, the behavior upon 0 to $2(S+1)-1$ losses repeats
itself after $2(S+1)$ losses. More generally, a bit flip error of
similar intensity occurs at any $\ell-\lp=S+1$ modulo $2(S+1)$.
Thus, as $\a\rightarrow\infty$, the QEC matrix $\e$ develops sparse
but large bit-flip error entries at each such $\ell,\lp$, implying
a large probability of undetectable errors. Therefore, $\cat$ codes
should work best for values of $\a$ at which the probability of losing
$\ell=S+1$ photons is small.

The reader should by now see that, for a given $\g$ and $S$, a $\cat$
code performs optimally at specific $\a_{\star}(\g)$. But how does
one determine the optimal $S$? We do not claim to answer this question
fully, only noting that (A) the optimal $S$ depends on $\g$ and
(B) our energy constraints limit the selection of $\a_{\star}$ to
choose from, which in turn limit the selection of $S$. For an example
of (A), notice in Fig.~\ref{f:nbar}(c) that $\nb_{\cat}(\a_{\star},S=1)<2$
while $\nb_{\cat}(\a_{\star},S>1)>2$, so the highest spacing achievable
with a sweet-spot is $S=1$. Similarly for (B), for $\nb_{\cat}\leq5$,
there are three available $\a_{\star}$, one for each $S=1,2,3$.
At $\g=0.095$, the $\fe$ is highest for $S=3$ {[}as shown in Fig.~\ref{f:nbar}(c){]}
while at smaller $\g$, $S=2$ is optimal (see Table~\ref{tab:numdet}).
In summary, $\cat$ codes cannot perform well at sufficiently large
$\a$ and instead are optimal for specific values of $\a_{\star}\apprle S$.

\section{Binomial codes\label{sec:bin-codes}}

In terms of Fock states, the logical states (defined here in the $X_{\bin}$-basis
\cite{bin}) are
\begin{equation}
|\m_{\bin}\ket=\frac{1}{\sqrt{2^{N+1}}}\sum_{m=0}^{N+1}\left(-1\right)^{\m m}\sqrt{{N+1 \choose m}}|\left(S+1\right)m\ket\,.\label{eq:bin}
\end{equation}
Their mean occupation number (\ref{eq:nbar}) is
\begin{equation}
\nb_{\bin}=\half\left(N+1\right)\left(S+1\right)\,.\label{eq:nbarbin}
\end{equation}
The non-negative integer $N$ governs the order in $\g$ to which
dephasing errors $\{\ph^{n}\}_{n=0}^{N}$ can be corrected exactly
and is similar to $\a$ in $\cat$ codes. The spacing $S$ is the
same as that of $\cat$ codes since $\bin(S)$ and $\cat(S)$ are
spanned by the same Fock states.$^{\ref{fn:shift}}$ Therefore, the
ability for $\bin(S)$ to perfectly detect $\ell\leq S$ loss events
using the check operator (\ref{eq:disrot}) is identical to that of
$\cat(S)$: $\e_{0\ell}^{\bin}=\pb E_{0}^{\dg}E_{\ell}\pb=0$ in the
language of the QEC matrix (\ref{eq:qec}). The difference lies in
the other parameter $N$, the discrete analogue of $\a$ for $\cat$.
Recall that for $\cat(S)$, the limit $\a\rightarrow\infty$ exponentially
suppresses all uncorrectable parts in $\e_{\ell\ell}^{\cat}$ for
all $\ell$. Similarly, in $\bin(S)$, tuning $N$ allows one to suppress
distortion due to loss events up to a given order $O(\g^{N})$. Therefore,
it should be no surprise that for a fixed $S$,
\begin{equation}
\bin(N\rightarrow\infty,S)\,\sim\,\cat(\a\rightarrow\infty,S)\,.\label{eq:limit}
\end{equation}
While the two codes coincide in this limit, the advantage of $\bin$
codes is that, unlike $\cat$, the suppression of distortion in $\e_{\ell\ell}^{\bin}$
occurs \textit{without oscillations}. While the oscillatory nature
of the normalization factors in $\e_{\ell\ell}^{\cat(S)}$ (\ref{eq:ecat})
allows for peaks as well as troughs in $\fe$ vs. $\nb$ {[}shown
in Figs.~\ref{f:nbar}(a-c){]}, there are no such oscillations in
$\e_{\ell\ell}^{\bin(S)}$. Note that this difference only shows up
at $S\apprge3$ since only then are there significant oscillations
in $\fe(\cat)$. While $\cat$ and $\bin$ perform about the same
at smaller $S$, there is an intermediate $\nb$ regime for larger
$S$ at which $\cat$ underperforms (due to being at a trough between
two sweet spots $\a_{\star}$) while $\bin$ continues to improve.
For example, observe the difference between $\cat$ and $\bin$ at
$S=3$ in Fig.~\ref{f:nbar}(a): $\fe(\bin)\approx\fe(\cat)$ for
all $\nb\leq15$ except at $\nb\approx6$.

In Sec.~\ref{subsec:Error-correction-criteria-for}, we delve into
the performance of $\bin$ codes from Figs.~\ref{f:num} and \ref{f:nbar}.
In Sec.~\ref{subsec:Removing-energy-constraints}, we comment on
their performance with no energy constraints. Switching gears in Sec.~\ref{subsec:Relation-to-spin-coherent},
we study their structure. We show that qubit $\bin$ codes are spin-coherent
states embedded into the $\{|(S+1)m\ket\}_{m=0}^{N+1}$ subspace of
the oscillator. In an alternative characterization in Appx.~\ref{appx:Qudit-binomial-codes},
we link $\bin$ codes to discrete-variable bit-flip codes. These formulations
extend to other sets of codes, summarized in Table~\ref{f:bin}(a),
and allow one to construct $\bin$ check operators for dephasing errors.
In Sec.~\ref{subsec:Error-correction-procedure-for}, we introduce
a scheme to detect and correct errors in $\bin$ using the check operators
from before.

\subsection{QEC matrix for binomial codes\label{subsec:Error-correction-criteria-for}}

Let us fix $S$ and compare $\bin(S)$ to $\cat(S)$. A $\bin(N,S)$
code protects against loss errors $\aa^{\ell\leq S}$ (due to spacing
$S$) and dephasing errors $\ph^{n\leq N}$ (due to the nature of
the binomial distribution). Since loss operators $E_{\ell}$ (\ref{eq:krausloss})
consist of superpositions of the two errors, we can readily read off
the leading order in $\g$ for which there are uncorrectable parts
in $\e_{0\ell}^{\bin}$ \textemdash{} $O(\g^{S+1})$ \textemdash{}
and distortion matrices $\e_{\ell\ell}^{\bin}$ \textemdash{} $O(\g^{N+1})$.
However, that is not the whole story.

We know that both $\bin$ and $\cat$ suppress all distortion errors
$z_{\ell\ell}$ with increasing $\nb$. The backaction distortion
$z_{\ell\ell}^{\bin}$ does not oscillate with $\nb$ (as opposed
to $z_{\ell\ell}^{\cat}$ oscillation with $\a$) and yields a quicker
suppression of errors than $z_{\ell\ell}^{\cat}$. We use the basis
of positive/negative superpositions $|\pm_{\bin}\ket$ of the $\bin$
states (\ref{eq:bin}) to calculate it,
\begin{equation}
\pb=|+_{\bin}\ket\bra+_{\bin}|+|-_{\bin}\ket\bra-_{\bin}|\,,
\end{equation}
in order to have the backaction-induced errors be along the $z$-axis
and match $z_{\ell\ell}^{\cat}$. Note that this amounts to the $Z_{\bin}$-basis
of the original paper \cite{bin}. The respective probabilities of
no loss and backaction distortion can be concisely expressed,\begin{subequations}
\begin{align}
c_{\ell\ell}^{\bin} & =\frac{\g^{\ell}}{\ell!}\frac{d^{\ell}}{dx^{\ell}}\left.\left(\frac{1+x^{S+1}}{2}\right)^{N+1}\right|_{x=1-\g}\\
z_{\ell\ell}^{\bin} & =\frac{\g^{\ell}}{\ell!}\frac{d^{\ell}}{dx^{\ell}}\left.\left(\frac{1-x^{S+1}}{2}\right)^{N+1}\right|_{x=1-\g}\,.
\end{align}
\end{subequations}For $\ell=0$, the above should be compared to
$c_{00}^{\cat}$ (\ref{eq:ccat}) and $z_{00}^{\cat}$ (\ref{eq:zcat}).
Clearly, $z_{00}^{\bin}$ does not oscillate vs. $N$. While the above
is still difficult to analyze analytically for $\ell>0$, we see numerically
that there are no oscillations in $\nb$ of the fidelity of $\bin(S)$,
leading to certain regimes of $\nb$ at which $\bin$ outperforms
$\cat$ for a given $\g$. Heuristically, as $\nb\rightarrow\infty$,
$z_{\ell\ell}^{\bin(S)}\rightarrow0$ order-by-order in $\g$ while
$z_{\ell\ell}^{\cat(S)}\rightarrow0$ as a damped oscillating function
with damping coefficient $(1-\g)\a^{2}$. The latter limit turns out
to be less controlled, leading to detrimental oscillations in $\fe^{\cat(S)}$.
For example, we plot the uncorrectable parts of $\e_{\ell\lp}^{\cat}$
and $\e_{\ell\lp}^{\bin}$ for $\g=0.095$, $\nb_{\cc}\approx6$,
and $S=3$ in Fig.~\ref{f:qec}(a) and (b), respectively. While the
uncorrectable parts $x_{0\ell}^{\bin},x_{0\ell}^{\cat}$ are comparable
in magnitude, one can see that $z_{\ell\ell}^{\bin}$ is visibly less
than $z_{\ell\ell}^{\cat}$. However, this effect is most prominent
only when $E_{0}$ is the only dominant error ($\g\nb_{\cc}\ll1$)
and when $\cat$ oscillations begin to have a detrimental effect ($S\apprge3$
and $\nb_{\cc}\geq5$). Inside those regimes, we can see that $\bin$
breaks away from $\cat$ {[}see Figs.~\ref{f:nbar}(a-c) and insets
in Figs.~\ref{f:num}(b,c){]} while outside of those regions, the
two codes perform quite similarly.

Another advantage of $\bin$ codes manifests itself at large $\nb_{\bin}$.
Studying $x_{0,\ell}^{\bin}$ and $y_{0,\ell}^{\bin}$ is quite difficult,
so we explain the advantage by studying subspaces that are mapped
to under errors. For $\cat$ codes, the undetectable error $\aa^{S+1}$
maps the code \textit{exactly} to the code subspace, $\aa^{S+1}\pc\propto\pc$.
For $\bin$ codes, the mapping is to a subspace that has a component
orthogonal to the code space. Quantum information in this component
(which may only be one-dimensional) can then be mapped back to the
code space, yielding an extra layer of approximate error correction.
The same is true for $\ell>S+1$ as long as $N$ is sufficiently high.
We consider two examples of this effect, one known and one new.

Let us consider the action of the undetectable error $\aa^{2}$ on
$\bin(N=1,S=1)$ and $\cat(\a\gg1,S=1)$ ($\a$ is large only for
simplicity; its value is irrelevant to the key point). The logical
states $|+_{\bin}\ket\propto|0\ket+|4\ket$ and $|-_{\bin}\ket=|2\ket$
are mapped to states $|2\ket$ and $|0\ket$, respectively. We see
that the latter error state, $\aa^{2}|-_{\bin}\ket\propto|0\ket$,
overlaps with $|0\ket-|4\ket$, a state orthogonal to the code space.
One can thus add an extra Kraus operator to the recovery that maps
any quantum information in this extra error subspace back to the code
space. This cannot be done for $\cat$, where the logical states $|+_{\cat}\ket\propto|\a\ket+\kkk{-\a}$
and $|-_{\cat}\ket\propto|i\a\ket+\kkk{-i\a}$ are mapped to $\pm|\pm_{\cat}\ket$
under $\aa^{2}$ (recall that $\aa|\a\ket=\a|\a\ket$), yielding a
completely uncorrectable phase flip. For $\bin(N=1,S=1)$, this extra
one-dimensional subspace is used to correct from the leading-order
backaction error \cite{bin}. However, there are enough such extra
subspaces for sufficiently high $N$ to correct for both backaction
(up to $\g^{N}$) \textit{and }some loss errors $\aa^{\ell\geq S+1}$.

Now consider $\bin(N=4,S=1)$. Recall that a $\bin(N=4)$ protects
from backaction up to $\g^{4}$, so our calculations are only to that
order. While the code exactly detects only one loss $E_{1}$, it turns
out there is an extra subspace allowing for approximate correction
of $E_{2}$ and even $E_{3}$. The two-dimensional code subspace $\text{Span}\{|\pm_{\bin}\ket\}$
is supported on the six-dimensional Fock subspace 
\begin{equation}
\h=\text{Span}\{|0\ket,|2\ket,|4\ket,|6\ket,|8\ket,|10\ket\}\,.
\end{equation}
Another two-dimensional subspace is reserved for correcting backaction
$E_{0}$. This leaves an extra two dimensions $\text{Span}\{Q_{0}E_{2}|\pm_{\bin}\ket\}\subset\H_{\text{no-loss}}$
for approximately correcting the loss error $E_{2}$, where the projection
$Q_{0}$ removes any overlap with the code space and the subspace
used for correcting backaction. Indeed, one can add an extra isometry
mapping such error states\begin{subequations}
\begin{align}
\!\!\!\!Q_{0}E_{2}|+_{\bin}\ket & \propto\sqrt{2}\eta^{2}|2\ket-(1+\eta^{2})|6\ket+\sqrt{10}|10\ket\\
\!\!\!\!Q_{0}E_{2}|-_{\bin}\ket & \propto\sqrt{10}\eta^{2}|0\ket-(1+\eta^{2})|4\ket+\sqrt{2}|8\ket
\end{align}
\end{subequations}back into the code space ($\eta\equiv1-\g$).

Similarly, this code can also approximately correct the next error
$E_{3}$. The relevant Fock subspace is now 
\begin{equation}
\ho=\text{Span}\{|1\ket,|3\ket,|5\ket,|7\ket,|9\ket\}\,.
\end{equation}
The two-dimensional subspace $\text{Span}\{E_{1}|\pm_{\bin}\ket\}$
is devoted to correcting $E_{1}$, leaving three extra dimensions.
Two of those dimensions are then used to correct against $E_{3}$.
Letting $Q_{1}$ be the projection on the remaining three-dimensional
subspace $\ho/\text{Span}\{E_{1}|\pm_{\bin}\ket\}$, one can construct
a mapping from the extra error subspace $\text{Span}\{Q_{1}E_{3}|\pm_{\bin}\ket\}$
back to the code space.

\begin{figure}[t]
\includegraphics[width=1\columnwidth]{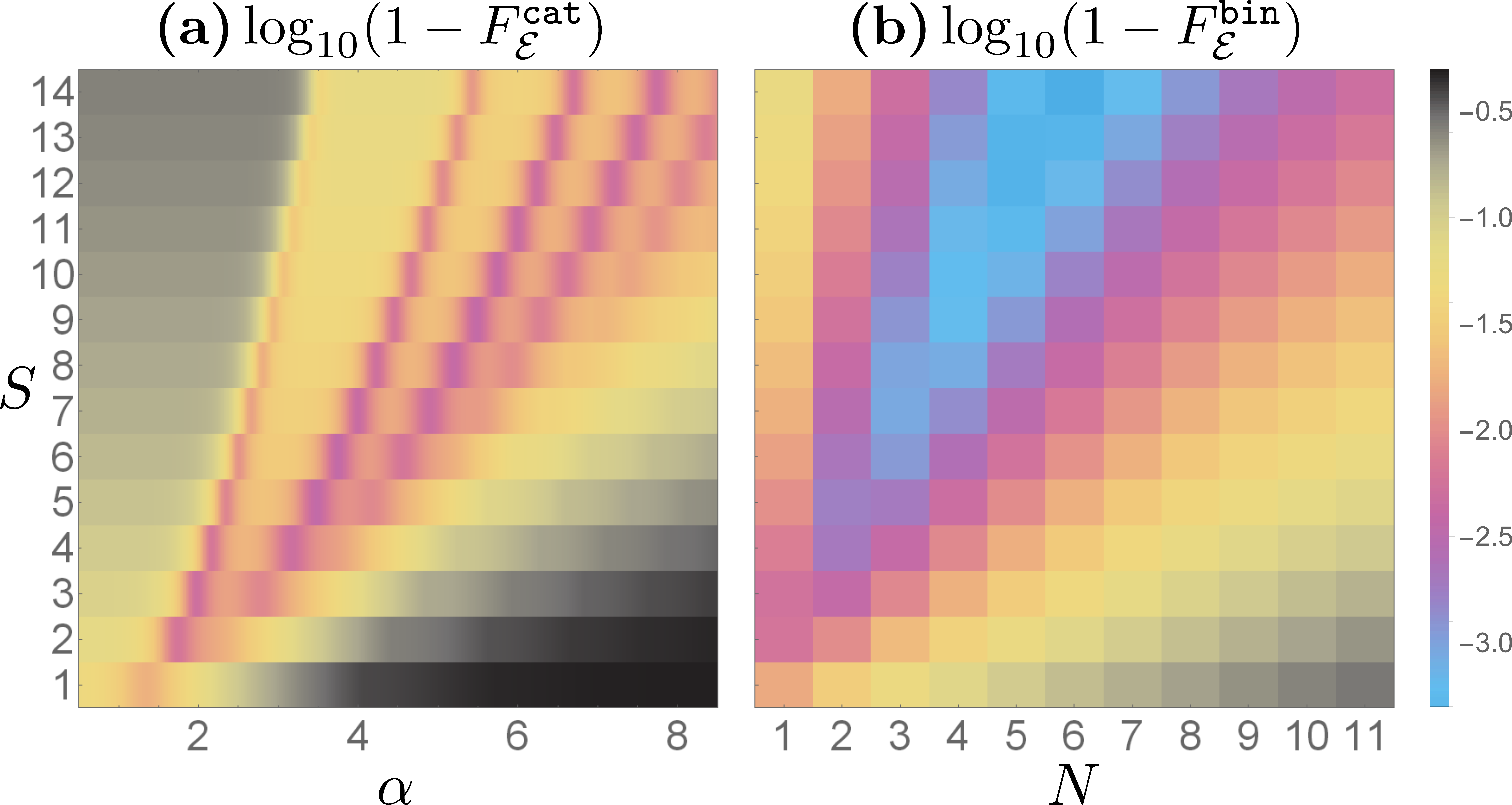}\caption{\label{f:binlarge}Log$_{10}$ plot of $1-\protect\fe(\protect\g=0.095)$
(\ref{eq:fid}) vs. \textbf{(a)} $\protect\cat$ code parameters $S$
and $\protect\a$ and \textbf{(b)} $\protect\bin$ code parameters
$S$ and $N$ (cf. Fig.~1 in Ref.~\cite{bin}). For a given $S$,
$\protect\cat$ achieves the best performance (purple) at multiple
$S$-dependent sweet-spots $\protect\a_{\star}$. While $\protect\fe(\protect\a_{\star})$
for $\protect\cat(S)$ does not increase with increasing $S$ for
the values we've sampled, performance of $\protect\bin(N,S\approx2N)$
clearly does (cyan). There is thus reason to believe that $\protect\bin$
outperforms $\protect\cat$ when no energy constraints are imposed.}
\end{figure}

\subsection{Removing energy constraints\label{subsec:Removing-energy-constraints}}

We have numerically investigated $\nb_{\bin}\geq10$ in order to see
whether a certain direction in the $N,S$ parameter space produces
increasing $\fe$ with increasing $\nb_{\bin}$ (\ref{eq:nbarbin}).
While it is unlikely that $\fe\rightarrow1$ as $\nb\rightarrow\infty$
for \textit{any} single-mode code, we have numerical evidence showing
that $\fe^{\bin(N)}$ for certain $S=\xi N$ monotonically increases
to some value $\fe^{\bin(\infty)}<1$, with $\xi\geq1$ dependent
on $\g$. For example, as shown in Fig.~\ref{f:binlarge}, the limit
$S\approx2N\rightarrow\infty$ does seem to be giving us monotonically
increasing $\fe$; we have verified this monotonic increase for $N\leq14$
and $S\leq36$ at $\g=0.1$ but it could very well be that the curves
eventually decrease for sufficiently high $N,S$.

The performance improvement of $\bin$ codes at large $\nb_{\bin}$
can be attributed to the advantage of having extra error subspaces,
discussed in the latter portion of Sec.~\ref{subsec:Error-correction-criteria-for}.
To quantify this advantage, one can cook up a recovery procedure consisting
of two sets of isometries. The \textit{first-level set} of recovery
isometries maps the \textit{correctable} ($\ell\in\{0,1\cdots,S\}$)
error-subspace code states $E_{\ell}|\pm_{\bin}\ket$ back to the
codespace. This part is similar to the $\cat$ code recovery scheme
from Ref.~\cite{Li2016} and to the $\bin$ scheme in Sec.~\ref{subsec:Error-correction-procedure-for}.
The \textit{second-level set} consists of isometries mapping extra
error subspaces $Q_{\ell}E_{\ell+S+1}|\pm_{\bin}\ket$ back into the
code space, where $Q_{\ell}$ project out the first-level error subspaces
$\text{Span}\{E_{\ell}|\pm_{\bin}\ket\}$ (and, in the case of $\ell=0$,
the codespace as well). Of course, such a multi-level recovery can
be extended to three and more levels. When implemented, the two-level
recovery yields a similar scaling with $N,S$ as the optimal recovery
in Fig.~\ref{f:binlarge}(b) for the $N,S$ we were able to sample.
Nevertheless, it does not explain why $S$ increases faster than $N$
for the best codes in that figure. While we have shown that increasing
both $S$ and $N$ allows $\bin$ to correct (at least) approximately
for more loss events, the intricate choice of which parameter to increase
faster to give the optimal fidelity remains an interesting open question.

\subsection{Relation to spin-coherent states\label{subsec:Relation-to-spin-coherent}}

We characterize all single-mode binomial codes, two-mode binomial
codes (closely related to $\noon$ codes \cite{Bergmann2016a}), and
multi-qubit permutation invariant codes \cite{Ouyang2014} using irreducible
representations (irreps) of the Lie algebra $\su(2)$. Interestingly,
$\bin(S=0)$ were of interest to the quantum optical community due
to their sub-Poissonian distribution \cite{Stoler1985} (see also
\cite{dodonov}, Ch.~5), and the connection to $\su(2)$ was noticed
first back then \cite{Dattoli1987,Hong-yi1994}.

Consider a spin-$J$ consisting of $2J+1$ levels and define the standard
spin operators
\begin{equation}
J_{z}\equiv\sum_{m=0}^{2J}(m-J)|J,m-J\ket\bra J,m-J|\,,\label{eq:jz}
\end{equation}
and similarly $J_{x}$ and $J_{y}$ (\cite{schiff}, Ch.~7). We can
then define its rotated version $J_{r}\left(\t,\phi\right)=R_{\t,\phi}J_{z}R_{\t,\phi}^{\dg}$,
where $R_{\t,\phi}$ is a rotation by azimuthal angle $\theta\in[0,\pi]$
and polar angle $\phi\in[0,2\pi)$ in the spherical coordinate system
parameterizing the spin's Bloch sphere. For each $\{\t,\phi\}$,
$J_{r}(\t,\phi)$ has eigenstates $|\t,\phi\ket_{J}$ with eigenvalue
$J$. These are called the $\su(2)$ or spin-coherent states \cite{Arecchi1972}
(see also \cite{puri}):
\begin{equation}
|\t,\phi\ket_{J}=\sum_{m=0}^{2J}\frac{(e^{i\phi}\tan\frac{\t}{2})^{m}}{(1+\tan^{2}\frac{\t}{2})^{J}}\sqrt{{2J \choose m}}|J,m-J\ket\,.\label{eq:spin}
\end{equation}
For each $J\in\{0,\half,1,\cdots\}$, $\{J_{x},J_{y},J_{z}\}$ form
an irrep of the $\su(2)$ algebra, satisfying the well-known angular
momentum commutation relations. The labeling by $J$ \textit{exhaustively
characterizes} the irreps of $\su\left(2\right)$, so every spin-coherent
state corresponds to some irrep $J$. We go through several codes
and show how they all correspond to the spin-coherent states 
\begin{equation}
\kkk{{\textstyle \frac{\pi}{2},\pi\m}}_{J}=\frac{1}{2^{J}}\sum_{m=0}^{2J}(-1)^{\m m}\sqrt{{2J \choose m}}|J,m-J\ket,\label{eq:spinspecific}
\end{equation}
whose basis elements $|J,m-J\ket$ are mapped either to Fock states
of an oscillator(s) or a multi-qubit system, summarized in Fig.~\ref{f:bin}(a).
Moreover, 
\begin{equation}
\frac{J_{x}}{J}\kkk{{\textstyle \frac{\pi}{2},\pi\m}}_{J}=\left(-1\right)^{\m}\kkk{{\textstyle \frac{\pi}{2},\pi\m}}_{J}\,,
\end{equation}
providing us with a logical $Z$-operator $J_{x}/J$ and a check operator
$(J_{x}/J)^{2}$ for all of the codes. In addition, since spin-coherent
states resolve the identity operator of the spin, they offer a way
to visualize the location of various states before and after certain
errors on a generalized Bloch sphere of the spin. Shown in Fig.~\ref{f:bin}(b),
$|{\textstyle \frac{\pi}{2},\pi\m}\ket_{J}$ correspond to spin-coherent
states at the antipodal points $(\frac{\pi}{2},0)$ and $(\frac{\pi}{2},\pi)$.

\subsubsection{Binomial codes}

Setting $2J=N+1$, the coefficients of $|\frac{\pi}{2},\pi\m\ket_{J}$
(\ref{eq:spinspecific}) match those of the $\bin$ states (\ref{eq:bin}).
If we additionally map the spin states $|J,m-J\ket$ into Fock states
$|m\ket$, then
\begin{equation}
|{\textstyle \frac{\pi}{2},\pi\m}\ket_{J=\frac{N+1}{2}}\rightarrow|\m_{\bin(N,S=0)}\ket\,.
\end{equation}
The operator $J_{z}$ (\ref{eq:jz}) is then mapped to $\ph-J$, revealing
the well-known Holstein-Primakoff mapping of a spin into a boson.
For larger spacing $S>0$, we map $|J,m-J\ket$ into Fock states $|(S+1)m\ket$.
Therefore, we have shown that $\bin$ codes correspond to single-mode
irreps of $\su(2)$ produced by the Holstein-Primakoff mapping.

For $\bin$ codes, the check operator $(J_{x}/J)^{2}$ is related
to the protection from dephasing errors (characterized by $N$), and
its non-destructive measurement can be used to detect such errors
(see Sec.~\ref{subsec:Error-correction-procedure-for}). Moreover,
the Bloch sphere picture offers a nice interpretation of why a $\bin(N,S)$
code protects against $k\leq N$ dephasing errors. Since $J_{z}=\ph-J$
in this irrep, the action of a dephasing error $\ph^{k}$ is directly
related to application of $J_{z}$ and the code states are eigenstates
of $J_{x}$ at antipodal parts of the Bloch sphere. Thus, one action
of $J_{z}$ raises (lowers) the expectation value of $J_{x}$ for
the logical zero (one) state, moving them closer together from their
antipodal positions at the equator {[}Fig.~\ref{f:bin}(b){]}. The
states only begin to overlap when a high enough power $(J_{z})^{k>N}$
has been applied, which corresponds to an unprotected dephasing error
$\ph^{k>N}$ in the bosonic representation.

\subsubsection{Permutation-invariant codes\label{subsec:Permutation-invariant-codes}}

\begin{figure}[t]
\begin{tabular}{c}
\textbf{\!\!\!\!\!\!\!\!\!\!\!(a)}\tabularnewline
$\phantom{D_{S}^{M}}$\tabularnewline
$\phantom{D_{S}^{M}}$\tabularnewline
$\phantom{D_{S}^{M}}$\tabularnewline
\tabularnewline
\end{tabular}\!\!%
\begin{tabular}{cc}
\toprule 
$\cc$ & basis\tabularnewline
\midrule
$\bin$ & Fock states $|(S+1)m\ket$\tabularnewline
$\pt$ & $M$-qubit Dicke states $|D_{(S+1)m}^{M\geq2J(S+1)}\ket$\tabularnewline
$\tw$ & ~~Fock states $|(S+1)(2J-m),(S+1)m\ket$\tabularnewline
\bottomrule
\end{tabular}~~~~~~

\medskip{}

\includegraphics[width=1\columnwidth]{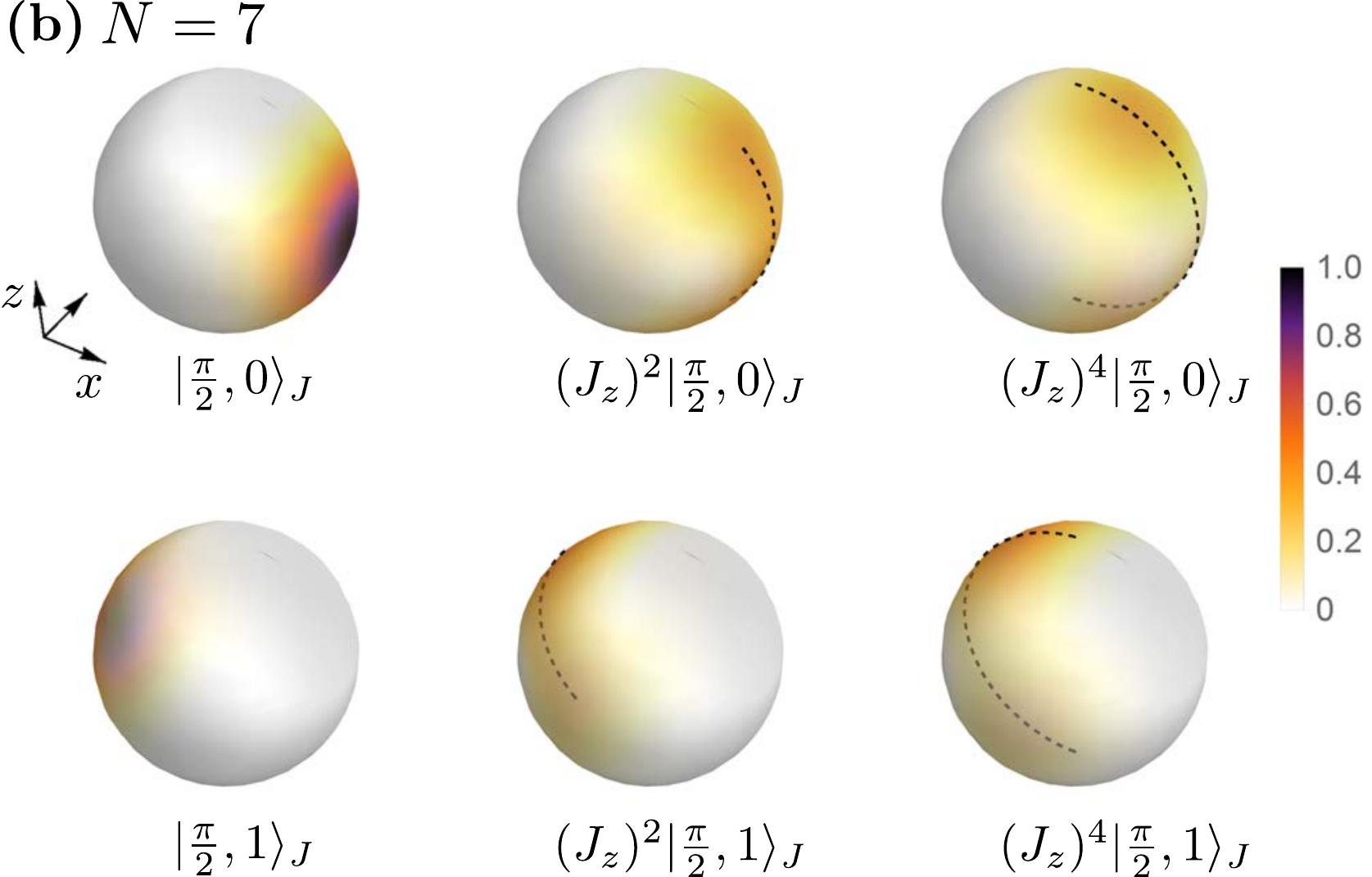}

\caption{\label{f:bin}\textbf{(a)} Table listing the basis elements used to
express $\protect\bin$omial (\ref{eq:bin}), $\protect\pt$utation-invariant
(\ref{eq:perm}), and two-mode binomial codes $\protect\bin2$ (\ref{eq:bin2})
(with $m\in\{0,1,\cdots,2J\}$). The coefficients next to these basis
states are those of spin-coherent states $|{\textstyle \frac{\pi}{2},\pi\protect\m}\protect\ket_{J}$
(\ref{eq:spin}) of a spin $J$. \textbf{(b)} Plots of overlaps $|_{J}\protect\bra\protect\t,\phi|\psi\protect\ket|^{2}$
vs. $\protect\t,\phi$ given a spin state $|\psi\protect\ket$. The
first (second, third) column shows normalized states $\{(J_{z})^{p}|{\textstyle \frac{\pi}{2},\pi\protect\m}\protect\ket_{J=4}\}_{\protect\m=0}^{1}$
for $p\in\{0,2,4\}$. These plots show that powers of the ``error''
$J_{z}$ cause $|{\textstyle \frac{\pi}{2},0}\protect\ket_{J}$ and
$|{\textstyle \frac{\pi}{2},1}\protect\ket_{J}$ to approach each
other in the Bloch sphere and eventually overlap at the two poles.
The dashed arcs connecting $\protect\t=\pm\frac{\pi}{2J}p$ serve
as a guide to the eye. Since $J_{z}$ is mapped to $\protect\ph-J$
under the Holstein-Primakoff transformation, this provides an interpretation
of dephasing errors for $\protect\bin$ codes, which are correctable
as long as $p<J=\protect\half(N+1)$. Here, $N=7$ and one can see
from the third column that $(J_{z})^{4}|{\textstyle \frac{\pi}{2},0}\protect\ket_{J}$
and $(J_{z})^{4}|{\textstyle \frac{\pi}{2},1}\protect\ket_{J}$ overlap.}
\end{figure}

These codes (denoted here as $\pt$) were introduced by Ouyang \cite{Ouyang2014}
(see also \cite{ouyangthesis}) to tackle single-qubit amplitude damping.
Given $M$ qubits and parameters $\{J,S\}$, the logical states are
\begin{equation}
|\m_{\pt}\ket=\frac{1}{2^{J}}\sum_{m=0}^{2J}\left(-1\right)^{\m m}\sqrt{{2J \choose m}}|D_{\left(S+1\right)m}^{M}\ket\,,\label{eq:perm}
\end{equation}
where the Dicke state $|D_{\left(S+1\right)m}^{N}\ket$ is the fully
symmetrized $M$-qubit state with $\left(S+1\right)m$ qubits in state
$|1\ket$ and the remaining qubits in $|0\ket$. Therefore, we need
to have $M\geq2J(S+1)$ in order to accommodate all of the required
Dicke states. If $2J=3S+1$ and $M=3S^{2}+5S$, these codes can protect
against qubit amplitude damping errors of weight $S$ \cite{Ouyang2014}.
In this context, the spacing $S$ (between excitations of the Dicke
states) quantifies a distance of the codes. We can already see the
resemblance to spin-coherent states, but here the irrep is more complicated.
For simplicity, let us set $S=0$ and $M=2J$; the $S>0$ case is
a straightforward extension whose basis elements are shown in Fig.~\ref{f:bin}(a).
For such cases, $|\m_{\pt}\ket$ are spin-coherent states of the largest
irrep of $\mathfrak{su}\left(2\right)$ arising from tensoring $M$
spin-$\nicefrac{1}{2}$ particles, with $J=\nicefrac{M}{2}$ playing
the role of a collective spin.

\subsubsection{Two-mode binomial codes}

There is another well-known $\su(2)$-related construct \textemdash{}
the Jordan-Schwinger mapping of a spin into two bosons. Letting $\aa_{1},\aa_{2}$
be the lowering operators of the two bosons and $X,Y,Z$ be the Pauli
matrices, we have
\begin{equation}
J_{[x,y,z]}=\half\sum_{j,k=0}^{1}\aa_{j}^{\dg}[X,Y,Z]_{jk}\aa_{k}\,.
\end{equation}
For example, $J_{z}=\half(\aa_{1}^{\dg}\aa_{1}-\aa_{2}^{\dg}\aa_{2})$.
The state space associated with each irrep of this type corresponds
to the subspace of fixed total occupation number, i.e., all two-mode
Fock states $|n_{1},n_{2}\ket$ such that $n_{1}+n_{2}=2J$ for an
irrep of spin $J$. One can thus see that the total spin is proportional
to the ``identity'' component on the subspace, $J=\half(\aa_{1}^{\dg}\aa_{1}+\aa_{2}^{\dg}\aa_{2})$.
Any code within a subspace of fixed $J$ thus consists of eigenstates
of the joint backaction $(1-\g)^{(\aa_{1}^{\dg}\aa_{1}+\aa_{2}^{\dg}\aa_{2})/2}$
in the two-mode pure-loss error operators (\ref{eq:krausloss}) (assuming
identical $\g$'s for both modes). Codes having this structure were
first considered in Ref.~\cite{Chuang1997}.

Spin-coherent states of these irreps correspond to a class of two-mode
binomial codes ($\tw$). Mapping the basis $|J,m-J\ket$ from Eq.~(\ref{eq:spinspecific})
to Fock states $|(S+1)(2J-m),(S+1)m\ket$ yields the $\tw$ code states
\begin{equation}
\!|\m_{\tw}\ket=\frac{1}{2^{J}}\sum_{m=0}^{2J}\left(-1\right)^{\m m}\sqrt{{2J \choose m}}|2J-(S+1)m,(S+1)m\ket.\label{eq:bin2}
\end{equation}
As with $\bin$, we can parameterize $\tw$ in terms of spacing $S$
and a dephasing error parameter $N$. Then, $J=\half(N+1)(S+1)$ in
order to fit all of the required two-mode Fock states. Note that $\tw(S=0)$
can be obtained by acting on the Fock state $|S+1,0\ket$ with a 50:50
beamsplitter \cite{Bergmann2016a} and were considered before in the
context of three-mode squeezing \cite{Shaterzadeh-Yazdi2008}. 

\subsubsection{Further generalizations}

We have covered four bases in which to embed a spin \textemdash{}
the spin's own basis $\{|J,m-J\ket\}_{m=0}^{2J}$, Fock states $\{|m\ket\}_{m=0}^{2J}$,
Dicke states $\{|D_{m}^{M}\ket\}_{m=0}^{2J}$, and two-mode Fock states
$\{|2J-m,m\ket\}_{m=0}^{2J}$. There are other relations between these
bases and further code extensions. First, we can go in reverse of
what was discussed above and embed any bosonic code into a multi-qubit
Hilbert space by mapping Fock states to Dicke states. While this produces
$\pt$ codes when the bosonic code is $\bin$, it produces previously
unexplored codes when the bosonic code is, e.g., $\cat$ or $\lat$
(although such states require $J\rightarrow\infty$ due their infinite-dimensional
support). Second, Dicke states $\{|D_{m}^{M}\ket\}_{m=0}^{2J}$ converge
to Fock states in the limit of fixed $J$ but large $M\gg2J$ \cite{Arecchi1972}.
This famous limit is equivalent to the south pole of the Bloch sphere
flattening out into ordinary bosonic phase space in the limit that
the Bloch sphere is infinitely large. In this limit,
\begin{equation}
\pt(M\rightarrow\infty,J,S)\rightarrow\bin(N=2J-1,S)\,.
\end{equation}
Third, the $\tw$ states can be tensored $2J$ times to construct
logical states for the $4J$-mode $\noon$ code \cite{Bergmann2016a},
$|\m_{\noon}\ket=|\m_{\tw}\ket^{\otimes2J}$. The same procedure can
of course be applied to $\bin$ codes. Offering an interesting alternative
to spacing, $\noon$ codes instead concatenate $\tw$ with a $2J$-block
bit-flip code to correct for up to $2J-1$ loss errors. Fourth, qubit
(qudit) $\bin$ codes can \textit{themselves} be thought of as bit-flip
codes when expressed in a basis of multi-qubit (multi-qudit) states
(see Appx.~\ref{appx:Qudit-binomial-codes}).

\subsection{Error-correction procedure for binomial codes\label{subsec:Error-correction-procedure-for}}

The existence of approximate error recovery maps for the various codes
does not explicitly suggest by what means these recovery maps can
be implemented nor whether fault-tolerant error recovery is possible
for these codes. For qubit stabilizer codes, the theory of fault-tolerant
error correction has been developed. For $\lat$, methods of fault-tolerant
quantum error correction \cite{Gottesman2001} are possible which
simply generalize the techniques of qudit ($d$-dimensional) stabilizer
codes to $d\rightarrow\infty$.

In this section, we investigate for the binomial qubit codes what
measurements of commuting check operators could give sufficient error
information to undo a set of errors. The recovery procedure we give
is not necessarily the optimal one obtained by optimization in Sec.~\ref{sec:Take-home-messages}.
The binomial code family $\bin(N,S)$ can correct against errors in
the error set $\err=\{I,\aa,\ldots,\aa^{L},\aa^{\dagger},\ldots,(\aa^{\dagger})^{G},\hat{n},\ldots\hat{n}^{D}\}$
with $S=L+G$ and $N={\rm max}(L,G,2D)$. We know from Eq.~(\ref{eq:spinspecific})
that the codewords correspond to antipodal spin-coherent states of
spin $J=\half(N+1)$. We will refer to the $N+1$-dimensional subspace
as the spin-space.

Imagine that one error from the set $\err$ has taken place on an
encoded state. The following procedure describes how to undo this
error. First, one non-destructively measures the eigenvalues of the
check operator $J_{x}^{2}$ which has eigenvalue $+J^{2}$ on all
states in the codespace. Here we assume that the operator $J_{x}$
only has support on the Fock states $|m(S+1)\ket$ and thus has zero
eigenvalues elsewhere. Of course the check operator $J_{x}$ is not
unique and any form of non-destructively learning the value $|m_{x}|$
is permitted. For odd $N$ (integer spin $J$), such a measurement
has outcomes $|m_{x}|^{2}$ with $|m_{x}|=0,1,\cdots,J$. The outcome
$m_{x}=0$ cannot have come about from one of the dephasing errors
of the form ${\hat{n}}^{k}$ since this error operator maps an initial
state with $|m_{x}|=J$ to a superposition of states with $|m_{x}|\geq J-k$
so that for $k\leq D$, one can not reach $|m_{x}|=0$. For even $N$
(half-integer spin $J$), $m_{x}$ will never be zero by the application
of a dephasing error. Hence if one finds the eigenvalue $m_{x}=0$,
one concludes that photon loss or photon gain errors of the form $\aa^{k}$,
$k\leq L$ and $(\aa^{\dagger})^{l}$, $l\leq G$ must have occurred.
In order to learn more about these photon loss and gain errors, one
then measures the photon parity check operator (\ref{eq:disrot}).
If one finds any another value of $|m_{x}|=k$ for the first measurement,
one rotates the two-dimensional $m_{x}=\pm k$ subspace back to the
two-dimensional $m_{x}=\pm J$ subspace by some unitary transformation.
For stabilizer (resp. $\lat$), this correction can be a Pauli operator
(resp. small displacement). Note that, unlike for stabilizer, $\cat$
or $\lat$ codes, it is necessary to physically apply the correction.
In other words, unlike the use of Pauli frames \cite{Knill2005},
we cannot just record the value of $|m_{x}|$ and keep the quantum
information in this error space with a lower value for $|m_{x}|$:
subsequent dephasing errors would lead to more laddering up and down
in the spin-space so the QEC conditions would no longer be met {[}see
Fig.~\ref{f:bin}(b){]}. In case $m_{x}=0$, one non-destructively
measures the eigenvalues of $C_{\bin}$ (\ref{eq:disrot}) (via phase
estimation, say), allowing one to learn the photon parity $k$ modulo
$S+1$. When $k\leq G$ (at most $G$ photons are gained), one applies
\begin{equation}
U_{k}^{+}=\sum_{\mu=0}^{1}\frac{|\mu_{\bin}\ket\bra\mu_{\bin}|\aa^{k}}{\sqrt{\bra\mu_{\bin}|\aa^{k}(\aa^{\dagger})^{k}|\mu_{\bin}\ket}}+V_{{\rm else}}^{+}\,,
\end{equation}
where $V_{{\rm else}}^{+}$ is chosen to make $U_{k}^{+}$ unitary.
When $k>G$ (at most $L$ photons lost), one applies 
\begin{equation}
U_{l}^{-}=\sum_{\mu=0}^{1}\frac{|\mu_{\bin}\ket\bra\mu_{\bin}|(\aa^{\dagger})^{l}}{\sqrt{\bra\mu_{\bin}|(\aa^{\dagger})^{l}\aa^{l}|\mu_{\bin}\ket}}+V_{{\rm else}}^{-}
\end{equation}
with $l=S+1-k$. These rotations are not a simple adding or subtracting
of photons since some Fock states in $|\mu_{\bin}\ket$ have been
annihilated.

This form of error correction unfortunately does not correct products
of dephasing and photon loss/gain errors, which are in principle errors
against which the code can correct \cite{bin}. Note also that Ref.~\cite{bin}
has shown that specifically for the photon loss channel with errors
as in Eq.~(\ref{eq:krausloss}), only the measurement of the rotation
operator $C_{\bin}$ (\ref{eq:disrot}) is required (since there is
one particular dephasing error associated with a particular number
of photon losses). The procedure above thus falls short of giving
a general prescription for error correction for the binomial codes.
However, for the binomial qubit code $\bin(N=2,S=1)$, one can give
a scheme which corrects all the errors which meet the quantum error
conditions. The code $\bin(N=2,S=1)$ can correct against the errors
in $\err=\{I,\aa,\hat{n}\}$. For these parameters, the code space
is inside the Fock space with a maximum of $6$ photons, $\mathcal{F}_{6}=\mathrm{Span}\{|0\ket,\cdots,|6\ket\}$.
We split this space into the direct sum of the even subspace and the
odd subspace, that is $\mathcal{F}_{6}=\mathcal{F}_{6}^{\text{even}}\oplus\mathcal{F}_{6}^{\text{odd}}=\mathrm{Span}\{|0\ket,|2\ket,|4\ket,|6\ket\}\oplus\mathrm{Span}\{|1\ket,|3\ket,|5\ket\}$.

The code space is inside $\mathcal{F}_{6}^{\text{even}}$. We identify
$\mathcal{F}_{6}^{\text{even}}$ with a spin $J=3/2$ via the mapping
\begin{eqnarray*}
|6\ket\equiv\kkk{m_{z}={\textstyle \frac{3}{2}}}, & \,\,\,\,\,\quad|4\ket\equiv\kkk{m_{z}={\textstyle \frac{1}{2}}},\quad\\
|2\ket\equiv\kkk{m_{z}=-{\textstyle \frac{1}{2}}}, & \,\,\,\,\,\quad|0\ket\equiv\kkk{m_{z}=-{\textstyle \frac{3}{2}}},
\end{eqnarray*}
so that $|0_{\bin}\ket$ and $|1_{\bin}\ket$ code states are the
highest and lowest eigenstates of $J_{x}(J=3/2)$, 
\[
|0_{\bin}\ket=\kkk{m_{x}={\textstyle \frac{3}{2}}},\qquad|1_{\bin}\ket=\kkk{m_{x}={\textstyle -\frac{3}{2}}}.
\]
One dephasing error leaves the code states inside $\mathcal{F}_{6}^{\text{even}}$.
More precisely, it is a linear combination of the identity and $J_{z}(J=3/2)$
such that \begin{subequations}
\begin{align}
\hat{n}|0_{\bin}\ket & =2\kkk{m_{x}={\textstyle \frac{1}{2}}}+3|0_{\bin}\ket,\\
\hat{n}|1_{\bin}\ket & =2\kkk{m_{x}=-{\textstyle \frac{1}{2}}}+3|1_{\bin}\ket.
\end{align}
\end{subequations}For convenience, we relabel the error states $\kkk{m_{x}=\pm{\textstyle \frac{1}{2}}}\equiv|n\pm\ket$.

Remarkably, one photon loss maps $\mathcal{F}_{6}^{\text{even}}$
onto $\mathcal{F}_{6}^{\text{odd}}$ in such a way that the code states
are mapped onto shifted$^{\ref{fn:shift}}$ code states for $N=1$.
This is in fact true for general $N$ and $S$: one photon loss maps
the code space $(N,S)$ to the code space $(N-1,S)$ but shifted by
$+S$ in the Fock basis \cite{bin}. With this in mind, we can identify
$\mathcal{F}_{6}^{\text{odd}}$ with a spin $J=1$ with the mapping
\[
|5\ket\equiv|m_{z}=1\ket,\quad|3\ket\equiv|m_{z}=0\ket,\quad|1\ket\equiv|m_{z}=-1\ket,
\]
and then the error states are the highest and lowest eigenstates of
$J_{x}(J=1)$, 
\[
\aa|0_{\bin}\ket\propto|m_{x}=1\ket,\quad\aa|1_{\bin}\ket\propto|m_{x}=-1\ket,
\]
which we rename for convenience: $|m_{x}=\pm1\ket\equiv|a\pm\ket$.
The third state of this spin-$1$ subspace is called the unknown state
as ending up in this state means the loss of logical information,
$|m_{x}=0\ket\equiv|?\ket$. To complete the description, one can
note that the action of $\hat{n}$ on $\mathcal{F}_{6}^{\text{odd}}$
is also a linear combination of $J_{z}(J=1)$ and the identity $I$,
i.e mapping $|a\pm\ket$ to a linear combination of itself and $|?\ket$,
and using the fact that $\aa\hat{n}=\hat{n}\aa-\aa$, one can note
that $|n\pm\ket$ is mapped by $\aa$ onto a linear combination of
$|a\pm\ket$ and $|?\ket$. These relations are summarized in Table~\ref{tab:threehalftable}.

\begin{table}
\begin{tabular}{c| c c c |c}
$J_x(J=3/2)$ & $\mathcal{F}_6^{\text{even}}$ & &$\mathcal{F}_6^{\text{odd}}$ & $J_x(J=1)$\\ 
\hline $\frac{3}{2}$ & $|\mu_\bin=0\ket$ & $\xrightarrow{\hat{a}}$ & $|a+\ket$ & 1\\[.5em]
& $\hat{n}\downarrow$ & & $\downarrow\hat{n}$ & \\[.5em] 
$\frac{1}{2}$ & $|n+\ket$ & \multirow{2}{*}{$\xrightarrow{\hat{a}}$} & \multirow{2}{*}{$|?\ket$}&\multirow{2}{*}{$0$}\\[1em]  $-\frac{1}{2}$ & $|n-\ket$ & & &\\[.5em]   
& $\hat{n}\uparrow$ & & $\uparrow\hat{n}$ & \\[.5em]  
$-\frac{3}{2}$ & $|\mu_\bin=1\ket$ & $\xrightarrow{\hat{a}}$ & $|a-\ket$ & $-1$\\[.5em]  \hline 
\end{tabular}

\caption{\label{tab:threehalftable}Relations between code states and error
states for $\protect\bin(N=2,S=1)$.}
\end{table}

One possible way to extract error information is then to measure (via
phase estimation) the eigenvalues of the following unitary 
\begin{equation}
U=\exp\left\{ \frac{2\pi i}{b}\left[a\left(J_{x}(3/2)\right)^{2}\oplus\left(J_{x}(1)\right)^{2}\right]\right\} ,
\end{equation}
where the two parameters $a$ and $b$ can be chosen to obtain good
spacing between different eigenvalues. For example one can choose
$a=8$ and $b=5$, leading to Table~\ref{tab:eigstruct}. 

To obtain the 4 eigenvalues of $U$ (via phase estimation), one needs
at least 2 qubit ancillas. A more direct method would be to first
measure photon parity. If odd, then correct for photon loss. If even,
then one measures the eigenvalues of $J_{x}^{2}$ by measuring $U=\exp(i32\pi J_{x}(3/2)^{2}/9)$,
which has eigenvalue $+1$ for the no-error case and eigenvalue $\exp(i32\pi/36)\approx-1$
in the dephasing error case.
\begin{table}[h]
\begin{tabular}{ccccc}
\hline 
Eigenstates & ~$|\mu=0\text{ or }1\ket$~ & $|n+\ket,|n-\ket$ & $|a+\ket,|a-\ket$ & $|?\ket$\tabularnewline
Eigenvalues & $\mathrm{e}^{i6\pi/5}$ & $\mathrm{e}^{i4\pi/5}$ & $\mathrm{e}^{i2\pi/5}$ & $1$\tabularnewline
Decoding & no error & dephasing & ~~photon loss~~ & failure\tabularnewline
\hline 
\end{tabular}\caption{Eigenstructure of the proposed unitary to be measured for error correction
of $\protect\bin(N=2,S=1)$.}
\label{tab:eigstruct} 
\end{table}

\section{GKP codes\label{sec:GKP-codes}}

While their error-correcting properties were first revealed in Ref.~\cite{Gottesman2001},
$\lat$ states have connections to quantum foundations \cite{Aharonov1969},
solid-state physics \cite{Zak1967}, and signal processing (where
their analogues are frequency combs). The ideal (i.e., infinite $\nb$)
square lattice $\gkp$ codespace, denoted by its projection $P_{\gkp}^{\text{ideal}}$,
is the simultaneous +1 eigenspace of the two commuting stabilizers
\begin{equation}
S_{\textbf{x}}=D_{\sqrt{2\pi}}\,\,\,\,\,\,\,\,\,\,\,\,\text{and}\,\,\,\,\,\,\,\,\,\,\,\,S_{\textbf{p}}=D_{i\sqrt{2\pi}}\,,\label{eq:stab}
\end{equation}
where $D_{\alpha}\equiv e^{\alpha\aa^{\dagger}-\alpha^{\star}\aa}$
is the displacement operator (note that $D_{\sqrt{2\pi}}=e^{-i2\sqrt{\pi}\hat{p}}$).
The projection onto the code can be constructed out of all of their
powers,
\begin{equation}
P_{\gkp}^{\text{ideal}}\equiv\left(\frac{1}{\sqrt{\pi}}\sum_{n\in\Z}S_{\textbf{x}}^{n}\right)\left(\frac{1}{\sqrt{\pi}}\sum_{n\in\Z}S_{\textbf{p}}^{n}\right)\equiv P_{\mathbf{x}}P_{\mathbf{p}}\,.
\end{equation}
Applying the Poisson summation formula allows us to express $P_{\mathbf{x}}$
($P_{\mathbf{p}}$) as a sum of projections onto eigenstates $|n\sqrt{\pi}\ket_{\hat{x}}$
($|n\sqrt{\pi}\ket_{\hat{p}}$) of $\hat{x}$ ($\hat{p}$). We demonstrate
this for $P_{\mathbf{p}}$:\begin{subequations}
\begin{align}
P_{\mathbf{p}} & =\frac{1}{\sqrt{\pi}}\sum_{n\in\Z}e^{i2\sqrt{\pi}n\hat{x}}\\
 & =\sum_{n\in\Z}\d\left(\hat{x}-\sqrt{\pi}n\right)\\
 & =\sum_{n\in\Z}|\sqrt{\pi}n\ket_{\hat{x}}\bra\sqrt{\pi}n|\,.
\end{align}
\end{subequations}These sets of positions and momenta makes up the
\textit{code lattice}, the lattice dual to the stabilizer lattice
(in the language of Ref.~\cite{Gottesman2001}) and generated by
the logical operators\begin{subequations}
\begin{align}
X_{\gkp} & =D_{\sqrt{\pi/2}}=S_{\textbf{x}}^{1/2}\label{eq:logicals}\\
Z_{\gkp} & =D_{i\sqrt{\pi/2}}=S_{\textbf{p}}^{1/2}\,.
\end{align}
\end{subequations}The maximally mixed state $\half P_{\gkp}$ reproduces
this lattice, shown in the fourth panel in Fig.~\ref{f:codes}.

\begin{figure}[t]
\includegraphics[width=1\columnwidth]{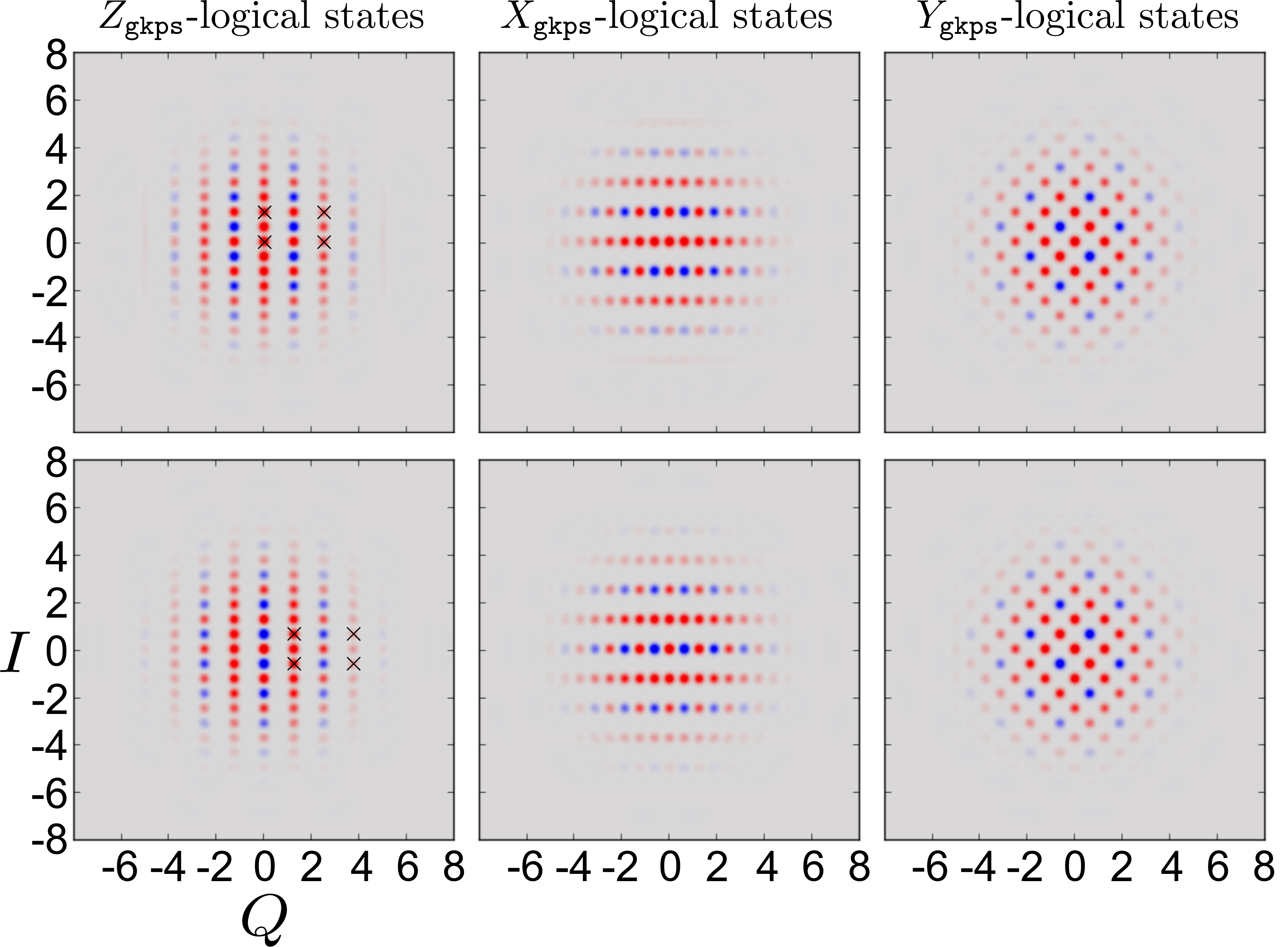}

\caption{\label{f:gkp}Wigner function sketches of the two $Z_{\protect\gkp}$-,
$X_{\protect\gkp}$-, $Y_{\protect\gkp}$-logical states. Comparing
to the fourth panel in Fig.~\ref{f:codes}, which shows that the
code lattice is square, here we see that the lattices formed by the
logical states may be square or rectangular, depending on which logical
operator is considered. The unit cell of the state lattices (\ref{eq:GKP in terms of coh})
is marked by ``$\times$'' in the two leftmost panels; the remaining
dots appear as a result of the coherences between different coherent
states.}
\end{figure}

Conventionally, $\gkp$ logical states are expressed in terms of squeezed
states,
\begin{align}
|\mu_{\gkp}^{\text{ideal}}\rangle & \propto\sum_{n\in\mathbb{Z}}|\sqrt{\pi}(2n+\mu)\rangle_{\hat{x}}\,.\label{GKP states in terms of position eigenstates}
\end{align}
One can obtain an equivalent (see Appx.~\ref{appx:Equivalence-between-squeezed})
representation in terms of coherent states by projecting the vacuum
state $|0\ket$ onto the code and the $\pm1$ eigenstates of $Z_{\gkp}$:
\begin{align}
|\mu_{\gkp}^{\text{ideal}}\rangle & \propto[I+(-1)^{\m}Z_{\gkp}]P_{\gkp}^{\text{ideal}}|0\rangle\label{eq:GKP in terms of coh}\\
 & =\sum_{\vec{n}\in\mathbb{Z}^{2}}D_{\sqrt{\frac{\pi}{2}}(2n_{1}+\mu)}D_{i\sqrt{\frac{\pi}{2}}n_{2}}|0\rangle\,,\nonumber 
\end{align}
where $\vec{n}\equiv(n_{1},n_{2})$. The above displacements generate
the two \textit{state lattices} for $Z_{\gkp}$-logical states, whose
horizontal spacing is twice that of the code lattice due to the $I\pm Z_{\gkp}$
term. In general, the state lattice depends on the logical basis used
(see Fig.~\ref{f:gkp}) while the code lattice is basis-independent.

The usual way to make the states (\ref{GKP states in terms of position eigenstates})
have finite $\nb$ (and therefore be normalizable) is to assume finite
squeezing for each position eigenstate and add a $\D^{2}$-dependent
Gaussian envelope, producing the $\gkp$ states in Eq.~(\ref{eq:gkpsqueezed})
below. Alternatively, one can add a Gaussian envelope to Eq.~(\ref{eq:GKP in terms of coh}),
yielding a representation in terms of coherent states, Eq.~(\ref{eq:gkpcoherent}).
A third finite-$\nb$ representation can be written in terms of $|\mu_{\gkp}^{\text{ideal}}\ket$
smeared by a Gaussian distribution of displacements \cite{Gottesman2001},
making contact with the errors that the codes are designed to correct.
This is the third equation below:\begin{subequations}
\begin{align}
\!\!\!|\mu_{\gkp}^{\Delta}\rangle & \propto\sum_{n\in\Z}e^{-\frac{\pi}{2}\D^{2}(2n+\m)^{2}}D_{\sqrt{\frac{\pi}{2}}\left(2n+\m\right)}S_{-\ln\D}|0\ket\label{eq:gkpsqueezed}\\
 & \sim\sum_{\vec{n}\in\mathbb{Z}^{2}}e^{-\frac{\pi}{2}\Delta^{2}[(2n_{1}+\m)^{2}+n_{2}^{2}]}D_{\sqrt{\frac{\pi}{2}}(2n_{1}+\mu)}D_{i\sqrt{\frac{\pi}{2}}n_{2}}|0\rangle\label{eq:gkpcoherent}\\
 & \sim\int d^{2}\alpha\frac{e^{-|\alpha|^{2}/\D^{2}}}{\sqrt{\pi\Delta^{2}/2}}D_{\alpha}|\mu_{\gkp}^{\text{ideal}}\rangle\,,\label{eq:gkpsmeared}
\end{align}
\end{subequations}where $\mu\in\lbrace0,1\rbrace$ and $S_{r}=e^{+\half r(\aa^{2}-\aa^{\dg2})}$
is the squeezing operator. We use $\D\in[0,1]$ for both the envelope
and squeezing parameters for simplicity. These representations numerically
converge to each other very quickly in the $\D\rightarrow0$ limit,
but there are visual differences between them for small envelopes.
A fourth representation in terms of Fock states is possible using
Eq.~(86) from Ref.~\cite{dodonov}. Note that $|0_{\gkp}^{\D}\ket$
and $|1_{\gkp}^{\D}\ket$ are non-orthogonal for nonzero $\D$, and
this source of error manifests itself in the QEC matrix.

Recall that $\gkp(\D\rightarrow0)$ states can protect against displacement
errors $D_{\alpha_{1}+i\alpha_{2}}$ in phase space, where $|\alpha_{1}|,|\alpha_{2}|<\sqrt{\nicefrac{\pi}{8}}$.
The representations can easily generalize to more tightly-packed lattices,
yielding a slightly larger volume of correctable displacements. To
construct the coherent-state representation of the $Z_{\lat}$-logical
states, one first constructs commuting stabilizers (following Ref.~\cite{Gottesman2001})
and repeats Eq.~(\ref{eq:GKP in terms of coh}). Adding an envelope,
this representation (\ref{eq:gkpcoherent}) is particularly simple
to express:
\begin{equation}
|\m_{\lat}^{\Delta}\ket\propto\sum_{\a\in L(\m)}e^{-\D^{2}\left|\a\right|^{2}}e^{-i\a_{1}\a_{2}}|\a\ket\,,\label{eq:gkp}
\end{equation}
where $\kkk{\a=\a_{1}+i\a_{2}}$ is a coherent state and $L(\m)$
is the state lattice for each code state $\m$. We considered both
these lattices and their shifted versions (\ref{eq:gkpnum}) for the
$\lat$ numerics (see Appx.~\ref{sec:details}). For all analytics
below, we use the finite-$\nb_{\gkp}$ unshifted square-lattice states
$|\mu_{\gkp}^{\Delta}\rangle$ (\ref{eq:gkpsmeared}), noting any
generalizations to other lattices.

We have calculated moments of the occupation number, yielding a geometric
(i.e., thermal) distribution:
\begin{equation}
\overline{n_{\gkp}^{\ell}}\equiv\frac{1}{2}\tr\{P_{\gkp}\ph^{\ell}\}\sim\ell!\nb_{\gkp}^{\ell}\,,\label{eq:moments}
\end{equation}
where the average occupation number is
\begin{equation}
\nb_{\gkp}\sim\frac{1}{2\D^{2}}-\half\,.
\end{equation}
As expected, the moments diverge as the states become unnormalizable
in the small $\D$ limit.

\subsection{QEC matrix for GKP codes}

Recall that any trace class bosonic operator $A$ (i.e., satisfying
$\tr\{A^{\dg}A\}<\infty$) can be expanded in terms of displacement
operators using the orthogonality condition of $D_{\alpha}$ at the
superoperator level \cite{Cahill1969}, 
\begin{equation}
\tr\{D_{\a}^{\dg}D_{\b}\}=\pi\d^{2}\left(\a-\b\right)\,.\label{eq:disportho}
\end{equation}
The expansion is then $A=\int\frac{d^{2}\alpha}{\pi}\tr\{D_{\alpha}^{\dagger}A\}D_{\alpha}$,
where the integral is over all of phase space and $\tr\{D_{\alpha}^{\dagger}A\}$
is the \textit{characteristic function} of $A$. Protection of $\lat$
against pure loss was previously discussed using an approximation
of $\aa$ in terms of a \textit{sum} of displacements instead of an
integral, at first very briefly \cite{Gottesman2001} and subsequently
taking into account the maximum number of photons in the oscillator
\cite{Terhal2016}. Here we calculate the QEC matrix $\e_{\ell\lp}^{\gkp}$
(\ref{eq:qec}) by expressing Kraus operators in terms of the full
integral expansion.

Unlike $\aa$, the error operator $E_{\ell}$ and its variants are
trace class due to the damping term, yielding
\begin{equation}
E_{\ell}^{\dagger}E_{\lp}=\int\frac{d^{2}\alpha}{\pi}e^{-\half(1-\gamma)|\alpha|^{2}}\langle\ell|D_{\alpha^{\star}}|\lp\rangle D_{\alpha\sqrt{\g}}\,,\label{EE in terms of displacement}
\end{equation}
where $\langle\ell|D_{\alpha^{\star}}|\lp\rangle$ are matrix elements
of the displacement operator $D_{\alpha}$ in the Fock state basis
(\ref{displacement in terms of Laguerre}). To obtain this, one can
express the trace in $\tr\{D_{\alpha}^{\dagger}E_{\ell}^{\dagger}E_{\lp}\}$
as a sum over Fock states, plug in Eq.~(\ref{displacement in terms of Laguerre}),
and use the generating function of Laguerre polynomials (\ref{eq:genfunc}).
Complementing the expansion of Gaussian noise $\{D_{\a}\}$ in terms
of photon creation and annihilation operators (e.g., Ref.~\cite{Hofer2016}),
the above equation completes the ``Rosetta stone'' expressing each
of the two primary noise models in the language of the other.

To gain a flavor of the calculations below, let us first examine how
we can calculate $c_{\ell\ell}^{\gkp}=\frac{1}{2}\tr\{P_{\gkp}E_{\ell}^{\dagger}E_{\ell}\}$.
Using Eq.~(\ref{EE in terms of displacement}), this calculation
boils down to determining $\tr\{P_{\gkp}D_{\a}\}$. Consider first
the infinite $\nb_{\gkp}$ limit, recalling from Fig.~\ref{f:codes}
that $P_{\gkp}$ is a sum of (unphysical) points of fixed position
and momentum arranged in a square code lattice. Then, $\tr\{P_{\gkp}D_{\a}\}$
will be nonzero only for those $\a$ which displace the lattice back
on top of itself, i.e., $\a$ displaces by a multiple of the lattice's
unit cell. For those cases, the overlap of each point with itself
will be infinite, and so the total result is $\sum_{\Lambda}\d^{2}(\a-\Lambda)$,
where the sum is over all displacements $\Lambda\in\sqrt{\pi/2}(n_{1},n_{2})$
(with integers $n_{1,2}$) preserving the code lattice. Coming back
to finite $\nb_{\gkp}$, a natural guess would be to substitute the
Gaussian representation $\frac{1}{\D}e^{-\frac{1}{2\Delta^{2}}|\alpha-\Lambda|^{2}}$
for the Dirac $\d$-function in the sum. This almost obtains the right
result, but there are two more steps. The first is normalization,
which cancels the $\frac{1}{\D}$ in front of the Gaussian representation
of the $\d$-function. The second is addition of the Gaussian envelope,
yielding
\begin{equation}
\frac{1}{2}\tr\{P_{\gkp}D_{\a}\}\sim\sum_{\Lambda}e^{-\frac{1}{2\Delta^{2}}|\alpha-\Lambda|^{2}}e^{-\frac{\D^{2}}{2}|\Lambda|^{2}}\,.\label{eq:trace}
\end{equation}
Notice that what used to be a Dirac $\d$ is now a Kronecker $\d_{\a,\Lambda}\sim e^{-\frac{1}{2\Delta^{2}}|\alpha-\Lambda|^{2}}$
in the small $\D$ limit. As a sanity check, setting $\a=0$ yields
unity in that limit.

Calculating these overlaps is more involved (see Appx.~\ref{subsec:Projecting-displacements-onto}),
but we can nevertheless use the above intuition to understand the
more general element (with $\n\in\{0,1\}$)
\begin{align}
\langle\mu_{\gkp}^{\D}|D_{\alpha}|\n_{\gkp}^{\D}\rangle & \sim\label{eq:gkpqec}\\
 & \!\!\!\!\!\!\!\!\!\!\!\!\!\!\!\!\!\!\!\!\sum_{\vec{n}\in\mathbb{Z}^{2}}e^{i\pi(n_{1}+\frac{\mu+\n}{2})n_{2}}e^{-\frac{1}{2\Delta^{2}}|\alpha-\Lambda_{\delta\mu}^{\vec{n}}|^{2}}e^{-\frac{\Delta^{2}}{2}|\Lambda_{\delta\mu}^{\vec{n}}|^{2}}\,,\nonumber 
\end{align}
where $\delta\mu\equiv\mu-\n$ and $\Lambda_{\delta\mu}^{\vec{n}}\equiv\sqrt{\frac{\pi}{2}}[(2n_{1}+\delta\mu)+in_{2}]$.
Since there are two different state lattices, there are extra phases
in the sum and the sum is over displacements $\a\in\Lambda_{\delta\mu}^{\vec{n}}$
which overlap the two lattices. We can now plug this into $E_{\ell}^{\dagger}E_{\lp}$
(\ref{EE in terms of displacement}) and proceed to calculate the
integral (see Appx.~\ref{subsec:QEC-criteria-for}), yielding
\begin{align}
\langle\mu_{\gkp}^{\D}|E_{\ell}^{\dagger}E_{\lp}|\n_{\gkp}^{\D}\rangle & \sim\sqrt{c_{\ell\ell}^{\gkp}c_{\lp\lp}^{\gkp}}\sum_{\vec{n}\in\mathbb{Z}^{2}}e^{-\frac{(1-\gamma)}{2\gamma}|\Lambda_{\delta\mu}^{\vec{n}}|^{2}}\nonumber \\
 & \!\!\!\!\!\!\!\!\!\!\!\!\!\!\!\!\!\!\!\!\times e^{i\pi(n_{1}+\frac{\mu+\nu}{2})n_{2}}e^{-\frac{\Delta^{2}}{2}|\Lambda_{\delta\mu}^{\vec{n}}|^{2}}\langle\ell|D_{(\Lambda_{\delta\mu}^{\vec{n}})^{\star}/\sqrt{\g}}|\lp\rangle,\label{GKP error correction criteria}
\end{align}
for $\g\nb_{\gkp}\rightarrow\infty$, where $c_{\ell\ell}^{\gkp}$
will turn out to be the probability of losing $\ell$ photons, 
\begin{equation}
c_{\ell\ell}^{\gkp}=\half\tr\{P_{\gkp}E_{\ell}^{\dg}E_{\ell}\}\sim\frac{(\gamma\nb_{\gkp})^{\ell}}{(\gamma\nb_{\gkp}+1)^{\ell+1}}.\label{eq:thermal}
\end{equation}
Thus, the photon loss distribution for $\gkp$ is a asymptotically
thermal with mean $\gamma\nb$. Ignoring the $\D^{2}$ envelope term
from now on, all $\D$-dependence of Eq.~(\ref{GKP error correction criteria})
is contained in $c_{\ell\ell}^{\gkp}$.

Notice that $|\langle\ell|D_{(\Lambda_{\delta\mu}^{\vec{n}})^{\star}/\sqrt{\g}}|\lp\rangle|\leq1$
because they are overlaps between two states. Thus, the only quantity
regulating the sum (\ref{GKP error correction criteria}) is $e^{-\frac{(1-\gamma)}{2\gamma}|\Lambda_{\delta\mu}^{\vec{n}}|^{2}}$.
Assuming $\g\ll1$, the ``on-site'' term ($\vec{n}=\vec{0}$) in
Eq.~(\ref{GKP error correction criteria}) is $\m$-independent and
thus contributes to 
\begin{align}
c_{\ell\lp}^{\gkp} & \sim c_{\ell\ell}^{\gkp}\delta_{\ell\lp}\,,
\end{align}
while the ``nearest-neighbor'' terms ($|\vec{n}|=1$) contribute
to the leading-order uncorrectable parts
\begin{align}
|z_{\ell\lp}^{\gkp}| & \sim\sqrt{c_{\ell\ell}^{\gkp}c_{\lp\lp}^{\gkp}}e^{-\frac{\pi}{4}\frac{1-\gamma}{\gamma}}\langle\ell|(D_{\sqrt{\frac{\pi}{2\gamma}}}+D_{\sqrt{\frac{\pi}{2\gamma}}}^{\dg})|\lp\rangle\label{eq:gkpz}
\end{align}
and $|x_{\ell\lp}^{\gkp}|=|z_{\ell\lp}^{\gkp}|$. (There is no $|y_{\ell\lp}^{\gkp}|$
to this order.) The uncorrectable parts are the same (up to sign)
due to the identical effect of position and momentum displacements
on the code. So, while the nearest neighbors $\vec{n}=(0,\pm1)$ contributed
to $z_{\ell\lp}^{\gkp}$ and $\vec{n}=(\pm1,0)$ contributed to $x_{\ell\lp}^{\gkp}$,
the two quantities have to be equal in magnitude due to this effect.
(Considering more general lattices can of course break this balance.)
We can also see another symmetry manifest itself \textemdash{} the
invariance of the lattice under parity $(-1)^{\ph}$. Since the sum
of displacements is even under parity, it does not connect even Fock
states to odd ones and guarantees that $z_{\ell\lp}^{\gkp}=0$ unless
$\ell-\lp$ is even. This means that technically $\gkp$ codes have
spacing $S=1$. However, this spacing disappears when the lattice
is slightly shifted and the symmetry lost, but the performance of
the codes remains. This should not be surprising since shifted $\gkp$
codespaces are akin to Pauli frames in the stabilizer formalism \cite{Terhal2016}.
The most striking result is that the reason for this high performance
is not due to the spacing, but to the suppression by the $\g$-dependent
exponential factor in Eq.~\eqref{eq:gkpz}. Namely, while $\e_{\ell\lp}^{\gkp}$
contains uncorrectable parts for all $\ell,\lp$ (modulo symmetry
constraints), \textit{all of these parts} are suppressed exponentially
by $e^{-\frac{\pi}{4}\frac{1-\gamma}{\gamma}}$ when $\g\ll1$. As
an example, we show the comparable strength of the exponential suppression
of uncorrectable parts for $\gkp(\nb\approx6)$ in Fig.~\ref{f:qec}.
Assuming that the infidelity $1-\fe$ to leading order in $\g$ is
polynomial in all uncorrectable parts, one expects $1-\fe$ to also
be exponentially suppressed by $\frac{1-\g}{\g}$. We proceed to show
this by bounding $1-\fe$ using an explicit recovery.

\begin{figure*}[t]
\includegraphics[width=0.9\textwidth]{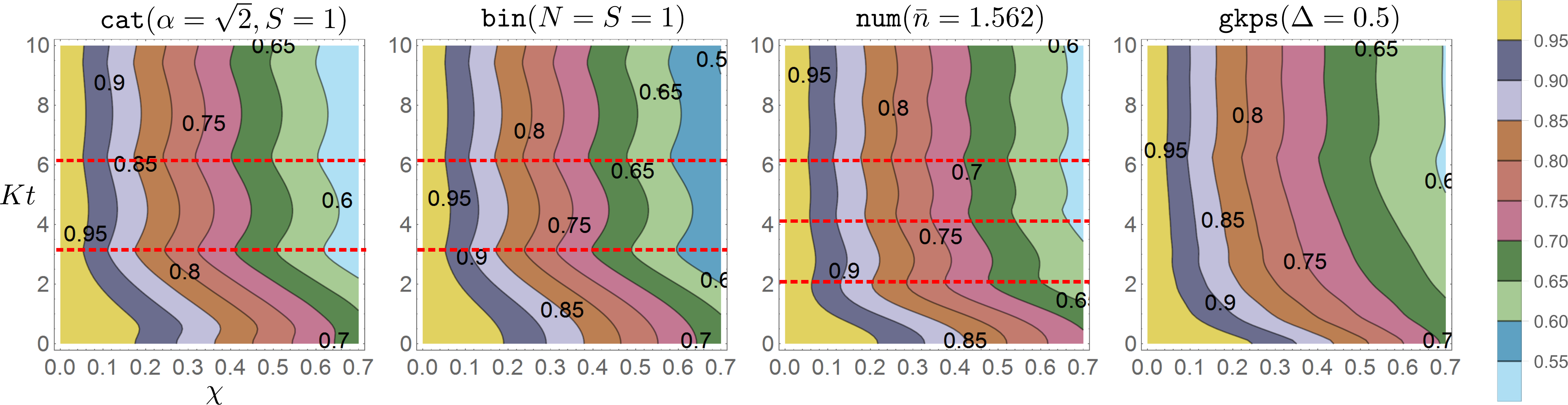}

\caption{\label{f:kerr}Contour plot of $\protect\fe$ vs. dimensionless Kerr
parameter $Kt$ and damping parameter $\chi\equiv\protect\k t$ for
$\protect\cat$, $\protect\bin$, $\protect\num$, and $\protect\gkp$
picked such that they all have $\protect\nb\approx2$. Here, $K$
is the strength of the Kerr Hamiltonian (\ref{eq:kerr}), $\protect\k$
is the cavity decay rate, and $t$ is time. Starting with a fixed
$\chi$ and tracking increasing $Kt$, we see that $\protect\fe$
quickly decreases to a constant for all codes considered, implying
a potentially universal failure of error-correction when $Kt$ is
large. However, $\protect\fe$ is minimal at high-symmetry points
of the codes (dashed red lines) and maximal in-between. For example,
$\protect\half P_{\protect\num}$ for the $\protect\num$ code is
three-fold symmetric (see Fig.~\ref{f:codes}), and $\protect\fe(\protect\num)$
is minimal at $Kt$ being the first few multiples of $\nicefrac{2\pi}{3}$.}
\end{figure*}

\subsection{Removing energy constraints\label{subsec:Removing-energy-constraints-1}}

In Fig.~\ref{f:num}, we have observed that $\fe^{\gkp}$ is significantly
higher than that for all other codes for most $\g$. While the non-trivial
exponential suppression of uncorrectable parts (\ref{eq:gkpz}) of
the QEC matrix hints at an analytical explanation, this still does
not tell us how $\fe$ scales with $\g$. For this, we need to consider
a specific analytically tractable recovery. Having investigated several
recoveries, the simplest one we found is based on the fact that the
combination of amplification and pure loss produces Gaussian (i.e.,
displacement) noise \cite{cvbook} \textemdash{} a channel which most
naturally fits the error-correction capabilities of $\gkp$.

Coming back to the formulation of $\L$ in terms of a beam-splitter
(\ref{eq:env}), consider amplifying the signal
\begin{equation}
\aa\rightarrow\sqrt{G}\aa+\sqrt{G-1}\bb^{\dg}
\end{equation}
after application of $\L$. Here, $G$ is the gain of the amplifier,
which we set to $G=e^{\chi}=\frac{1}{1-\g}$ to compensate the effect
of damping. Tracing out the $\bb$ mode, amplification is simply the
properly normalized transpose of pure loss,
\begin{equation}
\A(\cdot)=\left(1-\g\right)\L^{\dgt}(\cdot)\,,
\end{equation}
where ``$\dgt$'' is the adjoint in the matrix representation. The
Kraus operators of $\A$ are $\sqrt{1-\g}E_{\ell}^{\dg}$ for $\ell\in\{0,1,\cdots\}$
\{e.g., Ref.~\cite{Ivan2011}, Eq.~(5.5)\} and it is simple to verify
that $\A$ is indeed a channel: $\left(1-\g\right)\sum_{\ell=0}^{\infty}E_{\ell}E_{\ell}^{\dg}=I$. 

Upon amplification, the pure-loss channel $\L$ is transformed into
a Gaussian noise channel with variance $\frac{\gamma}{1-\gamma}$.
The noise comes from two parts: the intrinsic noise due to amplification
and the amplified noise due to pure loss. More explicitly, we can
apply Eq.~\eqref{EE in terms of displacement} to express $\A\L$
in terms of displacements and use displacement orthogonality (\ref{eq:disportho}):\begin{widetext}
\begin{align}
\A\L(\r) & =\left(1-\g\right)\int\frac{d^{2}\alpha}{\pi}\frac{d^{2}\beta}{\pi}e^{-\frac{1-\gamma}{2}(|\alpha|^{2}+|\beta|^{2})}D_{\frac{\a}{\sqrt{\g}}}\,\rho\,D_{\frac{\beta}{\sqrt{\g}}}\sum_{\ell,\lp=0}^{\infty}\langle\ell|D_{\alpha^{\star}}|\lp\rangle\langle\lp|D_{\beta^{\star}}|\ell\rangle\\
 & =\left(1-\g\right)\int\frac{d^{2}\alpha}{\pi}d^{2}\beta e^{-\frac{1-\gamma}{2}(|\alpha|^{2}+|\beta|^{2})}D_{\frac{\a}{\sqrt{\g}}}\,\rho\,D_{\frac{\beta}{\sqrt{\g}}}\delta^{2}(\alpha+\beta)=\frac{1-\g}{\gamma}\int\frac{d^{2}\alpha}{\pi}e^{-\frac{1-\gamma}{\gamma}|\alpha|^{2}}D_{\alpha}\,\rho\,D_{\alpha}^{\dagger}\,.\nonumber 
\end{align}
\end{widetext}

Appending amplification with the conventional $\lat$ recovery $\R^{\text{GKP}}$
which measures and corrects displacements within the correctable unit
cell \cite{Gottesman2001}, the total recovery we consider is
\begin{equation}
\R^{\rgkp}=\R^{\text{GKP}}\A\,.\label{eq:agkp}
\end{equation}

Note that the above derivation of Gaussian noise is exact for \textit{all
values} of $\g$. We bound $\fe$ by calculating the success probability
that $\R^{\rgkp}$ will succeed in correcting the above Gaussian noise
$\A\L$ starting with \textit{ideal} (i.e., infinite $\nb_{\gkp}$)
code states:
\begin{equation}
P_{\text{succ}}(\gamma)=\frac{1-\gamma}{\gamma}\int_{\blacksquare}\frac{d\alpha_{1}d\alpha_{2}}{\pi}e^{-\frac{1-\gamma}{\gamma}(\alpha_{1}^{2}+\alpha_{2}^{2})}\,,
\end{equation}
where the integration is over correctable displacements $|\alpha_{1}|,|\alpha_{2}|\le\sqrt{\nicefrac{\pi}{8}}$
denoted by $\blacksquare$. The channel infidelity $1-\frgkp$ can
then be estimated using the failure probability, which is the complementary
integral outside of the unit cell. Upper bounding that integral by
integrating the complement of the circle with radius $\sqrt{\nicefrac{\pi}{8}}$
(instead of the complement of the square with length $\sqrt{\nicefrac{\pi}{8}}$)
yields
\begin{align}
P_{\text{fail}}(\gamma) & <\frac{1-\gamma}{\gamma}\int_{|\alpha|\ge\sqrt{\frac{\pi}{8}}}\frac{d^{2}\alpha}{\pi}e^{-\frac{1-\gamma}{\gamma}|\alpha|^{2}}=e^{-\frac{\pi}{8}\frac{1-\gamma}{\gamma}}\,.
\end{align}
We remark that this bound can be improved to $e^{-\frac{\pi}{4\sqrt{3}}\frac{1-\gamma}{\gamma}}$
using ideal hexagonal $\lat$.

\section{Additional features\label{sec:Additional-nonlinearityfeatures}}

\subsection{Nonlinearity\label{sec:Addition-of-a}}

We have tried to address the effect of pure loss on our codes, but
real-world microwave cavities have undesired unitary evolution (i.e.,
coherent errors). In general, the joint effect of pure loss and a
unitary process on the state depends not only on how many losses have
occurred, but also their specific times. The purpose of this subsection
is to answer the following:
\[
\text{\textit{Does adding coherent errors reduce code performance? }}
\]
The answer to this, at least in our case, is a firm ``yes''. 

The coherent (i.e., unitary) error we add is generated by a Kerr nonlinearity
with Hamiltonian
\begin{equation}
H_{K}\equiv\half K\ph\left(\ph-1\right)=\half K\aa^{\dg2}\aa^{2}\,,\label{eq:kerr}
\end{equation}
with Kerr parameter $K$. Here, we show what happens when our codes
get exposed to the joint evolution of pure loss and Kerr, namely,
the channel 
\begin{equation}
\L_{\chi,Kt}(\cdot)=e^{-iKt\left[\half\ph\left(\ph-1\right),\cdot\right]+\c{\cal D}(\cdot)}\,,
\end{equation}
where ${\cal D}(\cdot)$ is the Lindbladian for the pure loss channel
from Sec.~\ref{subsec:Codes-and-error}. Since the Kerr nonlinearity
is prominent in cavities coupled with transmons (as opposed to optical
fibers), we use the excitation loss rate $\k$ to quantify the strength
of pure loss (recall that $\g=1-e^{-\k t}$). Thus, the two unitless
scales of the problem are $\chi\equiv\k t$ and $Kt$. An analytic
Kraus representation for $\L_{\chi,Kt}$ has yet to be obtained, but
various approaches have come close \cite{Klimov2003,Kunz2017}.

Figure \ref{f:kerr} plots $\fe(Kt,\chi)$ for four code families
at $\nb_{\cc}\approx2$. For $Kt=0$ (horizontal axis in each plot),
we see the same behavior in $\fe$ vs. pure loss strength as we saw
before. For the other extreme of $\chi=0$ (vertical axis), we see
unit $\fe$ since Kerr nonlinearity alone is a perfectly correctable
unitary process. Starting with a fixed nonzero $\chi$ and looking
up at the vertical line of increasing $Kt$, we see that $\fe$ quickly
decreases to a (roughly) constant value for all codes. Since the optimal
recovery is not able to ascertain exactly when photon loss events
occurred, Kerr evolution induces rotations of unknown angle between
those events, $e^{-iH_{K}t}\aa\propto\aa e^{i\ph Kt}e^{-iH_{K}t}$,
and thus destroys the quantum information. Metrology protocols are
also susceptible to this effect \cite{Duivenvoorden2017}. The value
to which $\fe$ decreases at large $Kt$ seems to be (roughly) universal
across all codes, so there might be a fundamental limit to correcting
large $\nb$-dependent coherent errors in the presence of incoherent
errors. However, it is still possible to use error-correction to our
advantage in, e.g., the $Kt\approx1$ regime (given $\nb\approx2$).
Incidentally, in that regime, $\cat$ shows an \textit{increase} in
$\fe$, implying that a slight amount of Kerr is actually helping
$\cat$-code performance. We are investigating this effect in a subsequent
publication.

Lastly, we want to mention that $\fe$ for a given $\cc$ is minimal
when $Kt$ is at an angle by which rotating $\half P_{\cc}$ leaves
the projection invariant. Recall that evolution under $e^{-iH_{K}t}$
causes a coherent state to transform into cat states at certain rational
$t$ \cite{Yurke1986a,Miranowicz1990,Tara1993}, but such recurrences
are quickly degraded under pure loss \cite{Milburn1986}. Nevertheless,
we see some periodicity in code performance, e.g., in the case of
$\num$ in the third panel in Fig.~\ref{f:kerr}. From Fig.~\ref{f:codes},
we know that $\half P_{\num}$ for this $\num$ code is three-fold
rotationally symmetric in phase space. Coincidentally, $\fe(\num)$
is minimal at $Kt$ being the first few multiples of $\nicefrac{2\pi}{3}$
(dashed red lines) and maximal in-between. We do not know the reason
for this effect, but one can see that it occurs for all codes to various
extent.

\subsection{Parity measurements}

Here, we briefly consider the question
\[
\text{\textit{What if we know how many photons were lost? }}
\]
While microwave cavities still do not have the capability to directly
count photons, one can perform non-demolition photon parity measurements
of microwave cavity modes \cite{Sun2014,Ofek2016} and vibrational
modes of trapped ions \cite{Lutterbach1997}. If we assume that (1)
we have a fixed-parity initial state and (2) we can measure parity
$(-1)^{\ph}$ during the loss portion of the channel $\E$ perfectly
and faster than any timescale of the system, then we can in principle
track every loss event $\aa$ without destroying the state. This results
in an \textit{unraveled} \cite{gregoratti2009} system, where part
of the knowledge reserved for the environment \textemdash{} the number
and times of the loss events \textemdash{} is now learned by the experimenter.
For example, a trajectory lasting time $t$ during which jumps occurred
at times $\tau_{2}\geq\tau_{1}$ would incur the conditional evolution
\begin{equation}
\widetilde{E}_{2}|\psi\ket\equiv e^{-\half\k(t-\tau_{2})\ph}\aa e^{-\half\k(\tau_{2}-\tau_{1})\ph}\aa e^{-\half\k\tau_{1}\ph}|\psi\ket\,,
\end{equation}
where $|\psi\ket$ is the initial state of the oscillator and we have
not yet renormalized the state. Defining $\widetilde{E}_{\ell}$ in
similar fashion and permuting all $\aa$'s to the left leaves us with
$\widetilde{E}_{\ell}=f_{\ell}E_{\ell}$, where $E_{\ell}$ (\ref{eq:krausloss})
is the Kraus operator for the pure-loss channel and $f_{\ell}$ is
a function of the jump times. Since $f_{\ell}$ is a scalar, knowledge
of jump times is irrelevant to error-correction against pure loss,
and we can ignore it from now on. We model this process using an extended
version of pure loss $\L$ \textemdash{} the quantum instrument \cite{wildebook}
\begin{equation}
\lt(\r)=\sum_{\ell=0}^{\infty}E_{\ell}\r E_{\ell}^{\dg}\otimes|\ell\ket\bra\ell|\,,\label{eq:lossknow}
\end{equation}
where $\r$ is a single-mode density matrix. The second tensor factor
represents our knowledge of $\ell$, making sure that each $E_{\ell}\r E_{\ell}^{\dg}$
is mapped into an orthogonal subspace of the extended Hilbert space.
The corresponding recovery $\R$ has Kraus operators $U_{\ell}^{\cc}\otimes\bra\ell|$,
with $U_{\ell}^{\cc}$ being a unitary mapping $E_{\ell}P_{\cc}$
into $P_{\cc}$.

\begin{figure}
\includegraphics[width=0.9\columnwidth]{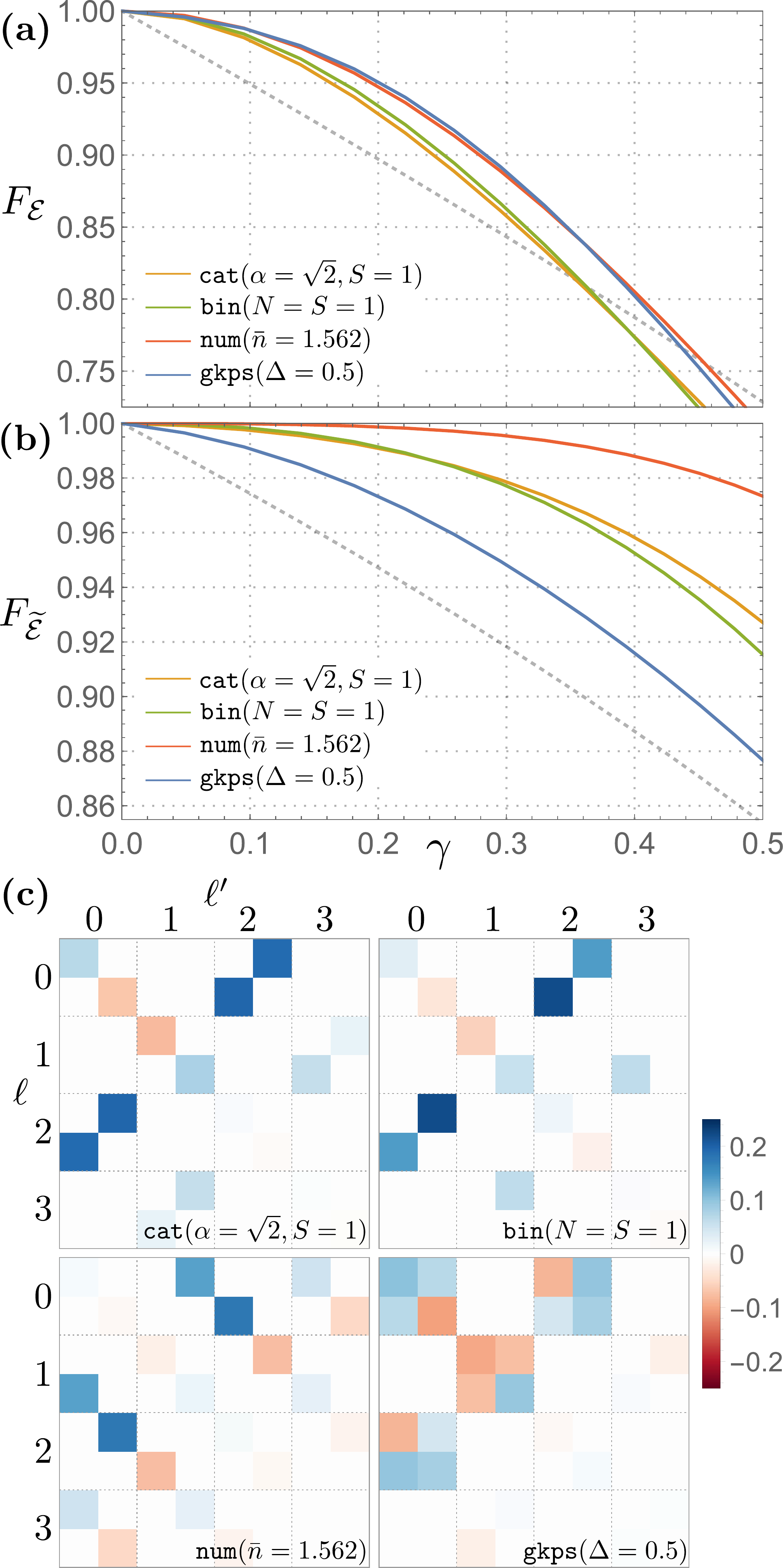}

\caption{\label{f:diags}Channel fidelity for $\protect\cat$, $\protect\bin$,
$\protect\num$, and $\protect\gkp$ picked such that they all have
$\protect\nb\approx2$, given \textbf{(a)} the pure-loss channel $\protect\L$
(\ref{eq:krausloss}) and \textbf{(b)} the pure-loss channel with
the capability of knowing the number of photons lost, modeled by the
quantum instrument $\protect\lt$ (\ref{eq:lossknow}). \textbf{(c)}
Uncorrectable parts of the QEC matrices, $\protect\e^{\protect\cc}-c^{\protect\cc}$
(see Sec.~\ref{sec:QEC matrix}), for the four codes (cf. Fig.~\ref{f:qec}).
Since uncorrectable parts in the diagonal blocks $\protect\e_{\ell\ell}^{\protect\cc}$
are all that matter in recovering from $\protect\lt$, we see that
$\protect\cat$, $\protect\bin$, and $\protect\num$ outperform $\protect\lat$
at this $\protect\nb$.}
\end{figure}

We compare $\fe$ (\ref{eq:fid}) with $\fet$, the fidelity given
the extended loss channel $\lt$ (\ref{eq:lossknow}), in Figs.~\ref{f:diags}(a)
and (b), respectively. Code performance improves for all codes, even
the naive $0/1$ Fock state encoding (dashed gray line). Note that,
given this extra knowledge, only the diagonal blocks $\e_{\ell\ell}^{\cc}$
of the QEC matrix (\ref{eq:qec}) are relevant. The $\gkp$ codes
have the largest uncorrectable parts in those blocks at this $\nb_{\gkp}$
{[}see Fig. \ref{f:diags}(c){]}, so their performance increases the
least out of all the codes. The $\num$ code winds up being the optimal
encoding in the eyeball norm, but the parity-tracking procedure described
above is not applicable to this code since its states (\ref{eq:sqrt17})
are not of fixed parity.

\section{Conclusion\label{sec:Conclusion}}

The results of this manuscript can be categorized into two parts:
one regarding code performance and one regarding code structure. Here,
we summarize these results and comment on future directions and open
questions.

\subsection{Code performance}

We gave a numerical and analytical performance comparison of the four
primary single-mode continuous-variable quantum codes \textemdash{}
$\cat$ \cite{Cochrane1999,Leghtas2013b,cats}, $\bin$omial \cite{bin},
$\num$erically optimized (\cite{bin} and here), and $\lat$ \cite{Gottesman2001}
codes \textemdash{} against the pure loss channel with loss rate $\g$.
For the numerical part, we compared the codes' ability to preserve
entanglement using channel-adapted error correction \cite{Fletcher2007}
subject to two caveats: \one~the encoding, recovery, and decoding
are all assumed perfect and \two~the codes are grouped by their mean
occupation number $\nb_{\cc}\leq2,5,10$. For the analytical part,
we calculated the quantum error-correction (QEC) conditions for $\cat$,
$\bin$, and $\lat$. We briefly discuss our results below, but encourage
the reader to peruse Sec.~\ref{sec:Take-home-messages} for more
details.

Even though $\cat$ and $\bin$ follow the traditional convention
of correcting exactly against a subset of errors, their performance
is significantly worse than that of $\lat$ for many $\g$. While
$\lat$ do not exactly correct against any errors, we find that the
violation of the QEC conditions for each error is insignificant compared
to the violation of the leading-order uncorrectable errors for $\cat$
and $\bin$ for most $\nb_{\cc}$. 

In the limit of vanishing $\g$, we observe the following order of
performance: $\lat<\cat<\bin<\num$. On the one hand, since $\cat$
and $\bin$ codes correct exactly against the first few errors, their
performance scales polynomially with $\g$. We further reveal the
regions in the $\{\g,\nb_{\cc}\}$ parameter space in which $\bin$
codes outperform $\cat$ codes. On the other hand, we analytically
show that $\lat$ code entanglement infidelity is $O(e^{-\frac{c}{\g}})$
(with $c$ a constant dependent on the type of $\lat$ code). 

As $\g$ increases, $\lat$ quickly overpowers the rest of the codes
and their performance persists even for high $\g$. For example, optimal
recovery of a $\lat$ state with 10 photons on average yields a fidelity
of 99.5\% given $\g\approx20\%$, compared to a fidelity of 96.9\%
for $\bin$ and 96.6\% for $\cat$. At $\g\approx30\%$, where about
one in every three photons are lost, the fidelity of $\lat$ is still
95\%, which is 5\% higher than that of $\cat$ and $\bin$. At such
high $\g$, we observed the following order of performance: $\bin\apprle\cat<\num<\lat$. 

We extended our analysis of entanglement preservation to determine
achievable rates of quantum communication using these codes, where
we saw similar orders of performance. We also show that sending a
$\lat$ state with an average occupation number of two photons produces
a higher communication rate than distributing two photons among four
modes using the smallest encoding protecting against one loss error.

Relaxing the $\nb_{\cc}$ constraint, we have numerical evidence showing
that performance of $\lat$ codes and a subset of $\bin$ codes increases
with increasing $\nb_{\cc}$. Since the ideal $\lat$ states indeed
have infinite $\nb_{\lat}$, it is reasonable that their performance
increases monotonically as they become more ideal. To back this claim
analytically, we cook up a simple recovery procedure involving phase-insensitive
amplification that converts the pure-loss channel into Gaussian noise;
this procedure can also be used in multimode extensions of $\lat$
codes \cite{Noh2018}. The $\bin$ increase in performance can be
justified by showing they have a larger set of approximately correctable
errors than previously thought; we do so in Secs.~\ref{subsec:Error-correction-criteria-for}-\ref{subsec:Removing-energy-constraints}.

We added a unitary error in the form of a Kerr nonlinearity $K$ in
order to see how code performance is changed. We observed that $\cat$
code performance increases slightly at small $K$ and that all code
performance oscillates with periods depending on their symmetries;
these are subjects of future investigation. We also observed that,
at sufficiently large $K$, the performance of all codes fails at
about the same rate, signaling the need to keep such coherent errors
low in a real device. We also briefly addressed changes in code performance
if one is able to learn how many photons were lost.

There are obvious generalizations of this analysis to other multi-mode
codes mentioned in the introduction, storing multiple qubits worth
of information; we are currently pursuing some of them. Another direction
has to do with having $\lat$ and $\bin$ codes catch up to $\cat$
codes in terms of experimental realizability. While $\lat$ codes
may have been considered by some to be unphysical in the past, recent
technological advances in, e.g., microwave cavity \cite{Heeres2016},
atomic ensemble \cite{Chen2015,vuleticnature}, or trapped ion \cite{Kienzler2017}
control, suggest that making these states may be within reach. In
fact, there are recent theoretical proposals related to making and
maintaining $\lat$ states in two of the aforementioned technologies
\cite{Terhal2016,Motes2017} (see Refs.~\cite{Travaglione2002,Pirandola2004,Pirandola2006,Pirandola2006a,Vasconcelos2010,Weigand2017}
for other proposals) and a related trapped-ion experiment \cite{Fluhmann2017}.
While the comparison offered here is completely free from consideration
of experimental imperfections, we hope that our conclusions will motivate
the community to pursue quantum information processing and communication
schemes with $\bin$ and $\lat$ states.

\subsection{Code structure\label{subsec:Code-structure}}

We discussed a connection between $\bin$ codes and spin-coherent
states and used it to characterize related two-mode binomial codes
as well as multi-qubit permutation-invariant codes \cite{Ouyang2014}.
This connection yields a check operator for $\bin$ dephasing errors,
and we discussed an error-correction scheme that utilizes this operator.
This connection was also extended to qudit versions of the aforementioned
codes, yielding a generalization of spin-coherent states and a check
operator for qudit codes. 

By mapping the coefficients of the qudit $\bin$ code into a particular
subspace of multiple qudits, we introduced a multi-qudit (i.e., discrete
variable) code that extends the multi-qubit permutation-invariant
codes. The extension, which we call $\pa$ in order to differentiate
it from another extension \cite{Ouyang2016}, turns out to be nothing
but a multi-qudit bit-flip code when expressed in the basis of products
of the individual qudit states. However, when expressed in terms of
a qudit extension of Dicke states, the coefficients next to those
states match those of the qudit binomial codes. This relates the protection
of the continuous-variable binomial codes to that of a discrete-variable
bit-flip code. A similar bit-flip-like trick was used for another
code \textemdash{} the $\noon$ code \cite{Bergmann2016a} \textemdash{}
where two-mode $\noon$ states and their multi-mode generalizations
were tensored together to form codes protecting against pure loss.
Such intriguing connections between discrete- and continuous-variable
codes should be investigated further. In addition, the generalization
of spin-coherent states introduced here may be useful in experimental
settings such as atomic ensembles (e.g., \cite{vuleticnature,Hu2015})
and magnetometry (e.g., \cite{Budker2007,Taylor2008,Acosta2009}).
\begin{acknowledgments}
The authors acknowledge Steven T. Flammia, David Poulin, Saikat Guha,
Richard Kueng, Mazyar Mirrahimi, John Preskill, R. J. Schoelkopf,
Matti Silveri, Murphy Yuezhen Niu, and Bei Zeng for enlightening discussions.
V.V.A. thanks Misha Guy and the Yale Center for Research Computing
for resources and support and acknowledges support from the Walter
Burke Institute for Theoretical Physics at Caltech. V.V.A., K.N.,
C.S., L.L., and L.J. acknowledge support through the ARL-CDQI, ARO
(Grants No. W911NF-14-1-0011 and No. W911NF-14-1-0563), ARO MURI (W911NF-16-1-0349),
NSF (EFMA-1640959), AFOSR MURI (FA9550-14-1-0052 and FA9550-15-1-0015),
the Alfred P. Sloan Foundation (BR2013-049), the Packard Foundation
(2013-39273). K.D., C.V., and B.M.T. acknowledge support through the
ERC Consolidator Grant (682726). S.M.G. acknowledges support through
the NSF (DMR-1609326) and ARO (W911NF1410011).
\end{acknowledgments}

\appendix

\section{The many faces of channel fidelity\label{sec:The-many-faces}}

A well-known property of $\fe$ is the relation to the average input-output
fidelity of $\E$ \cite{Horodecki1999} (see also \cite{Nielsen2002}),
\begin{equation}
\int d\psi\bra\psi|\E(|\psi\ket\bra\psi|)|\psi\ket=\frac{d\fe+1}{d+1}\,,
\end{equation}
where $d$ is the dimension of the system. Above, $\bra\psi|\E(|\psi\ket\bra\psi|)|\psi\ket$
is the input-output fidelity for some initial state $|\psi\ket$ of
the source qubit and $d\psi$ is a uniform distribution over all pure
states. Due to the above equality, one should not be surprised that
the capacity of entanglement transmission determined by $\fe$ is
equivalent to the capacity of pure state preservation determined by
the input-output fidelity \cite{Barnum2000}. In addition, since $\fe$
is a fidelity between two states, it gives rise to a metric, is stable
under addition of ancillary systems, and satisfies the chaining property
(meaning that it can be used to provide a bound on the error of a
larger quantum computation). These properties can be proven using
Ref.~\cite{Gilchrist2005}, where $\fe=F_{\text{pro}}(\E,\id)$.

The channel fidelity can be related to the worst-case input-output
fidelity $\min_{|\psi\ket}\bra\psi|\E(|\psi\ket\bra\psi|)|\psi\ket$
\cite{Ng2010},
\begin{equation}
1-d\sqrt{1-\fe^{2}}\leq\min_{|\psi\ket}\bra\psi|\E(|\psi\ket\bra\psi|)|\psi\ket\,.\label{eq:avrg}
\end{equation}
The dependence of the bound on the dimension as well as the square
of $\fe$ suggests that $\fe$ is not a good measure of the worst
case scenario. There is indeed a discrepancy between average and worst-case
behavior for channels that contain a combination of coherent and incoherent
noise \cite{Kueng2016,Wallman2015}. Such an example here is pure
loss with an additional Kerr nonlinearity, considered in Sec.~\ref{sec:Additional-nonlinearityfeatures}.
However, the pure loss channel alone contains only incoherent noise
and so $\fe$ is a reasonable marker of even worst-case behavior;
moreover, one can prove that the dimension dependence goes away entirely
\cite{ampdamp}. Note that Eq.~(\ref{eq:avrg}) was derived starting
from the worst-case infidelity $1-\min_{|\psi\ket}\bra\psi|\E(|\psi\ket\bra\psi|)|\psi\ket$,
applying the Fuchs\textendash van de Graaf inequalities \cite{Fuchs1999}
to convert the infidelity to the maximum trace distance $\max_{|\psi\ket}\|(\E-\id)(|\psi\ket\bra\psi|)\|_{\text{tr}}$
(with $\|A\|_{\text{tr}}\equiv\tr\{\sqrt{A^{\dg}A}\}$), upper-bounding
said trace distance by its stabilized version $\max_{|\psi\ket}\|(\E\otimes\id-\id^{\otimes2})(|\psi\ket\bra\psi|)\|_{\text{tr}}$
(i.e., the diamond norm), upper-bounding the diamond norm by the trace
norm $d\|\r_{\E}-\r_{\id}\|_{\text{tr}}$ between the Choi matrices
of $\E$ and $\id$ using Lemma 7 from Refs.~\cite{Wallman2014,Wallman2016}
(this is where the $d$-dependence comes in), and once again applying
Fuchs\textendash van de Graaf to turn the trace norm into $d\sqrt{1-\fe^{2}}$.

An information-theoretic property of $\fe$ is its presence in the
quantum Fano inequality \cite{Sch96b} (see also \cite{nielsen_chuang},
Thm. 12.9). This is an upper bound on the von Neumann entropy of $\r_{\E}$,
\begin{equation}
H(\r_{\E})\leq H(\{\fe,1-\fe\})+(1-\fe)\log_{2}(d^{2}-1)\,,\label{eq:bound}
\end{equation}
where $H(\r_{\E})=-\tr\{\r_{\E}\log_{2}\r_{\E}\}$. The entropy $H(\r_{\E})$
is also called the entropy exchange since it quantifies the entropy
gained by the environment responsible for the non-unitary nature of
$\E$. There is also the anti-Fano inequality \cite{Caves1999}, a
lower bound on $\fe$ in terms of the entropy,
\begin{equation}
\fe\geq e^{-2H(\r_{\E})}\,.
\end{equation}

A similar relation to information-theoretic quantities can be made
regarding a specific error map \textemdash{} the erasure channel.
Let us divide $\bo$ into two regions, $\bo_{1}$ and $\bo_{2}$,
and trace over $\bo_{2}$. Then, the $\fe$ given the optimal recovery
which reconstructs $\bo_{2}$ using only $\bo_{1}$ satisfies
\begin{equation}
\fe\geq e^{-\half I(\al:\bo_{2}|\bo_{1})}\,,
\end{equation}
where $I(\al:\bo_{2}|\bo_{1})$ is the conditional mutual information
quantifying correlations between $\al$ and $\bo_{2}$ given information
from $\bo_{1}$ \cite{Fawzi2015}. In this context, $\fe$ is also
called the fidelity of recovery \cite{Seshadreesan2015}.

From yet another information theory perspective (\cite{Konig2009},
Thm. 2), the optimal $\fe$ yields the equality
\begin{equation}
d\fe=2^{-H_{\text{min}}\left(\bo|\al\right)_{\L\S}}\,,
\end{equation}
where $H_{\text{min}}\left(\bo|\al\right)_{\L\S}$ is the conditional
min entropy of the Choi matrix $\r_{\L\S}$ (\ref{eq:choi}) of the
encoding and loss portions of $\E$. This inequality can be adapted
from the equation below Eq.~(4.5) in Ref.~\cite{tomamichel} by
noting that $\R^{\dgt}\S$ is a unital map. In this context, $\al$
and $\bo$ (of dimensions $\infty$ and two, respectively) share a
state $\r_{\L\S}=\L\S\otimes\id(|\varPsi\ket\bra\varPsi|)$, and $H_{\text{min}}\left(\bo|\al\right)_{\L\S}$
is the most conservative way to quantify the uncertainty about the
state of $\bo$ after the state of $\al$ is sampled.

\section{Numerical benchmarking details\label{sec:details}}

\begin{table*}[t]
\begin{tabular}{cccccccc} \toprule  ~~~~~~~$\g$~~~~~~~ & \multicolumn{7}{c}{$\nb_\cc\leq2$}\tabularnewline \hline  & \multicolumn{2}{c}{$\cat$} & \multicolumn{2}{c}{$\bin$} & \multicolumn{1}{c}{$\gkp$} & \multicolumn{2}{c}{$\lat$}\tabularnewline    & $\a$ & $S$\: & $N$ & $S$ & $\D$ & $\D$ & $a$\tabularnewline  
\hline
0.0124 & 1.440 & 1\: & 1 & 1 & 0.481 & 0.477 & 1.550\tabularnewline 0.0247 & 1.440 & 1\: & 1 & 1 & 0.481 & 0.477 & 1.618\tabularnewline 0.0488 & 1.396 & 1\: & 1 & 1 & 0.481 & 0.477 & 1.618\tabularnewline 0.0723 & 1.369 & 1\: & 1 & 1 & 0.481 & 0.477 & 1.618\tabularnewline 0.0952 & 1.351 & 1\: & 1 & 1 & 0.481 & 0.477 & 1.618\tabularnewline 0.1175 & 1.332 & 1\: & 1 & 1 & 0.481 & 0.477 & 1.618\tabularnewline 0.1393 & 1.508 & 2\: & 1 & 1 & 0.481 & 0.477 & 1.618\tabularnewline 0.1605 & 1.508 & 2\: & 1 & 1 & 0.481 & 0.477 & 1.618\tabularnewline 0.1813 & 1.508 & 2\: & 1 & 1 & 0.481 & 0.477 & 1.618\tabularnewline 0.2015 & 1.508 & 2\: & 1 & 1 & 0.481 & 0.477 & 1.618\tabularnewline 0.2212 & 1.508 & 2\: & 1 & 1 & 0.481 & 0.477 & 1.618\tabularnewline 0.2404 & 1.508 & 2\: & 1 & 1 & 0.481 & 0.477 & 1.618\tabularnewline 0.2592 & 1.508 & 2\: & 1 & 1 & 0.481 & 0.477 & 1.618\tabularnewline 0.2775 & 1.508 & 2\: & 1 & 1 & 0.481 & 0.477 & 1.618\tabularnewline 0.2953 & 1.508 & 2\: & 1 & 1 & 0.481 & 0.477 & 1.618\tabularnewline 0.3127 & 1.508 & 2\: & 1 & 1 & 0.481 & 0.477 & 1.618\tabularnewline 0.3297 & 1.508 & 2\: & 1 & 1 & 0.481 & 0.477 & 1.618\tabularnewline 0.3462 & 1.508 & 2\: & 1 & 1 & 0.481 & 0.477 & 1.618\tabularnewline 0.3624 & 1.508 & 2\: & 1 & 1 & 0.481 & 0.477 & 1.618\tabularnewline 0.3781 & 1.194 & 1\: & 0 & 0 & 0.481 & 0.477 & 1.618\tabularnewline 0.3935 & 1.183 & 1\: & 0 & 0 & 0.481 & 0.477 & 1.525\tabularnewline 0.4084 & 1.173 & 1\: & 0 & 0 & 0.481 & 0.477 & 1.525\tabularnewline 0.4231 & 1.173 & 1\: & 0 & 0 & 0.481 & 0.510 & 1.450\tabularnewline 0.4373 & 1.162 & 1\: & 0 & 0 & 0.500 & 0.659 & 1.350\tabularnewline 0.4512 & 0 & 0\: & 0 & 0 & 0.535 & 0.659 & 1.350\tabularnewline 0.4647 & 0 & 0\: & 0 & 0 & 0.577 & 0.913 & 1.250\tabularnewline 0.4780 & 0 & 0\: & 0 & 0 & 0.632 & 0.933 & 1.200\tabularnewline 0.4908 & 0 & 0\: & 0 & 0 & 0.632 & 0.953 & 1.150\tabularnewline 0.5034 & 0 & 0\: & 0 & 0 & 0.632 & 0.976 & 1.100\tabularnewline
\toprule
\end{tabular}~~~~~\begin{tabular}{ccccccc} \toprule   \multicolumn{7}{c}{$\nb_\cc\leq5$}\tabularnewline \hline \multicolumn{2}{c}{$\cat$} & \multicolumn{2}{c}{$\bin$} & $\gkp$ & \multicolumn{2}{c}{$\lat$}\tabularnewline 
 $\a$ & $S$\: & $N$ & $S$ & $\D$ & $\D$ & $a$\tabularnewline  
\hline
1.739 & 2\: & 2 & 2 & 0.309 & 0.309 & 1.650\tabularnewline 1.746 & 2\: & 2 & 2 & 0.309 & 0.309 & 1.650\tabularnewline 1.962 & 3\: & 1 & 2 & 0.309 & 0.309 & 1.650\tabularnewline 1.969 & 3\: & 1 & 3 & 0.309 & 0.309 & 1.700\tabularnewline 1.975 & 3\: & 1 & 3 & 0.309 & 0.309 & 1.700\tabularnewline 1.981 & 3\: & 1 & 3 & 0.309 & 0.309 & 1.700\tabularnewline 1.987 & 3\: & 1 & 3 & 0.309 & 0.309 & 1.700\tabularnewline 1.994 & 3\: & 1 & 3 & 0.309 & 0.309 & 1.700\tabularnewline 1.994 & 3\: & 1 & 3 & 0.309 & 0.309 & 1.700\tabularnewline 2.000 & 3\: & 1 & 3 & 0.309 & 0.309 & 1.700\tabularnewline 2.000 & 3\: & 1 & 3 & 0.309 & 0.309 & 1.700\tabularnewline 2.000 & 3\: & 1 & 3 & 0.309 & 0.309 & 1.700\tabularnewline 2.000 & 3\: & 1 & 3 & 0.309 & 0.309 & 1.700\tabularnewline 1.994 & 3\: & 1 & 3 & 0.309 & 0.309 & 1.700\tabularnewline 1.994 & 3\: & 1 & 3 & 0.309 & 0.309 & 1.700\tabularnewline 1.987 & 3\: & 1 & 3 & 0.309 & 0.309 & 1.700\tabularnewline 1.981 & 3\: & 1 & 3 & 0.309 & 0.309 & 1.700\tabularnewline 1.975 & 3\: & 1 & 3 & 0.309 & 0.309 & 1.700\tabularnewline 1.969 & 3\: & 1 & 3 & 0.309 & 0.309 & 1.700\tabularnewline 1.643 & 2\: & 1 & 3 & 0.309 & 0.309 & 1.700\tabularnewline 1.636 & 2\: & 1 & 2 & 0.316 & 0.312 & 1.732\tabularnewline 1.628 & 2\: & 1 & 2 & 0.392 & 0.394 & 1.650\tabularnewline 1.612 & 2\: & 1 & 2 & 0.471 & 0.510 & 1.450\tabularnewline 1.162 & 1\: & 0 & 0 & 0.500 & 0.659 & 1.350\tabularnewline
0 & 0\: & 0 & 0 & 0.535 & 0.659 & 1.350\tabularnewline 
0 & 0\: & 0 & 0 & 0.577 & 0.913 & 1.250\tabularnewline 
0 & 0\: & 0 & 0 & 0.632 & 0.933 & 1.200\tabularnewline 
0 & 0\: & 0 & 0 & 0.632 & 0.953 & 1.150\tabularnewline 
0 & 0\: & 0 & 0 & 0.632 & 0.976 & 1.100\tabularnewline\toprule
\end{tabular}~~~~~\begin{tabular}{ccccccc} \toprule   \multicolumn{7}{c}{$\nb_\cc\leq10$}\tabularnewline \hline \multicolumn{2}{c}{$\cat$} & \multicolumn{2}{c}{$\bin$} & $\gkp$ & \multicolumn{2}{c}{$\lat$}\tabularnewline 
$\a$ & $S$\: & $N$ & $S$ & $\D$ & $\D$ & $a$\tabularnewline   
\hline
2.890 & 3\: & 3 & 4 & 0.221 & 0.221 & "\tabularnewline 
2.890 & 3\: & 3 & 4 & 0.221 & 0.221 & "\tabularnewline 
3.162 & 4\: & 2 & 4 & 0.221 & 0.221 & "\tabularnewline 
3.162 & 4\: & 2 & 5 & 0.221 & 0.221 & "\tabularnewline 
1.975 & 3\: & 2 & 5 & 0.221 & 0.221 & "\tabularnewline 
1.981 & 3\: & 2 & 5 & 0.221 & 0.221 & 1.725\tabularnewline 
1.987 & 3\: & 2 & 5 & 0.221 & 0.221 & 1.725\tabularnewline 
1.994 & 3\: & 2 & 5 & 0.221 & 0.221 & 1.725\tabularnewline 
1.994 & 3\: & 2 & 5 & 0.221 & 0.221 & 1.725\tabularnewline 2.000 & 3\: & 2 & 5 & 0.221 & 0.221 & 1.725\tabularnewline 2.000 & 3\: & 2 & 5 & 0.221 & 0.221 & 1.725\tabularnewline 2.000 & 3\: & 1 & 3 & 0.221 & 0.221 & 1.725\tabularnewline 2.000 & 3\: & 1 & 3 & 0.221 & 0.221 & 1.725\tabularnewline 1.994 & 3\: & 1 & 3 & 0.221 & 0.221 & 1.725\tabularnewline 1.994 & 3\: & 1 & 3 & 0.221 & 0.221 & 1.725\tabularnewline 1.987 & 3\: & 1 & 3 & 0.221 & 0.221 & 1.725\tabularnewline 1.981 & 3\: & 1 & 3 & 0.221 & 0.221 & 1.725\tabularnewline 1.975 & 3\: & 1 & 3 & 0.221 & 0.221 & 1.725\tabularnewline 1.969 & 3\: & 1 & 3 & 0.221 & 0.221 & 1.725\tabularnewline 1.643 & 2\: & 1 & 3 & 0.246 & 0.243 & 1.700\tabularnewline 1.636 & 2\: & 1 & 2 & 0.316 & 0.312 & 1.732\tabularnewline 1.628 & 2\: & 1 & 2 & 0.392 & 0.394 & 1.650\tabularnewline 1.612 & 2\: & 1 & 2 & 0.471 & 0.510 & 1.450\tabularnewline 1.162 & 1\: & 0 & 0 & 0.500 & 0.659 & 1.350\tabularnewline 
0 & 0\: & 0 & 0 & 0.535 & 0.659 & 1.350\tabularnewline 
0 & 0\: & 0 & 0 & 0.577 & 0.913 & 1.250\tabularnewline 
0 & 0\: & 0 & 0 & 0.632 & 0.933 & 1.200\tabularnewline 
0 & 0\: & 0 & 0 & 0.632 & 0.953 & 1.150\tabularnewline 
0 & 0\: & 0 & 0 & 0.632 & 0.976 & 1.100\tabularnewline
\toprule
\end{tabular}




\caption{\label{tab:numdet}Code parameters for the code giving the highest
$\protect\fe$ (\ref{eq:fid}) out of all codes of a given code family
with the constraint $\protect\nb_{\protect\cc}\leq2,5,10$ and a given
loss rate $\protect\g$ (\ref{eq:gamma}). The first column lists
the $\protect\g$'s sampled while the next three sets of seven columns
give the code parameters for $\protect\cat(\protect\a,S)$ (\ref{eq:cat}),
$\protect\bin(N,S)$ (\ref{eq:bin}), $\protect\gkp(\protect\D)$
(\ref{eq:gkp}), and $\protect\lat(\protect\D,a)$ (\ref{eq:gkpnum}).
Each of the three sets corresponds to one of the three energy constraints.
Optimal code values below $\protect\g\leq0.0124$ do not change significantly
and so are not shown. For $\protect\g\geq0.4512$, a small $\protect\nb$
is preferable for all codes and optimal $\protect\cat$ and $\protect\bin$
switch to encoding into the first two Fock states. For $\protect\nb_{\protect\lat}\leq10$
and $\protect\g\leq0.05$, denoted with the " symbol in the last
column, we bound $\protect\fe$ with the channel fidelity $\protect\fe^{\text{QR}}$
that uses the quadratic recovery $\protect\R^{\protect\qr}$ \cite{Tyson2010}
(see also \cite{Beny2010,Beny2011}) for unshifted hexagonal $\protect\lat$
(\ref{eq:gkp}) because $\protect\fe$ decreases significantly around
that regime due to numerical precision limitations of the optimization.
We also use $\protect\fe^{\text{QR}}$ to bound $\protect\fe$ in
Sec.~\ref{subsec:Removing-energy-constraints}.}
\end{table*}

The parameters for the specific members of the $\cat/\bin/\gkp/\lat$
code families that optimize $\fe(\g)$ in Fig.~\ref{f:num} are given
in Table~\ref{tab:numdet}. 

There are a total of five $\num$ codes, organized by their approximate
mean occupation number
\begin{equation}
\nb_{\num}\in\{1.562,2.696,2.770,4.149,4.336\}\,.\label{eq:num}
\end{equation}
Interestingly, the first code \textemdash{} the $\sqrt{17}$ code
\cite{bin} \textemdash{} can be expressed as\begin{subequations}
\begin{align}
|0_{\num}^{\nb\approx1.562}\ket & ={\textstyle \frac{1}{\sqrt{6}}\left(\sqrt{7-\sqrt{17}}|0\ket+\sqrt{\sqrt{17}-1}|3\ket\right)}\,,\label{eq:sqrt17}\\
|1_{\num}^{\nb\approx1.562}\ket & ={\textstyle \frac{1}{\sqrt{6}}\left(\sqrt{9-\sqrt{17}}|1\ket-\sqrt{\sqrt{17}-3}|4\ket\right)}\,.
\end{align}
\end{subequations}All five code states are listed in the ancillary
\textsc{Mathematica} notebook accompanying this manuscript on the
\textsc{arXiv}. The $\nb_{\num}\in\{1.562,2.696,4.149\}$ codes are
from Ref.~\cite{bin} while the $\nb_{\num}\in\{2.770,4.336\}$ codes
were obtained here using a different optimization routine, described
as follows. In order to find logical states $|\m_{\num}\ket$ for
$\m\in\{0,1\}$ which allow for the correction of error operators
$e_{\ell}\in\{I,\aa,\aa^{2}\}$, we create a cost function from the
QEC matrix $f_{\m\n\ell\lp}=\bra\m_{\num}|e_{\ell}^{\dagger}e_{\lp}|\n_{\num}\ket$,
$c_{1}=\sum_{\ell,\lp}|f_{00\ell\lp}-f_{11\ell\lp}|^{2}+|f_{01\ell\lp}|^{2}.$
In order to prefer lower occupation, the penalty $c_{2}=\lambda_{\bar{n}}\nb_{\num}$
is introduced with $\lambda_{\bar{n}}=10^{-3}$. Code words are produced
by numerically optimizing the total cost over complex unit vectors:
\begin{equation}
\underset{|\psi_{0}\ket,|\psi_{1}\ket\in\mathbb{C}^{\nmax}}{\text{minimize}}\;\;c_{1}+c_{2}\,,
\end{equation}
where $\nmax$ is the Fock space cutoff. The $\sqrt{17}$ code is
the only code below $\nb_{\num}=2$. For $\nb_{\num}\leq5,10$, the
best-performing code for $\g\leq0.3935$ is $\num(4.149)$, for $\g=0.4084$
is $\num(2.770)$, and for $\g\geq0.4231$ is the $\sqrt{17}$ code.
We were unable to find good codes with $\nb_{\num}>5$ due to the
prominence of shallow local minima.

For the numerical comparison, we swept all values of the code parameters
for $\cat(\a,S)$, $\bin(N,S)$, $\num(\nb)$, and $\gkp(\D)$ subject
to the energy constraints. For $\lat(\D,a)$, we only considered values
of $\D$ which gave $\nb_{\lat}\approx2$, $5$, or $10$ (since we
knew from the $\gkp$ calculations that increasing $\nb$ generally
increased $\fe$ for all but the largest values of $\g$). We also
did not consider all possible non-rectangular lattices, but instead
implemented a subset of them by sweeping the lattice parameter $a\in[1,2]$
in the following coherent-state representation for the shifted non-square
$\lat$ code states:\begin{widetext}

\begin{equation}
|\m_{\lat}^{\text{shift}}\ket\propto\sum_{\vec{n}\in\Z}\left(-1\right)^{\m n_{1}}e^{-i\frac{\pi}{2}n_{2}(2n_{1}+\m)}e^{-\frac{\pi a}{4}\D^{2}\left[(2n_{1}+\m)^{2}+(\frac{2}{a}n_{2})^{2}\right]}\kkk{\frac{\sqrt{\pi a}}{2}\left(2n_{1}+\m+i\frac{2}{a}n_{2}\right)}\,.\label{eq:gkpnum}
\end{equation}
\end{widetext}The $\D\rightarrow0$ states are stabilized by
\begin{align*}
S_{\textbf{x}} & =-D_{\sqrt{\frac{\pi a}{2}}\left(1+i\frac{2}{a}\right)}\,\,\,\,\,\,\,\,\,\text{and}\,\,\,\,\,\,\,\,\,S_{\textbf{p}}=D_{4i\sqrt{\frac{\pi}{2a}}}\,.
\end{align*}
The resulting code lattice formed by $\half P_{\lat}^{\text{shift}}$
is shifted, retaining its error-correcting properties but not having
a lattice point at the origin. Both shifted and unshifted lattices
are used in the numerics, but only unshifted lattices are used for
analytical calculations in Appx.~\ref{appx:-analytical-calculations}.
Note that $|0_{\lat}^{\text{shift}}\ket$ is odd while $|1_{\lat}^{\text{shift}}\ket$
is even under parity $(-1)^{\ph}$, so there is no spacing $S$.

Since $\cat$ and $\lat$ code states are formally superpositions
of an infinite number of Fock states, we have to truncate them and
use a finite Fock state superposition $\sum_{n=0}^{\nmax}c_{n}|n\ket$
for each logical state. We picked $\nmax$ such that $\sum_{n=0}^{\nmax}|c_{n}|^{2}\geq0.99999$
for both logical states. For $\gkp/\lat$ code states, we used the
coherent state representation and picked only the lattice points values
$s,t\leq\left\lfloor 4/\D\right\rfloor $. The codes were generated
with \textsc{Mathematica} while the semidefinite program was executed
using the \textsc{CVX} package \cite{cvx} in \textsc{MATLab}, with
the \textsc{MATLink} add-on \cite{matlink} for \textsc{Mathematica}
acting as the bridge. Helpful routines were borrowed from Toby S.
Cubitt \cite{cubitt}.

\begin{figure}[t]
\includegraphics[width=1\columnwidth]{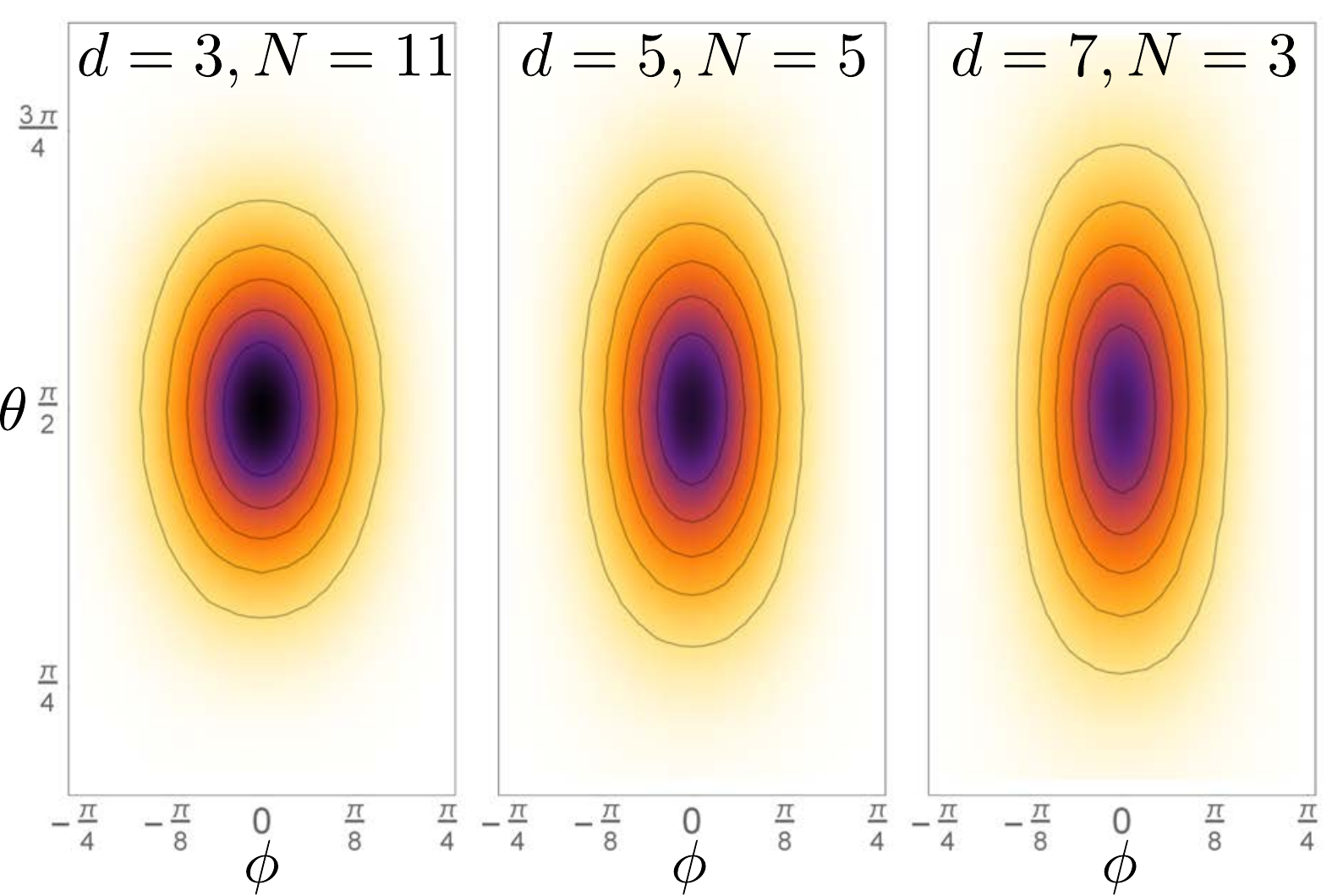}

\caption{\label{f:bin-2}Overlap $|_{J}\protect\bra\protect\t,\phi|{\textstyle \frac{\pi}{2},\frac{2\pi}{d}\protect\m}\protect\ket_{J,d}|^{2}$
vs. $\protect\t,\phi$ (in radians) for qudit states $|{\textstyle \frac{\pi}{2},\frac{2\pi}{d}\protect\m}\protect\ket_{J,d}$
(\ref{eq:extbin}) with $\protect\m=0$ and $d,N$ picked such that
the spin $J=\protect\half(d-1)(N+1)=12$. These states resemble spin-squeezed
states and characterize the qu$d$it $\protect\bin$ and $\protect\tw$
codes (see Appx.~\ref{appx:Qudit-binomial-codes}). For fixed $J$,
the degree of squeezing increases with $d$.}
\end{figure}

\section{Qudit $\protect\bin$ and $\protect\tw$ codes\label{appx:Qudit-binomial-codes}}

Here we extend the analogy from Sec.~\ref{subsec:Relation-to-spin-coherent}
between $\bin$ codes and spin-coherent states to qu$d$it $\bin$
codes, introducing a new multi-qudit code $\pa$ in the process and
eventually yielding a logical $X_{\cc}$-operator and a check operator
for the qu$d$it $\bin$ \cite{bin} and $\tw$ codes. We consider
the following generalization of spin-coherent states (\ref{eq:spin}),
\begin{equation}
|\t,\phi\ket_{J,d}=\sum_{m=0}^{2J}\frac{\sqrt{{N+1 \choose m}_{d}}\,e^{i\phi m}\tan^{2m}\frac{d-1}{d}\t}{\sqrt{\left(\sum_{\m=0}^{d-1}\tan^{2\m}\frac{d-1}{d}\t\right)^{N+1}}}|J,m-J\ket\,,\label{eq:extbin}
\end{equation}
where $|J,m-J\ket$ is the spin basis of a spin $J=\half\left(d-1\right)\left(N+1\right)$,
and ${N+1 \choose m}_{d}$ are extended binomial coefficients \cite{Neuschel2014}
(also called polynomial coefficients \cite{Caiado2007}), defined
by
\begin{equation}
\left(1+x+\cdots+x^{d-1}\right)^{N+1}=\sum_{m=0}^{\left(d-1\right)\left(N+1\right)}{N+1 \choose m}_{d}x^{m}\,.
\end{equation}
(Note a similar generalization in the proof of Ref.~\cite{Ouyang2016},
Thm. 1.2). For $d=2$, $\{|{\textstyle \frac{\pi}{2},\pi\m}\ket_{J,d=2}\}_{\m=0}^{1}$
reduce to the two antipodal spin-coherent states discussed in the
main text. In general, the $d$ states $\{|{\textstyle \frac{\pi}{2},\frac{2\pi}{d}\m}\ket_{J,d}\}_{\m=0}^{d-1}$
are similar to squeezed spin-coherent states equidistantly distributed
along the equator of the Bloch sphere. For a fixed $J$, the amount
of squeezing increases with increasing $d$, as shown in an example
in Fig.~\ref{f:bin-2}. This is sensible since increasing $d$ for
fixed $J$ means fitting more quantum information in the same amount
of space.

Mapping the basis $\kkk{J,m-J}$ to Fock states $|\left(S+1\right)m\ket$
or two-mode Fock states 
\[
|\left(S+1\right)[\left(d-1\right)\left(N+1\right)-m],\left(S+1\right)m\ket
\]
yields qudit versions of $\bin$ and $\tw$, respectively, for general
parameters $N,S$. It was shown in Ref.~\cite{bin} that qudit $\bin$
codes can protect to order $O(\g^{N})$ against dephasing and $O(\g^{S})$
against loss. We do not prove this for $\tw$ here, but anticipate
this to also be the case for those codes. Observing the protection
offered by $\noon$ codes \cite{Bergmann2016a}, it is also reasonable
to believe that tensor products $|\frac{\pi}{2},\frac{2\pi}{d}\m\ket_{J,d}^{\otimes M}$
(with $|J,m-J\ket$ mapped to one- or two-mode Fock states) will yield
yet another class of multi-mode codes. Regarding the multi-qubit mapping,
we introduce $\pa$ \textemdash{} a new extension of qubit $\pt$
codes to qudits. These codes can be obtained by mapping $\kkk{J,m-J}$
to a qudit generalization of Dicke states $|D_{m}^{N+1}\ket$ that
we denote as the \textit{extended binomial states} $|E_{m}^{d-1,N+1}\ket$,
i.e.,
\begin{align}
|\m_{\pa}\ket & =\sum_{m=0}^{\left(d-1\right)\left(N+1\right)}\frac{e^{i\frac{2\pi}{d}\m m}}{\sqrt{d^{N+1}}}\sqrt{{N+1 \choose m}_{d}}|E_{m}^{d-1,N+1}\ket\,.\label{eq:qudi}
\end{align}
The states $|E_{m}^{d-1,N+1}\ket$ are defined as normalized equal
superpositions of all multi-qudit states having a total of $m$ excitations
distributed over $N+1$ qudits,
\begin{equation}
|E_{m}^{d-1,N+1}\ket=\frac{1}{\sqrt{{N+1 \choose m}_{d}}}\sum_{_{\,\,\,\,\,\,\sum_{i}v_{i}=m}^{v_{1},\cdots,v_{N+1}=0}}^{d-1}|v_{1},\cdots,v_{N+1}\ket\,.
\end{equation}
The normalization of these states happens to be exactly the extended
binomial coefficient because ${N+1 \choose m}_{d}$ is, by definition,
the number of ways of obtaining $m$ as the sum of $N+1$ independent
random variables which take values from 0 to $d-1$ \cite{Caiado2007}.

The code $\pa$ is \textit{different} from the qudit $\pt$ codes
\cite{Ouyang2016} because those utilize a different generalization
of qubit Dicke states. For example, extended binomial states for $d=3$
and $N=1$ are\begin{subequations}
\begin{align}
|E_{0}^{2,2}\ket & =|00\ket,\,\,\,\,\,\,\,\,\,\,|E_{1}^{2,2}\ket=\frac{1}{\sqrt{2}}\left(|01\ket+|10\ket\right),\\
|E_{2}^{2,2}\ket & =\frac{1}{\sqrt{3}}\left(|02\ket+|11\ket+|20\ket\right),\\
|E_{3}^{2,2}\ket & =\frac{1}{\sqrt{2}}\left(|21\ket+|12\ket\right),\,\,\,\,\,\,\,\,\,\,|E_{4}^{2,2}\ket=|22\ket\,.
\end{align}
\end{subequations}In contrast, qudit Dicke states \cite{Ouyang2016}
are superpositions of a multi-qudit state which has a fixed number
of excitations \textit{for each qudit} and all of that state's permutations.
For the above case, the qudit Dicke states are $|E_{0}^{2,2}\ket,|E_{1}^{2,2}\ket,|E_{3}^{2,2}\ket,|E_{4}^{2,2}\ket$
along with $\frac{1}{\sqrt{2}}(|02\ket+|20\ket)$ and $|11\ket$.
In the general case of $N+1$ qu$d$its, there are ${N+d \choose d-1}$
qudit Dicke states while only $(d-1)\left(N+1\right)+1$ extended
binomial states. While the qudit Dicke states span the entire fully
symmetric $N+1$-qudit subspace, extended binomial states span only
a subspace of that subspace. After introduction of a spacing $S\neq0$
in similar fashion to $\pt$ codes (see Sec.~\ref{subsec:Permutation-invariant-codes}),
it may be that $|\m_{\pa}\ket$ protects against multi-qubit amplitude
damping, but such properties have yet to be proven. 

We conclude this section by relating $\pa$ to eigenstates of an $N$-qu$d$it
generalization of an $N$-qubit collective spin operator. This reveals
that such codes are closely related to bit-flip codes and provides
a check operator for qudit $\bin$ and $\tw$ codes.

\subsection{Relating $\protect\pa$ codes to bit-flip codes}

We start with the spin-coherent states from Sec.~\ref{sec:bin-codes}
(i.e., $d=2$) written in the irrep for which $J_{x}$ is a collective
operator for a $2J=M$-qubit system with $S=0$. In other words, $J_{x}=\frac{1}{2}\sum_{k=1}^{M}X_{k}$,
where $X_{k}$ is the Pauli matrix of the $k$th qubit. For those
parameters, the qubit states $|\m_{\pt}\ket$ in this irrep are simply
tensor products of eigenstates $|\left(-1\right)^{\m}\ket_{k}$ of
$X_{k}$,
\begin{equation}
|\m_{\pt}\ket=\bigotimes_{k=1}^{M}|\left(-1\right)^{\m}\ket_{k}=\frac{1}{\sqrt{2^{M}}}\bigotimes_{k=1}^{M}\left[|0\ket_{k}+\left(-1\right)^{\m}|1\ket_{k}\right]
\end{equation}
with $X_{k}|\left(-1\right)^{\m}\ket_{k}=\left(-1\right)^{\m}|\left(-1\right)^{\m}\ket_{k}$.
Proving this is simple if one writes out $\bigotimes_{k=1}^{M}|\left(-1\right)^{\m}\ket_{k}$
in terms of the Dicke states $\{|D_{m}^{M}\ket\}_{m=1}^{M}$. Observe
that, after performing all tensor products, $|\m_{\pt}\ket$ will
consist of an equal linear superposition of multi-qubit states (denoted
by binary strings) with coefficients $\pm1/\sqrt{2^{M}}$. To change
basis to Dicke states, we group multi-qubit states by their total
number of excitations (i.e., the number of $1$'s in each binary string).
For $m$ excitations out of $M$ qubits, the number of such states
is ${M \choose m}$. Moreover, since each additional excitation brings
about an additional factor of $-1$, all states with the same number
of excitations have matching coefficients. Thus, we can group each
superposition of states with fixed excitations into unnormalized Dicke
states. Multiplying and dividing each unnormalized Dicke state by
${M \choose m}^{-1/2}$ yields the original form of $|\m_{\pt}\ket$
in Eq.~(\ref{eq:perm}).

We can now generalize the above setup to qudits. Consider $M$ qudits
of dimension $d$ and let
\begin{equation}
X=\sum_{\n=0}^{d-1}|\n\ket\bra\n+1\text{ mod }d|
\end{equation}
now be the shift operator for a qudit (i.e., defined such that $X|d-1\ket=|0\ket$).
This $X$ has $d$ eigenstates 
\begin{equation}
|e^{i\frac{2\pi}{d}\m}\ket=\frac{1}{\sqrt{d}}\sum_{\n=0}^{d-1}e^{i\frac{2\pi}{d}\m\n}|\n\ket
\end{equation}
(with $\m\in\{0,1,\cdots,d-1\}$) whose eigenvalues are $e^{i\frac{2\pi}{d}\m}$.
Using the same procedure as above, one can consider tensor products
of $|e^{i\frac{2\pi}{d}\m}\ket$,
\begin{equation}
|\m_{\pa}\ket=\left(|e^{i\frac{2\pi}{d}\m}\ket\right)^{\otimes M}\,,
\end{equation}
and express them in the extended binomial basis. Now, $|\m_{\pa}\ket$
consists of equal superpositions of multi-qudit states with coefficients
$\{e^{i\frac{2\pi}{d}\m\n}/\sqrt{d^{M}}\}_{\n=0}^{d-1}$, but the
coefficients in front of multi-qudit states of fixed total excitation
match. The normalization of $|E_{m}^{d-1,M}\ket$ is the square root
of the number of multi-qudit states in $|E_{m}^{d-1,M}\ket$, which
we have already defined to be ${M \choose m}_{d}$. This yields the
$\pa$ code states from Eq.~(\ref{eq:qudi}) with $M=N+1$.

An important consequence of the above description is that now all
qudit $\texttt{codes}\in\{\bin,\tw\}$ admit a logical operator
\begin{equation}
X_{\cc}=\frac{1}{M}\sum_{k=1}^{M}X_{k}\,,
\end{equation}
where $X_{k}$ is $X$ for the $k$th qudit, and a corresponding check
operator $(X_{\cc})^{M}$. It is implied that both of these are projected
only onto the subspace spanned by the extended binomial states $\{|E_{m}^{d-1,M}\ket\}_{m=0}^{\left(d-1\right)M}$
(vs. the full permutation-symmetric subspace spanned by the qudit
Dicke states discussed above). Thus, the logical operator will be
a matrix of dimension $\left(d-1\right)M$ satisfying
\begin{equation}
X_{\cc}|\m_{\pa}\ket=e^{i\frac{2\pi}{d}\m}|\m_{\pa}\ket\,.
\end{equation}
Mapping the extended binomial state basis to the corresponding Fock
states thus creates analogous check operators for $\bin$ and $\tw$.
One can also consider products of $X_{k}$'s and form other check
operators. Such check operators should prove useful in experimental
realizations of the error correcting procedures of these codes.

\section{Calculations for $\protect\lat$ codes\label{appx:-analytical-calculations}}

\subsection{Useful identities}

Throughout the text, we have used the following standard identities
for coherent states $D_{\a}|0\ket=|\a\ket$ and $|\b\ket$:\begin{subequations}
\begin{align}
e^{x\ph}|\a\ket & =e^{-\half\left|\a\right|^{2}(1-|e^{x}|^{2})}|\a e^{x}\ket\label{eq:iden3}\\
D_{\a}D_{\b} & =e^{\half\left(\a\b^{\star}-\a^{\star}\b\right)}D_{\a+\b}\\
\bra\a|\b\ket & =e^{-\half\left(\left|\alpha\right|^{2}+\left|\beta\right|^{2}\right)+\alpha^{\star}\beta}\,.
\end{align}
\end{subequations}We also use the Fock space matrix elements of $D_{\a}$
\cite{vogel}, 
\begin{align}
\langle\ell|D_{\alpha}|\lp\rangle & =e^{-\frac{|\alpha|^{2}}{2}}\sqrt{\frac{\lp!}{\ell!}}L_{\lp}^{(\ell-\lp)}(|\alpha|^{2})\alpha^{\ell-\lp}\label{displacement in terms of Laguerre}
\end{align}
for $\ell\ge\lp$ and $\langle\ell|D_{\alpha}|\lp\rangle=(\langle\lp|D_{-\alpha}|\ell\rangle)^{\star}$
for $\ell<\lp$, where $L_{n}^{(a)}(x)$ is the generalized Laguerre
polynomial. The generating function of these polynomials is 
\begin{align}
\sum_{p=0}^{\infty}\frac{(m+p)!}{p!}t^{p}L_{m+p}^{(\alpha)}(x) & =\frac{m!e^{-\frac{tx}{1-t}}}{(1-t)^{m+\alpha+1}}L_{m}^{(\alpha)}\left(\frac{x}{1-t}\right)\,.\label{eq:genfunc}
\end{align}
Finally, we use the Poisson summation formula; for a function $f\left(x\right)$,
\begin{equation}
\sum_{n\in\Z}f\left(n\right)=\sum_{n\in\Z}\int_{-\infty}^{\infty}dxe^{2\pi inx}f\left(x\right)\,.\label{eq:poisson}
\end{equation}

\subsection{Equivalence between squeezed and coherent state representations for
$\protect\lat$\label{appx:Equivalence-between-squeezed}}

We sketch a derivation of Eq.~\eqref{GKP states in terms of position eigenstates}
from Eq.~\eqref{eq:GKP in terms of coh}. Writing the displacements
in Eq.~\eqref{eq:gkpcoherent} in terms of position and momentum
operators $\hat{x}$ and $\hat{p}$, inserting a resolution of the
identity in terms of position eigenstates between the displacements,
and using $e^{-ix_{2}\hat{p}}|x_{1}\rangle_{\hat{x}}=|x_{1}+x_{2}\rangle_{\hat{x}}$
yields
\begin{align}
|\mu_{\gkp}^{\text{ideal}}\rangle & \propto\sum_{\vec{n}\in\mathbb{Z}^{2}}\int dx|x+\sqrt{\pi}(2n_{1}+\mu)\rangle_{\hat{x}}\,_{\hat{x}}\langle x|e^{i\sqrt{\pi}n_{2}\hat{x}}|\text{vac}\rangle\,,
\end{align}
where $\vec{n}=(n_{1},n_{2})$, $|\text{vac}\ket$ is the Fock state
$|0\ket$, and we use ``$\propto$'' to ignore normalization and
any constant pre-factors that we obtain throughout the calculation.
Now, we recall that $_{\hat{x}}\langle x|\text{vac}\rangle\propto\exp(-\frac{1}{2}x^{2})$
and apply the Poisson summation (\ref{eq:poisson}) to the sum over
$n_{2}$, yielding a sum over Dirac $\d$-functions. We can then easily
evaluate the integral over $x$, yielding
\begin{align}
|\m_{\gkp}^{\text{ideal}}\ket & \propto\sum_{\vec{n}\in\Z_{2}}e^{-2\pi n_{2}^{2}}|\sqrt{\pi}(2n_{1}+2n_{2}+\mu)\rangle_{\hat{x}}\,.
\end{align}
Finally, we can redefine indices and evaluate one of the new sums
to yield 
\begin{align}
|\mu_{\gkp}^{\text{ideal}}\rangle & \propto\sum_{n\in\mathbb{Z}}|\sqrt{\pi}(2n+\mu)\rangle_{\hat{x}}\,.
\end{align}

\begin{widetext}

\subsection{Projecting displacements onto the $\protect\lat$ code space\label{subsec:Projecting-displacements-onto}}
\begin{widetext}
We have utilized all three representations (\ref{eq:gkpsqueezed}-c)
to verify the calculations below, initially calculating overlaps $\bra\m_{\gkp}^{\D}|\aa^{\dg p}\aa^{q}|\m_{\gkp}^{\D}\ket$
and summing them up to yield the QEC matrix $\e_{\ell\lp}^{\gkp}$.
We will not report on these calculations, noting that they are cumbersome,
but do yield the right answers.

We evaluate matrix elements $\langle\m_{\gkp}^{\text{ideal}}|D_{\alpha}|\n_{\gkp}^{\text{ideal}}\rangle$
of the displacement operator for ideal $\gkp$ states (written in
terms of position eigenstates $|\sqrt{\pi}(2n+\m)\rangle_{\hat{x}}$)
from Eq.~(\ref{GKP states in terms of position eigenstates}). We
can split $D_{\a}=D_{\a_{1}+i\a_{2}}$ into a shift by $\sqrt{2}\a_{1}$
in position and by $\sqrt{2}\a_{2}$ in momentum. The latter translates
$|\sqrt{\pi}(2n+\m)\rangle_{\hat{x}}$ while the former turns into
a phase since $|\sqrt{\pi}(2n+\m)\rangle_{\hat{x}}$ are eigenstates
of $\hat{x}$. We can then use the orthogonality of position eigenstates,
$_{\hat{x}}\bra x_{1}|x_{2}\ket_{\hat{x}}=\d(x_{1}-x_{2})$, and change
indices to obtain 
\begin{align}
\langle\m_{\gkp}^{\text{ideal}}|D_{\alpha}|\n_{\gkp}^{\text{ideal}}\rangle & =\sqrt{\frac{2}{\pi}}\sum_{n_{1},n_{2}\in\mathbb{Z}}e^{-i\alpha_{1}\alpha_{2}}e^{-i\sqrt{2\pi}(2n_{2}+\mu)\alpha_{2}}\delta\left(\alpha_{1}-\sqrt{\frac{\pi}{2}}(2n_{1}+\delta\mu)\right)\,,\label{Appendix muDnu start}
\end{align}
where we multiplied each $|\m_{\gkp}^{\text{ideal}}\ket$ by $(\frac{2}{\sqrt{\pi}})^{1/2}$
to remove constants in front of the sum (\ref{eq:idealme}) below.
We now apply the Poisson summation formula (\ref{eq:poisson}) to
turn the sum of $n_{2}$-dependent phases into another sum of Dirac
$\d$-functions for $\a_{2}$, yielding
\begin{align}
\langle\m_{\gkp}^{\text{ideal}}|D_{\alpha}|\n_{\gkp}^{\text{ideal}}\rangle & =\sum_{\vec{n}\in\mathbb{Z}}e^{i\pi(n_{1}+\frac{\mu+\nu}{2})n_{2}}\delta^{2}\left(\alpha-\Lambda_{\delta\mu}^{\vec{n}}\right)\,,\label{eq:idealme}
\end{align}
where $\Lambda_{\delta\mu}^{\vec{n}}=\sqrt{\frac{\pi}{2}}[(2n_{1}+\delta\mu)+in_{2}]$. 

Now let us consider finite $\gkp$ states in the smeared representation
(\ref{eq:gkpsmeared}) and calculate
\begin{align}
\langle\m_{\gkp}^{\D}|D_{\alpha}|\n_{\gkp}^{\D}\rangle & =\int\frac{d^{2}\beta d^{2}\gamma}{\pi\Delta^{2}/2}e^{-\frac{1}{\Delta^{2}}(|\beta|^{2}+|\gamma|^{2})}\langle\m_{\gkp}^{\text{ideal}}|D_{-\beta}D_{\alpha}D_{\gamma}|\n_{\gkp}^{\text{ideal}}\rangle\,.
\end{align}
We add the displacements and use Eq.~(\ref{eq:idealme}), whose $\d$-functions
allow us to immediately evaluate one of the integrals. The remaining
Gaussian integral is also simply evaluated to yield
\begin{align}
\langle\m_{\gkp}^{\D}|D_{\alpha}|\n_{\gkp}^{\D}\rangle & =\sum_{\vec{n}\in\mathbb{Z}^{2}}e^{i\pi(n_{1}+\frac{\mu+\nu}{2})n_{2}}e^{-\frac{1}{2\Delta^{2}}|\alpha-\Lambda_{\delta\mu}^{\vec{n}}|^{2}}e^{-\frac{\Delta^{2}}{8}|\alpha+\Lambda_{\delta\mu}^{\vec{n}}|^{2}}\,.\label{Appendix PDP gkps final}
\end{align}
We can then substitute $\a\sim\Lambda_{\delta\mu}^{\vec{n}}$ into
the envelope function in the $\D\rightarrow0$ limit, yielding Eq.~(\ref{eq:gkpqec}).
\end{widetext}

\subsection{QEC matrix for $\protect\lat$ codes\label{subsec:QEC-criteria-for}}
\begin{widetext}
To compute the QEC matrix for $\gkp$, let us sandwich both sides
of Eq.~(\ref{EE in terms of displacement}) by $\langle\m_{\gkp}^{\D}|$
and $|\n_{\gkp}^{\D}\rangle$:
\begin{align}
\langle\m_{\gkp}^{\D}|E_{\ell}^{\dagger}E_{\lp}|\n_{\gkp}^{\D}\rangle & =\int\frac{d^{2}\alpha}{\pi}e^{-\frac{(1-\gamma)}{2}|\alpha|^{2}}\langle\ell|D_{\alpha^{\star}}|\lp\rangle\langle\m_{\gkp}^{\D}|D_{\sqrt{\gamma}\alpha}|\n_{\gkp}^{\D}\rangle\,.\label{Appendix QEC criteria starting point}
\end{align}
Plugging in Eq.~\eqref{Appendix PDP gkps final} with $\a\sim\Lambda_{\delta\mu}^{\vec{n}}$
in the $\D^{2}$-dependent envelope and switching the sum and integral,
one obtains
\begin{align}
\langle\m_{\gkp}^{\D}|E_{\ell}^{\dagger}E_{\lp}|\n_{\gkp}^{\D}\rangle & \sim\sum_{\vec{n}\in\mathbb{Z}^{2}}e^{i\pi(n_{1}+\frac{\mu+\nu}{2})n_{2}}e^{-\frac{\Delta^{2}}{2}|\Lambda_{\delta\mu}^{\vec{n}}|^{2}}\int\frac{d^{2}\alpha}{\pi}e^{-\frac{(1-\gamma)}{2}|\alpha|^{2}}\langle\ell|D_{\alpha^{\star}}|\lp\rangle e^{-\frac{\gamma}{2\Delta^{2}}|\alpha-\Lambda_{\delta\mu}^{\vec{n}}/\sqrt{\gamma}|^{2}}\,.
\end{align}
Next, we evaluate the integral by changing to polar coordinates $\a=\left|\a\right|e^{i\t}$
and evaluating the angular integral first. This integral turns out
to be integral representation of the modified Bessel function of the
first kind, $I_{n}(z)=\int_{0}^{\pi}\frac{d\theta}{\pi}e^{z\cos\theta}\cos(n\theta)$.
Recalling that $\langle\ell|D_{\alpha^{\star}}|\lp\rangle$ \eqref{displacement in terms of Laguerre}
contain Laguerre polynomials, the remaining integral over $\left|\a\right|$
contains both $L_{\ell_{\min}}^{(|\ell-\lp|)}$ and $I_{|\ell-\lp|}$.
Luckily, it can be evaluated using Ref.~\cite{prudnikov}, Sec.~2.19.12,
Eq.~(6):
\begin{equation}
\int_{0}^{\infty}dxx^{\frac{\lambda}{2}}e^{-px}I_{\lambda}(2b\sqrt{x})L_{n}^{(\lambda)}(x)=b^{\lambda}\frac{(p-1)^{n}}{p^{\lambda+n+1}}e^{\frac{b^{2}}{p}}L_{n}^{(\lambda)}\left(\frac{b^{2}}{p(p-1)}\right)\,.
\end{equation}
The pre-factor $\frac{(p-1)^{n}}{p^{n+1}}$ eventually gives the thermal
weights in the QEC coefficients $c_{\ell\ell}^{\gkp}$ \eqref{eq:thermal}.
The resulting Laguerre polynomials can then be re-expressed in terms
of displacement matrix elements \eqref{displacement in terms of Laguerre}.
During this simplification, we take the limit
\begin{equation}
\gamma\left(\frac{1}{2\Delta^{2}}-\frac{1}{2}\right)\sim\gamma\nb_{\gkp}\gg1\,,
\end{equation}
relating $\g$ to $\nb_{\gkp}$. This yields the QEC matrix elements
\begin{equation}
\langle\m_{\gkp}^{\D}|E_{\ell}^{\dagger}E_{\lp}|\n_{\gkp}^{\D}\rangle\sim\frac{(\gamma\nb_{\gkp})^{\frac{\ell+\lp}{2}}}{(\gamma\nb_{\gkp}+1)^{\frac{\ell+\lp}{2}+1}}\sum_{\vec{n}\in\mathbb{Z}^{2}}e^{-\frac{(1-\gamma)}{2\gamma}|\Lambda_{\delta\mu}^{\vec{n}}|^{2}}e^{i\pi(n_{1}+\frac{\mu+\nu}{2})n_{2}}e^{-\frac{\Delta^{2}}{2}|\Lambda_{\delta\mu}^{\vec{n}}|^{2}}\langle\ell|D_{(\Lambda_{\delta\mu}^{\vec{n}})^{\star}/\sqrt{\gamma}}|\lp\rangle\,,
\end{equation}
where we can once again let $\D\rightarrow0$ to produce Eq.~\eqref{GKP error correction criteria}.
\end{widetext}

\end{widetext}

\bibliographystyle{apsrev4-1t}
\bibliography{C:/Users/Victor/Dropbox/THESIS/library}

\end{document}